\newif\ifjcp
\jcptrue  

\ifjcp
    \documentclass[twocolumn,aip,jcp,reprint,floatfix]{revtex4-2}
\else
    \documentclass[journal=jacsat,manuscript=article]{achemso}
    \setkeys{acs}{articletitle = true}
    \newcommand{\onlinecite}[1]{\hspace{-1 ex} \nocite{#1}\citenum{#1}}
    \SectionNumbersOn
\fi

\usepackage{float}
\usepackage{xcolor}
\usepackage{graphicx}
\usepackage{dcolumn}
\usepackage{bm}
\usepackage{amsmath}
\usepackage{amssymb}
\usepackage{braket}
\usepackage{dsfont} 
\usepackage{hyperref}
\usepackage{caption}
\usepackage{subcaption}
\usepackage{appendix}

\newcommand{\toadd}[1]{{#1}}

\newcommand{\mat}[1]{\mathbf{#1}}

\newcommand{\setinfo}{%
    \title{A `moment-conserving' reformulation of GW theory}
    \author{Charles J.~C. Scott}%
    \email{charles.j.scott@kcl.ac.uk}%
    \affiliation{Department of Physics, King's College London, Strand, London WC2R 2LS, U.K.}%
    \author{Oliver J. Backhouse}%
    \email{oliver.backhouse@kcl.ac.uk}%
    \affiliation{Department of Physics, King's College London, Strand, London WC2R 2LS, U.K.}%
    \author{George~H.~Booth}%
    \email{george.booth@kcl.ac.uk}%
    \affiliation{Department of Physics, King's College London, Strand, London WC2R 2LS, U.K.}%
    \date{\today}
}

\ifjcp
\else
\fi

\ifjcp\else
    \setinfo
\fi
\begin{document}
\ifjcp
    \setinfo
\fi

\begin{abstract}
We show how to construct an effective Hamiltonian whose dimension scales linearly with system size, and whose eigenvalues systematically approximate the excitation energies of $GW$ theory. This is achieved by rigorously expanding the self-energy in order to exactly conserve a desired number of frequency-independent moments of the self-energy dynamics.
Recasting $GW$ in this way admits a low-scaling $\mathcal{O}[N^4]$ approach to build and solve this Hamiltonian, with a proposal to reduce this further to $\mathcal{O}[N^3]$. This relies on exposing a novel recursive framework for the density response moments of the random phase approximation (RPA), where the efficient calculation of its starting point mirrors the low-scaling approaches to compute RPA correlation energies. The frequency integration of $GW$ which distinguishes so many different $GW$ variants can be performed without approximation directly in this moment representation. Furthermore, the solution to the Dyson equation can be performed exactly, avoiding analytic continuation, diagonal approximations or iterative solutions to the quasiparticle equation, with the full-frequency spectrum obtained from the complete solution of this effective static Hamiltonian. We show how this approach converges rapidly with respect to the order of the conserved self-energy moments, and is applied across the $GW100$ benchmark dataset to obtain accurate $GW$ spectra in comparison to traditional implementations. We also show the ability to systematically converge all-electron full-frequency spectra and high-energy features beyond frontier excitations, as well as avoiding discontinuities in the spectrum which afflict many other $GW$ approaches.
\end{abstract}

\ifjcp
    \maketitle
\fi

\section{Introduction}
Despite the phenomenal success of density functional theory (DFT) in electronic structure, its standard approach is both conceptually and (often) practically ill-suited for an accurate description of the energy levels in a material or chemical system \cite{doi:10.1021/cr200107z}. These quantities are however essential for predictions of fundamental bandgaps and other charged excitation properties which govern the photo-dynamics, transport and response properties of a system. Into this, {\em GW} theory has grown in popularity, first for materials and more recently for molecular systems,
as a post-mean-field approach to obtain charged excitation spectra in a principled diagrammatic fashion, free from empiricism\cite{Onida2002, Hybertsen1985, Aryasetiawan1998, Hedin1999, Aulbur2000, Friedrich2006, Kutepov2009, Ke2011, Bruneval2012, VanSetten2013, VanSetten2015, Reining2017, Golze2019}. 

The $GW$ approach is based on Hedin's equations\cite{Hedin1965, Hedin1970}, and in its most common formulation builds a self-energy to dress a reference description of the quasi-particles of a system (generally from DFT or Hartree--Fock (HF)) with an infinite resummation of all `bubble' diagrams. These diagrams make up the random phase approximation (RPA)\cite{Langreth1977, Chong1995, Heselmann2010, Ren2012},
and physically describes all collective quantum charge fluctuations in the electron density from the reference state arising from their correlated mutual Coulomb repulsion. This dynamically screens the effective interaction between the constituent quasi-particles of the system, whose physics generally dominates in small gapped semi-conducting systems. The use of this RPA screened interaction in $GW$ has therefore become widespread, correcting many of the failures of DFT for spectral properties.

At the core of $GW$ theory is a convolution, between the Green's function of the system, $G(\omega)$, and the screened Coulomb interaction $W_p(\omega)$, obtained (in general) at the RPA level of theory. This provides the dynamical part of the self-energy, $\Sigma(\omega)$, formally written as $\Sigma(\omega)= (i/2\pi) \int d \omega' e^{i \eta \omega'} G(\omega + \omega') W_p (\omega')$. There are many different variants of $GW$ theory\cite{Leeuwen09, Hybertsen1985, Reining2017, Golze2019, Foerster2011, Bruneval2013, PhysRevB.94.165109, Knight2016, Maggio2017, Bruneval2021, vonBarth1996, Holm1998, Schone1998, GarciaGonzalez2001, Faleev2004, VanSchilfgaarde2005, Stan2006, Kotani2007, Shishkin2007, Caruso2012, Bruneval2016, Kaplan2016, Jin2019, Duchemin2020, Duchemin2021, PhysRevB.106.235104,Ren_2012},
which primarily differ due to i) the choice (or absence) of self-consistency conditions on $G(\omega)$ and/or $W_p(\omega)$\cite{Leeuwen09}, ii) the approach to find the quasi-particle energies once $\Sigma(\omega)$ is obtained (i.e. application of Dyson's equation), and iii) different approximations to perform the frequency integration in the convolution itself. A numerically exact formulation of this convolution entails an $\mathcal{O}[N^6]$ scaling step, required to find the entire set of poles in $W_p(\omega)$ from the RPA. However, there are a number of techniques to approximate this frequency integration which can reduce this scaling (generally down to $\mathcal{O}[N^3-N^4]$) based on plasmon pole approximations, analytic continuation, contour deformation and explicit grid resolved approaches for the dynamics of these quantities, amongst others. All of these approaches compress and approximate the dynamical resolution of the key quantities in order to simplify the resulting convolutional integral.

Another key difference between the approaches rests on how the quasi-particle energies are updated from their mean-field molecular orbital (MO) energy starting point, once the self-energy has been constructed. This is formally an application of Dyson's equation, but is commonly approximated via a self-consistent solution to a quasi-particle (QP) equation entailing a diagonal approximation to the self-energy. This is valid when the quasiparticle energies are far from the self-energy poles, thereby asserting that the $GW$ largely just provides a shift in the original MO energies, rather than introducing significant quasiparticle renormalization, additional satellite peaks from state splitting or relaxation of the mean-field electron density. These assumptions can however break down (especially in more correlated systems), while the numerical solution of the QP equation can also converge to different (`spurious') results based on the specifics of how it is solved. A thorough study of the discrepancies due to different approaches to both the QP equation solution and the frequency integral in the convolution can be found in Ref.~\nocite{VanSetten2015}\citenum{VanSetten2015}.

In this work, we introduce a new approach to this frequency integration and reformulate $GW$ theory with a number of desirable properties, while retaining a low scaling. The key step is that the self-energy is not represented as an explicitly dynamical quantity, but instead in terms of a series of static matrices representing the \emph{moments} of its frequency-dependence up to a given order. These can be directly obtained, and from them a compressed representation of the full self-energy can be algebraically constructed which only has a number of poles which scales linearly with system size, but nonetheless exactly preserves the moment distribution of the exact $GW$ self-energy dynamics up to the desired order\cite{PhysRevB.44.13356, Backhouse2020b, Backhouse2021, Backhouse2022, Sriluckshmy2021}. This order can be systematically increased to more finely resolve the full dynamical dependence of the $GW$ self-energy. The dynamical information is therefore implicitly recast into a small number of {\em static} matrices, each of which can be obtained in $\mathcal{O}[N^4]$ time (with a proposed $\mathcal{O}[N^3]$ algorithm also given). This removes the need for the definition of any frequency or time grids in which to resolve dynamical quantities, spectral broadening, finite temperatures, Fourier transforms or analytic continuation, with all dynamics implicitly represented by this series of static quantities. 

Furthermore, once these spectral moments of the self-energy are obtained, the QP equation and restrictions to a diagonal self-energy representation can be entirely removed, with an exact application of Dyson's equation possible in this `moment' representation. This leads to the self-energy represented as a small number of explicit poles at specific energies which taken together have exactly the moment distribution in their dynamics as described. This allows for a simple construction of the full frequency dependence of the resulting quasi-particle spectrum (including any additional emergent satellite structures from the correlations) via diagonalization in an `upfolded' and explicit Hamiltonian representation. Moment expansions have a long history in the representation of dynamical quantities, with use in numerical approaches \cite{HAYDOCK198011,PhysRevB.44.13356}, characterizing sum rules \cite{ADACHI1988445,Karlsson2016PartialSA}, and as physical observables in their own right \cite{doi:10.1080/00268976.2010.508753, VANCAILLIE2000446, KALUGINA201620}.

In Sec.~\ref{sec:Mom_GW} we show how these spectral moments of the self-energy can be directly constructed from moments of the Green's function and the two-point density-density response function from RPA, without any further approximation. We show how these can be used to directly obtain the full-frequency $GW$ spectrum without the requirement of an explicit grid. In Sec.~\ref{sec:DD_RPA} we show how the RPA can be fully reformulated as a series expansion of moments of the \toadd{density-density (dd)} response, and in Sec.~\ref{sec:EfficientEval} show how they can be efficiently obtained in $\mathcal{O}[N^4]$ cost, based on ideas from the seminal work in 2010 by Furche and collaborators on low-scaling approaches for the RPA correlation energy \cite{Furche2010}. Furthermore, in Sec.~\ref{sec:CubicScaling} we propose an approach to further reduce the scaling of the whole algorithm directly (rather than asymptotically) to cubic cost with system size, without invoking screening or locality assumptions. We then apply the approach in Sec.~\ref{sec:Results} to the commonly used molecular $GW100$ test set frequently used to benchmark $GW$ implementations, demonstrating a rapid convergence of the moment expansion, and accurate and efficient results across this test set for the $G_0W_0$ level of theory.

\section{Moment-truncated GW theory} \label{sec:Mom_GW}

In $GW$ theory, the dynamical part of the self-energy, obtained as the convolution of the $G(\omega)$ and $W_p(\omega)$, can be formally expanded as a sum over the $\mathcal{O}[N^2]$ neutral excitations of RPA theory (representing the poles of the screened Coulomb interaction) and the charged excitations of the reference Green's function. In the absence of self-consistency (the most common `$G_0W_0$' formulation of the method, which we exclusively consider in this work), the Green's function is just given from the $\mathcal{O}[N]$ mean-field molecular orbital energies. This allows the self-energy to be explicitly evaluated in the frequency-domain as
\begin{align}
\Sigma_{pq}&(\omega)=\sum_{\nu} \sum_{ia,jb,k} \frac{ (pk|ia) (X_{ia}^{\nu} + Y_{ia}^{\nu})(X_{jb}^{\nu} + Y_{jb}^{\nu}) (qk|jb)}{\omega - (\epsilon_k - \Omega_{\nu})-i0^+} \nonumber \\
&+ \sum_{\nu} \sum_{ia,jb,c} \frac{ (pc|ia) (X_{ia}^{\nu} + Y_{ia}^{\nu})(X_{jb}^{\nu} + Y_{jb}^{\nu}) (qc|jb)}{\omega - (\epsilon_c + \Omega_{\nu})+i0^+} . \label{eq:fullSE}
\end{align}
In these expressions, $\Omega_\nu$ are the neutral excitation energies from RPA theory, and $X_{ia}^\nu$ and $Y_{ia}^\nu$ are the excitation and de-excitation amplitude components associated with this excitation. These are expanded in the basis of hole and particle spin-orbitals of the reference mean-field, represented by the indices $i,j,k$ ($a, b, c$) of dimension $o$ ($v$) respectively, and with orbital energies denoted by $\epsilon_x$.  The bare two-electron integrals are denoted by $(pk|ia)$ in standard Mulliken (`chemists') notation, which are therefore screened by the RPA reducible density response. More details on these quantities are given in Sec.~\ref{sec:DD_RPA}. The first term in Eq.~\ref{eq:fullSE} therefore represents the `lesser' part, and the second term the `greater' part of the full $G_0W_0$ self-energy. 

Exact evaluation of the RPA excitations ($\Omega_\nu$) scales as $\mathcal{O}[N^6]$, rendering it unsuitable for large-scale implementations. However, in this work, we are only interested in evaluating the spectral moments of the resulting self-energy, and finding a resulting compressed representation of the self-energy with fewer poles, but which by construction matches the spectral moments up to a desired order. These frequency-independent spectral moments are defined separately for the greater and lesser parts, and represent the $n^\mathrm{th}$-order moments of the resulting dynamical distributions, as
\begin{align}
\Sigma^{(n,<)}_{pq} &= -\frac{1}{\pi} \int_{-\infty}^{\mu} \mathrm{Im}[\Sigma(\omega)_{pq}] \omega^n d\omega \\
&= (-1)^n 
 \left. \frac{d^n \Sigma(\tau)_{pq}}{d \tau^n} \right|_{\tau=0^+} ,
\end{align}
and similarly
\begin{align}
\Sigma^{(n,>)}_{pq} &= \frac{1}{\pi} \int_{\mu}^{\infty} \mathrm{Im}[\Sigma(\omega)_{pq}] \omega^n d\omega \\
&= (-1)^n \left. \frac{d^n \Sigma(\tau)_{pq}}{d \tau^n} \right|_{\tau=0^-} ,
\end{align}
where $\mu$ represents the chemical potential of the system.
This exposes the relationship of these spectral moments to a Taylor expansion of the short-time dynamics of the greater and lesser parts of the self-energy, with the moments defining the integrated weight, mean, variance, skew, and higher-order moments of the dynamical distribution of each element of the self-energy in the frequency domain.

Applied to the $GW$ self-energy of Eq.~\ref{eq:fullSE}, the moments can be constructed as 
\begin{align}
\Sigma_{pq}^{(n,<)} = \sum_{\nu} & \sum_{ia,jb} \sum_k \left[ (pk|ia)(X_{ia}^\nu + Y_{ia}^{\nu}) \right. \nonumber \\
	& \left. (\epsilon_k-\Omega_{\nu})^n (X_{jb}^\nu + Y_{jb}^\nu) (qk|jb) \right] , \label{eq:SEMom1_less}
\end{align}
\begin{align}
\Sigma_{pq}^{(n,>)} = \sum_{\nu} & \sum_{ia,jb} \sum_c \left[ (pc|ia)(X_{ia}^\nu + Y_{ia}^{\nu}) \right. \nonumber \\
	&\left. (\epsilon_c+\Omega_{\nu})^n (X_{jb}^\nu + Y_{jb}^\nu) (qc|jb) \right] . \label{eq:SEMom1_great}
\end{align}
The moment distribution of a convolution of two quantities can be expressed via the binomial theorem as a sum of products of the moments of the individual quantities. This enables us to split apart the expressions above into products of the individual Green's function and density-density response moments. Defining the $n^\mathrm{th}$-order spectral moments of the RPA density-response, summed over both particle-hole excitation and de-excitation components, as
\begin{equation}
\eta^{(n)}_{ia,jb} = \sum_\nu (X_{ia}^\nu + Y_{ia}^\nu) \Omega_{\nu}^n (X_{jb}^{\nu} + Y_{jb}^{\nu}) , \label{eq:eta_def}
\end{equation}
we can rewrite Eqs.~\ref{eq:SEMom1_less} and \ref{eq:SEMom1_great} as
\begin{align}
\Sigma_{pq}^{(n,<)}&=\sum_{ia,jb,k} \sum_{t=0}^n \binom{n}{t} (-1)^{t} \epsilon_{k}^{n-t} (pk|ia) \eta_{ia,jb}^{(t)} (qk|jb) \label{eq:SEMoms2_less} \\
\Sigma_{pq}^{(n,>)}&=\sum_{ia,jb,c} \sum_{t=0}^n \binom{n}{t} \epsilon_{c}^{n-t} (pc|ia) \eta_{ia,jb}^{(t)} (qc|jb) . \label{eq:SEMoms2_great}
\end{align}
Evaluating the self-energy spectral moments of Eqs.~\ref{eq:SEMoms2_less}-\ref{eq:SEMoms2_great} up to a desired order $n$, represents the central step of the proposed `moment-conserving' $GW$ formulation, defining the convolution between $G(\omega)$ and $W_p(\omega)$ in this moment expansion of the dynamics. In Sec.~\ref{sec:DD_RPA} we show how the RPA can be reformulated to define specific constraints on the relations between different orders of the RPA density-response moments, $\eta^{(n)}_{ia,jb}$. These relations are subsequently used in Sec.~\ref{sec:EfficientEval} to demonstrate how the self-energy moments can be evaluated in $\mathcal{O}[N^4]$ scaling, with Sec.~\ref{sec:CubicScaling} going further to propose a cubic scaling algorithm for their evaluation (and therefore full $GW$ algorithm). In addition to these moments representing the dynamical part of the self-energy, we also require a static (exchange) part of the self-energy, $\mat{\Sigma}_\infty$, which can be calculated as
\begin{equation}
\mat{\Sigma}_{\infty} = \mat{K}[\mat{D}] - \mat{V}_{\mathrm{xc}} ,
\end{equation}
where $\mat{K}[\mat{D}]$ is the exchange matrix evaluated with the reference density matrix. This reference density matrix is found from a prior mean-field calculation via a self-consistent Fock or Kohn--Sham single-particle Hamiltonian, $\mat{f}[\mat{D}]$, with $\mat{V}_{\mathrm{xc}}$ being the exchange-correlation potential used in $\mat{f}$. Note that for a Hartree--Fock reference, this static self-energy contribution is zero. 

\subsection{Full $GW$ spectrum from self-energy moments}

Once the moments of the self-energy are found, it is necessary to obtain the resulting dressed $GW$ excitations and spectrum. While this is formally an application of Dyson's equation, the most common approach is to find each $GW$ excitation explicitly via a self-consistent solution (or linearized approximation) of the quasiparticle equation, while assuming a diagonal self-energy in the MO basis\cite{VanSetten2015}. This assumption neglects physical effects due to electron density relaxation and mixing or splitting of quasiparticle states in more strongly correlated systems. In this work, we allow for an exact invocation of Dyson's equation, which can be achieved straightforwardly in this moment domain of the effective dynamics, allowing extraction of quasi-particle weights associated with transitions, and a full matrix-valued form of the resulting $GW$ Green's function over all frequencies, with all poles obtained analytically without artificial broadening.

This is achieved by constructing an `upfolded' representation of an effective Hamiltonian, consisting of coupling between a physical and `auxiliary' space (with the latter describing the effect of the moment-truncated self-energy). Specifically, we seek an effective static Hamiltonian,
\begin{equation}
\mat{\tilde{H}} = 
\begin{bmatrix}
\mat{f} + \mat{\Sigma}_{\infty} & \mat{\tilde{W}} \\
\mat{\tilde{W}}^\dagger & \mat{\tilde{d}}
\end{bmatrix} , \label{eq:effH}
\end{equation}
whose eigenvalues are the charged excitation energies at the level of the moment-truncated $GW$, with quasiparticle weights and Dyson orbitals explicitly obtained from the projection of the corresponding eigenvectors into the physical (MO) space. The full Green's function can therefore be constructed as
\begin{equation}
\mat{G}(\omega) = \left(\omega \mat{I} - \mat{f} + \mat{\Sigma}_\infty - \mat{\tilde{W}}(\omega \mat{I} - \mat{\tilde{d}})^{-1} \mat{\tilde{W}}^\dagger \right)^{-1} . \label{eq:GFfromEffH}
\end{equation}
Such upfolded representations have been considered previously in diagrammatic theories, in a recasting of GF2 theory in terms of its moments \cite{Backhouse2020a, Backhouse2020b, Backhouse2021} as well as more recently to $GW$ amongst others\cite{Rebolini2016, Loos2018, doi:10.1063/5.0130837, Bintrim2021, Bintrim2022, Backhouse2022, Tolle2022}.
For `exact' $G_0W_0$, this auxiliary space (i.e. the dimension of $\mat{d}$) must scale as $\mathcal{O}[N^3]$ \cite{Bintrim2021,doi:10.1063/5.0130837}.
However, in the moment truncation, $\mat{\tilde{W}}$ and $\mat{\tilde{d}}$ can be directly constructed such that their effect exactly matches that of a truncated set of conserved $GW$ self-energy moments (separately in the particle and hole sectors), yet rigorously scales in dimension as $\mathcal{O}[n N]$, where $n$ is the number of conserved self-energy moments. This allows for a complete diagonalization of $\mat{\tilde{H}}$, obtaining all excitations in a single shot, and a reconstruction of the full $GW$ Green's function from its Lehmann representation in $\mathcal{O}[(nN)^3]$ computational effort, avoiding the need for any grids or iterative solutions once $\mat{\tilde{H}}$ is found.

To find this effective upfolded representation of the moment-conserving dynamics, we modify the block Lanczos procedure to ensure the construction a $\mat{\tilde{H}}$ of minimal size, whose effective hole and particle self-energy moments 
exactly match the ones from Eqs.~\ref{eq:SEMoms2_less}-\ref{eq:SEMoms2_great}. We first proceed by splitting the auxiliary space into a space denoting the effect of the hole (lesser) and particle (greater) self-energy, and consider each in turn. Focusing on the lesser self-energy, we can construct an {\em exact} upfolded self-energy representation\cite{Tolle2022}, via inspection from Eq.~\ref{eq:fullSE}, with
\begin{align}
\mat{W}_{p,k \nu} &= \sum_{ia} (pk|ia)(X_{ia}^{\nu} + Y_{ia}^{\nu}) \label{eq:ExactCouplings} \\
\mat{d}_{k\nu, l\nu'} &= (\epsilon_k - \Omega_\nu) \delta_{k,l} \delta_{\nu,\nu'} ,
\end{align}
where we remove the tilde above upfolded auxiliary quantities when denoting the exact upfolded $GW$ self-energy components.
We now consider the projection of the exact $GW$ upfolded matrix representation into a truncated block tridiagonal form, as
\begin{align}
    \nonumber
    \tilde{\mathbf{H}}_{\mathrm{tri}}
    &=
    \tilde{\mathbf{q}}^{(j), \dagger}
    \begin{bmatrix}
        \mathbf{f} + \boldsymbol{\Sigma}_{\infty} & \mathbf{W} \\
        \mathbf{W}^{\dagger} & \mathbf{d}
    \end{bmatrix}
    \tilde{\mathbf{q}}^{(j)}
    \\
    \label{eq:block_lanczos_hamiltonian}
    &=
    \begin{bmatrix}
        \mathbf{f} + \boldsymbol{\Sigma}_{\infty} & \mathbf{L} & & & & \mathbf{0} \\
        \mathbf{L}^{\dagger} & \mathbf{M}_{1} & \mathbf{C}_{1} & & & \\
        & \mathbf{C}_{1}^{\dagger} & \mathbf{M}_{2} & \mathbf{C}_{2} & & \\
        & & \mathbf{C}_{2}^{\dagger} & \mathbf{M}_{3} & \ddots & \\
        & & & \ddots & \ddots & \mathbf{C}_{j-1} \\
        \mathbf{0} & & & & \mathbf{C}_{j-1}^{\dagger} & \mathbf{M}_{j}
    \end{bmatrix}
    ,
\end{align}
where we define $\tilde{\mathbf{q}}^{(j)}$ as
\begin{align}
    \label{eq:block_lanczos_vectors}
    \tilde{\mathbf{q}}^{(j)}
    &=
    \begin{bmatrix}
        \mathbf{I} & \mathbf{0} \\
        \mathbf{0} & \mathbf{q}^{(j)}
    \end{bmatrix}
    .
\end{align}
The $\mathbf{q}^{(j)}$ are block Lanczos vectors of depth $j$, which form a recursive Krylov space as $
\mathbf{q}^{(j)}
=
\begin{bmatrix}
    \mathbf{q}_{1} &
    \mathbf{q}_{2} &
    \cdots &
    \mathbf{q}_{j}
\end{bmatrix}
$, ensuring that when taken together, they project to a block tridiagonal representation of the upfolded self-energy with $j$ on-diagonal blocks as shown.
The action of this block Lanczos tridiagonalization of the upfolded (hole or particle) self-energy is to exactly conserve these spectral moments of the self-energy\cite{Meyer1989, Weikert1996}. This block tridiagonal representation is equivalent to a truncated continued fraction form\cite{HAYDOCK198011,Haydock_1985}, widely used in the representation of dynamical quantities\cite{PhysRevB.46.15812,doi:10.1080/13642810208222682}, and even as an expansion previously considered within $GW$ theory \cite{PhysRevB.44.13356}. We therefore seek to reformulate the Lanczos recursion in terms of just these moments, rather than the action of the full upfolded Hamiltonian which we seek to avoid due to its scaling.

The initial couplings $\mathbf{L}$ and Lanczos vectors $\mathbf{q}_{1}$
can be found via a QR factorisation of the exact $GW$ couplings $\mathbf{W}$, as
\begin{align}
    \label{eq:coupling_qr}
    \mathbf{W}^{\dagger}
    &=
    \mathbf{q}_{1}
    \mathbf{L}^{\dagger}
    .
\end{align}
However, this will scale poorly, and so we can rewrite this to directly compute $\mat{L}$ from the computed self-energy moments, rather than requiring manipulations of the full auxiliary space. Via the Cholesky QR algorithm\cite{Fukaya2014, Fukaya2020}, we can relate $\mat{L}$ to the zeroth order self-energy moment as
\begin{align}
    \label{eq:copuling_qr_cholesky}
    \mathbf{L}^{\dagger}
    &=
    \left(
        \mathbf{W}
        \mathbf{W}^{\dagger}
    \right)^{\frac{1}{2}}
    =
    \left(
        \boldsymbol{\Sigma}^{(0)}
    \right)^{\frac{1}{2}}
    ,
\end{align}
where the indication of the sector of the self-energy has been dropped, with this process considered independently for the hole and particle (lesser and greater respectively) parts of the self-energy.
The initial Lanczos vector can then be computed as
$\mathbf{q}_{1} = \mathbf{W}^{\dagger} \mathbf{L}^{-1, \dagger}$.
Subsequent block Lanczos vectors can then be defined according to the standard
three-term recurrence
\begin{align}
    \label{eq:block_lanczos_three_term_recurrence}
    \mathbf{q}_{i+1}
    \mathbf{C}_{i}^{\dagger}
    &=
    \left[
        \mathbf{d} \mathbf{q}_{i}
        -
        \mathbf{q}_{i} \mathbf{M}_{i}
        -
        \mathbf{q}_{i-1} \mathbf{C}_{i-1}
    \right]
\end{align}
where the on-diagonal blocks are defined as
\begin{align}
    \label{eq:block_lanczos_on_diagonal_blocks}
    \mathbf{M}_{i}
    &=
    \mathbf{q}_{i}^{\dagger}
    \mathbf{d}
    \mathbf{q}_{i}
    .
\end{align}
In order to recast this process in terms of the self-energy moments directly,
we wish to express the block Lanczos recurrence in terms of the inner space
of the Lanczos vectors rather than spanning the large auxiliary space.
The choice of initial vectors in
Eq.~\ref{eq:coupling_qr}
permits the definition
\begin{align}
    \nonumber
    \mathbf{S}_{1, 1}^{(n)}
    &=
    \mathbf{q}_{1}^{\dagger}
    \mathbf{d}^{n}
    \mathbf{q}_{1}
    \\
    \label{eq:block_lanczos_recurrence_init}
    &=
    \mathbf{L}^{-1}
    \boldsymbol{\Sigma}^{(n)}
    \mathbf{L}^{-1, \dagger}
    ,
\end{align}
where the subscript indices on $\mat{S}$ indicate the projection into the block Lanczos space on the left and the right respectively. These provide the initialisation of the recurrence terms,
and have a dimension which scales linearly with system size (the same as the input self-energy moments). These $\mat{\Sigma}^{(n)}$ matrices are therefore the input to the procedure, defined by Eqs.~\ref{eq:SEMoms2_less}-\ref{eq:SEMoms2_great}.
One can then use the definition of the three-term recurrence in
Eq.~\ref{eq:block_lanczos_three_term_recurrence}
to express all definitions in terms of these moments, without formal reference to the large auxiliary space quantities ($\mat{d}$ or $\mat{W}$), as
\onecolumngrid
\begin{align}
    \label{eq:block_lanczos_recurrence_single}
    \mathbf{S}_{i+1, i}^{(n)}
    &=
    \mathbf{q}_{i+1}^{\dagger}
    \mathbf{d}^{n}
    \mathbf{q}_{i}
    =
    \mathbf{C}_{i}^{-1}
    \left[
        \mathbf{S}_{i, i}^{(n+1)}
        -
        \mathbf{M}_{i}
        \mathbf{S}_{i, i}^{(n)}
        -
        \mathbf{C}_{i-1}^{\dagger}
        \mathbf{S}_{i-1, i}^{(n)}
    \right]
    ,
    \\
    \nonumber
    \mathbf{S}_{i+1, i+1}^{(n)}
    &=
    \mathbf{q}_{i+1}^{\dagger}
    \mathbf{d}^{n}
    \mathbf{q}_{i+1}
    \\
    \label{eq:block_lanczos_recurrence_double}
    &=
    \mathbf{C}_{i}^{-1}
    \left[
        \mathbf{S}_{i, i}^{(n+2)}
        +
        \mathbf{M}_{i}
        \mathbf{S}_{i, i}^{(n)}
        \mathbf{M}_{i}
        +
        \mathbf{C}_{i-1}^{\dagger}
        \mathbf{S}_{i-1, i-1}^{(n)}
        \mathbf{C}_{i-1}
        -
        P(
            \mathbf{S}_{i, i}^{(n+1)}
            \mathbf{M}_{i}
        )
        +
        P(
            \mathbf{M}_{i}
            \mathbf{S}_{i, i-1}^{(n)}
            \mathbf{C}_{i-1}
        )
        -
        P(
            \mathbf{S}_{i, i-1}^{(n+1)}
            \mathbf{C}_{i-1}
        )
    \right]
    \mathbf{C}_{i}^{-1, \dagger}
    ,
\end{align}
\twocolumngrid
where the permutation operator $P$ is defined as
$P(\mathbf{Z}) = \mathbf{Z} + \mathbf{Z}^{\dagger}$.
For a Hermitian theory we can write
$\mathbf{S}_{i, j} = \mathbf{S}_{j, i}^{\dagger}$,
and assume the zeroth order Lanczos vector to be zero i.e.\
$
\mathbf{S}_{i, 0}^{(n)}
=
\mathbf{S}_{0, j}^{(n)}
=
\mathbf{S}_{0, 0}^{(n)}
=
\mathbf{0}
$.
Additionally,
the orthogonality of the Lanczos vectors requires that
$\mathbf{S}_{i, j}^{(0)} = \delta_{ij} \mathbf{I}$.
By considering again the Cholesky QR algorithm,
the off-diagonal $\mathbf{C}$ matrices can therefore be computed as
%
%
\begin{align}
    \mathbf{C}_{i}^{2} = &\left[ \mathbf{S}_{i, i}^{(2)} + \mathbf{M}_{i}^{2}
        + \mathbf{C}_{i-1}^{\dagger} \mathbf{C}_{i-1} \right. \nonumber \\ 
        & - \left. P(\mathbf{S}_{i, i}^{(1)} \mathbf{M}_{i}) -
        P(\mathbf{S}_{i, i-1}^{(1)} \mathbf{C}_{i-1}) \right] , \label{eq:block_lanczos_off_diagonal_blocks_recur}
\end{align}
%
%
and the on-diagonal $\mathbf{M}$ matrices can be found using
Eq.~\ref{eq:block_lanczos_on_diagonal_blocks}
\begin{align}
    \label{eq:block_lanczos_on_diagonal_blocks_recur}
    \mathbf{M}_{i}
    &=
    \mathbf{q}_{i}^{\dagger}
    \mathbf{d}
    \mathbf{q}_{i}
    =
    \mathbf{S}_{i, i}^{(1)}
    .
\end{align}
These recurrence relations allow the calculation of the on- and off-diagonal
blocks resulting in the block tridiagonal form of the Hamiltonian in
Eq.~\ref{eq:block_lanczos_hamiltonian}.
Despite the apparent complexity of the recurrence relations,
this algorithm contains no step scaling greater than
$\mathcal{O}[N^{3}]$, by eliminating the explicit reference to the full upfolded Hamiltonian, whilst still conserving the exact self-energy moments by construction. 

It can be seen from Eq.~\ref{eq:ExactCouplings} that the auxiliary space formally couples to both the hole and particle physical sectors (both occupied and virtual MOs). 
\toadd{To allow for this coupling (which is critical for generating effective higher order diagrams and to avoid a `non-Dyson' approximation\cite{Bintrim2021, Schirmer1998, Schirmer18, doi:10.1063/1.5032314, doi:10.1063/5.0023862, doi:10.1063/5.0130837, Dempwolff2019, Trofimov2005}),
the solution of the Dyson equation using the compressed self-energy,
i.e.\ diagonalisation of Eq.~\ref{eq:effH},
requires the combination of the the block tridiagonal Hamiltonian in Eq.~\ref{eq:block_lanczos_hamiltonian}
resulting from both the hole and particle self-energy moments.} This is constructed as
\begin{align}
    \label{eq:concatenated_h}
    \tilde{\mathbf{H}}
    &=
    \begin{bmatrix}
        \mathbf{f} + \boldsymbol{\Sigma}_{\infty} & \tilde{\mathbf{W}} \\
        \tilde{\mathbf{W}}^{\dagger} & \tilde{\mathbf{d}} \\
    \end{bmatrix}
    =
    \begin{bmatrix}
        \mathbf{f} + \boldsymbol{\Sigma}_{\infty} & \tilde{\mathbf{W}}^{<} & \tilde{\mathbf{W}}^{>} \\
        \tilde{\mathbf{W}}^{<, \dagger} & \tilde{\mathbf{d}}^{<} & \mathbf{0} \\
        \tilde{\mathbf{W}}^{>, \dagger} & \mathbf{0} & \tilde{\mathbf{d}}^{>} \\
    \end{bmatrix}
    ,
\end{align}
where $\tilde{\mathbf{W}}$ are equal to the $\mathbf{L}$ matrix padded by zeros
\begin{align}
    \label{eq:compressed_w}
    \tilde{\mathbf{W}}^{\lessgtr} =
    \begin{bmatrix}
        \mathbf{L}^{\lessgtr} & \mathbf{0} & \cdots & \mathbf{0}
    \end{bmatrix}
    ,
\end{align}
and $\tilde{\mathbf{d}}$ are defined by the block tridiagonal elements
\begin{align}
    \label{eq:compressed_d}
    \tilde{\mathbf{d}}^{\lessgtr}
    &=
    \begin{bmatrix}
        \mathbf{M}_{1}^{\lessgtr} & \mathbf{C}_{1}^{\lessgtr} & & \mathbf{0} \\
        \mathbf{C}_{1}^{\lessgtr, \dagger} & \mathbf{M}_{2}^{\lessgtr} & \ddots \\
        & \ddots & \ddots & \mathbf{C}_{j-1}^{\lessgtr} \\
        \mathbf{0} & & \mathbf{C}_{j-1}^{\lessgtr, \dagger} & \mathbf{M}_{j}^{\lessgtr}
    \end{bmatrix}
    .
\end{align}
This ensures conservation of both the separate hole and particle moments of the self-energy,
as well as conservation in the central moments according to their sum. 

The compressed Hamiltonian can be returned to a diagonal representation of the self-energy by diagonalising $\tilde{\mathbf{d}}$ and
appropriately rotating $\tilde{\mathbf{W}}$ into this basis.
The eigenvalues of $\tilde{\mathbf{H}}$ are moment-conserving approximations to those of the exact upfolded Hamiltonian,
and the corresponding eigenvectors $\mathbf{u}$ can be transformed into Dyson orbitals via $\mathbf{L} \mathbf{P} \mathbf{u}$,
where $\mathbf{P}$ is 
a projection into the physical space, and the $\mathbf{L}$ is required to transform the physical component of the eigenvectors back to the MO representation.
This process conserves exactly the first $2j$ hole and particle self-energy moments.
Commonly,
a notation referring to the number of iterations of the block Lanczos recurrence
$n_{\mathrm{iter}}$ is used;
in this notation the $n_{\mathrm{iter}} = 0$ calculation corresponds to the
inclusion of only a single on-diagonal block $\mathbf{M}_{1}$,
with modified couplings $\mathbf{L}$ to the physical space.
As such,
in this notation the number of conserved moments equals $2 n_{\mathrm{iter}} + 2$,
i.e.\ up to and including the $2 n_{\mathrm{iter}} + 1$ order moment.
This is the same number of moments as required as input to the recurrence relations,
and therefore the algorithm conserves all the moments used as input, which should be up to an odd order.
After $n_{\mathrm{iter}}$ applications of this algorithm to both the lesser and greater self-energy sectors,
this results in $N (2 n_\mathrm{iter} + 3)$ quasiparticle states (demonstrating the potential to capture satellite features with these additional poles).
As such,
application of this algorithm becomes theoretically equivalent to a full diagonalisation of the exact upfolded Hamiltonian in the limit of
$n_{\mathrm{iter}} \sim N^{2}$.

\section{Density response moments in the RPA} \label{sec:DD_RPA}

Having described the overall approach in Sec.~\ref{sec:Mom_GW}, what remains for a practical implementation is to ensure that the $GW$ self-energy moments described in Eqs.~\ref{eq:SEMoms2_less}-\ref{eq:SEMoms2_great} can be computed efficiently. As a first step towards this, in this section we show how the RPA can be motivated from the perspective of the two-point density-density (dd) response moments of Eq.~\ref{eq:eta_def}, which are central quantities to obtain in this approach to $GW$ theory. We find that we can reformulate RPA entirely in terms of these dd-moments of the system and a strict recursive form for their inter-relation\cite{PhysRevB.104.245114}. This recursive relation between the moments is a direct result of the fact that the RPA can be written as a quadratic Hamiltonian in Bosonic operators \cite{PhysRevB.1.471, PhysRev.178.1097, Lundqvist69, Hedin1999, Tolle2022}. This effectively ensures that all information required to build the 2-point RPA dd-response is contained in the first two spectral moments, analogous to how all the information on the density of states in mean-field theory (quadratic in Fermionic operators) is contained in the first two Green's function moments (i.e. the one-body density matrix and Fock matrix).

We start from the Casida formulation of RPA \cite{Hesselmann2011,doi:10.1146/annurev-physchem-040215-112308}, as a generalized eigenvalue decomposition
\begin{equation} \label{eq:Casida-eq}
\begin{bmatrix}
    \mat{A} & \mat{B} & \\
    \mat{-B} & \mat{-A} & 
\end{bmatrix} \begin{bmatrix} \mat{X} & \mat{Y} \\ \mat{Y} & \mat{X} \end{bmatrix} = 
\begin{bmatrix}
\mat{X} & \mat{Y} \\ \mat{Y} & \mat{X}
\end{bmatrix} \begin{bmatrix} \mat{\Omega} & \mat{0} \\ \mat{0} & -\mat{\Omega} \end{bmatrix} ,
\end{equation}
where the left and right eigenvectors form the biorthogonal set as
\begin{equation}
\begin{bmatrix}
\mat{X} & \mat{Y} \\ \mat{Y} & \mat{X}
\end{bmatrix}^T
\begin{bmatrix}
\mat{X} & \mat{Y} \\ \mat{-Y} & \mat{-X}
\end{bmatrix} = \begin{bmatrix}
\mat{I} & \mat{0} \\ \mat{0} & \mat{-I}
\end{bmatrix}. \label{eq:RPAbiorthog}
\end{equation}
This biorthogonality ensures an inverse relationship between $(\mat{X}+\mat{Y})$ and $(\mat{X}-\mat{Y})^T$, as
\begin{equation}
(\mat{X}+\mat{Y})(\mat{X}-\mat{Y})^T = (\mat{X}+\mat{Y})^T(\mat{X}-\mat{Y}) = \mat{I} \label{eq:inverse} .
\end{equation}
The $\mat{A}$ and $\mat{B}$ matrices are defined as
\begin{align}
A_{ia,jb} &= \left(\epsilon_a - \epsilon_i\right) \delta_{ij}\delta_{ab} + \mathcal{K}_{ia,bj}  \label{eq:Casida_def_A} \\
B_{ia,jb} &= \mathcal{K}_{ia,jb}. \label{eq:Casida_def_B}
\end{align}
Here, $\mathcal{K}$ is an interaction kernel which couples particle-hole excitations and de-excitations. In the traditional RPA (without second-order exchange), this coupling is taken to be the same for excitations, de-excitations and their coupling, given by the static, bare Coulomb interaction, $\mathcal{K}_{ia,jb} = (ia|jb) = \mathcal{K}_{ia,bj}$. Hole and particle orbital energies are given by $\epsilon_i$ and $\epsilon_a$ respectively, defining the irreducible polarizability of the system from the reference state in $\mat{A}$. Upon diagonalization, the eigenvectors defined by $X_{ia,\nu}$ and $Y_{ia,\nu}$ define the coefficients of the RPA excitations in the particle-hole and hole-particle basis, with energies $\Omega_{\nu}$, with $\mat{\Omega}$ therefore a diagonal matrix of the positive (neutral) RPA excitation energies.

These neutral excitations define the poles of the full RPA reducible density-density (dd) response function, which can be constructed as
\begin{equation}
\mat{\chi}(\omega) =
\begin{bmatrix}
\mat{X} & \mat{Y} \\
\mat{Y} & \mat{X}
\end{bmatrix}
\begin{bmatrix} \omega \mat{I}-\mat{\Omega} \hspace{10pt} & \mat{0} \\
\mat{0} \hspace{10pt} & -\omega \mat{I} - \mat{\Omega} \end{bmatrix}^{-1}
\begin{bmatrix}
\mat{X} & \mat{Y} \\
\mat{Y} & \mat{X}
\end{bmatrix}^T. \label{eq:rpaddresponse}
\end{equation}
Note that this matrix is formally equivalent to the alternative directly dynamical construction from $\mat{\chi}(\omega) = (\mat{P}(\omega)^{-1} - \mathcal{K})^{-1}$, where $\mat{P}(\omega)$ is the irreducible polarizability of the reference state. Considering the positive-frequency part of the dd-response (noting that the negative frequency part is symmetric due to the bosonic-like symmetry of Eq.~\ref{eq:rpaddresponse}), we can write a more compact form of the dd-response as
\begin{equation}
\eta(\omega) = (\mat{X}+\mat{Y}) (\omega \mat{I} - \mat{\Omega})^{-1} (\mat{X}+\mat{Y})^T , \label{eq:dd-res}
\end{equation}
which sums contributions from particle-hole and hole-particle fluctuations together, and from which optical properties such as dynamic polarizabilities can be computed \cite{doi:10.1063/1.456413}.
However, in this work we are interested in the order-by-order moments of the spectral distribution of Eq.~\eqref{eq:dd-res} over all RPA excitation energies, which is given as
\begin{equation}
\eta^{(n)}_{ia,jb} = -\frac{1}{\pi} \int_{0}^{\infty} \textrm{Im}[\eta_{ia,jb}(\omega)] \omega^n d\omega . \label{eq:eta_from_dd}
\end{equation}
The non-negative integer index $n$ denotes the order of this static dd spectral moment information.
Performing this integration results in the direct construction of the dd-moments as defined in Eq.~\ref{eq:eta_def}, which can be written more compactly in matrix form as
\begin{equation}
\mat{\eta}^{(n)} = (\mat{X}+\mat{Y}) \mat{\Omega}^{n} (\mat{X}+\mat{Y})^T , \label{eq:formal_moms}
\end{equation}
and constitutes a central object of interest in this work, required for the GW self-energy moment construction of Eqs.~\ref{eq:SEMoms2_less}-\ref{eq:SEMoms2_great}.
\footnote{
We also note that a related expansion and recursive relation can also be constructed in terms of the inverse of these moments courtesy of the equivalence $(\mat{X+Y})^{-1}=(\mat{X - Y})^T$, that is
$(\mat{\eta}^{(n)})^{-1} = (\mat{X}-\mat{Y}) \mat{\Omega}^{-n} (\mat{X}-\mat{Y})^T$.
}

We now show that the RPA can be entirely reformulated in terms of the dd-moments, Eq.~\eqref{eq:formal_moms}, without loss of information, and expose constraints on the relationship between the moments of different order at the RPA level. Firstly, we note that from the definition of the original eigendecomposition of Eq.~\ref{eq:Casida-eq}, along with insertion of a resolution of the identity via Eq.~\ref{eq:inverse}, we find a relation between the first two dd-moments as
\begin{align}
\mat{\eta}^{(1)} &= (\mat{X} + \mat{Y})\mat{\Omega}(\mat{X}+\mat{Y})^T = (\mat{A} - \mat{B}) , \label{eq:one_mom} \\
&= \mat{\eta}^{(0)} (\mat{A}+\mat{B}) \mat{\eta}^{(0)} , \label{eq:zero_mom}
\end{align}
noting that $(\mat{A}+\mat{B})=(\mat{X}-\mat{Y})\mat{\Omega}(\mat{X}-\mat{Y})^T$.
By taking the sum and difference of the two Casida equations of Eq.~\ref{eq:Casida-eq}, we also find
\begin{align}
\left(\mat{A}+\mat{B}\right)\left(\mat{X}+\mat{Y}\right) &= \left(\mat{X}-\mat{Y}\right)\mat{\Omega} \label{eq:ApB_def} \\
\left(\mat{A}-\mat{B}\right)\left(\mat{X}-\mat{Y}\right) &= \left(\mat{X}+\mat{Y}\right)\mat{\Omega} \label{eq:AmB_def} ,
\end{align}
from which an equation for the square of the RPA excitations can be found as
\begin{equation}
\left(\mat{A}-\mat{B}\right)\left(\mat{A}+\mat{B}\right)\left(\mat{X}+\mat{Y}\right)=\left(\mat{X}+\mat{Y}\right)\mat{\Omega}^2,
\end{equation}
which has previously appeared in the RPA literature \cite{Furche2001}. By right-multiplying by $(\mat{X}+\mat{Y})^T$, this leads to a relation between the zeroth and second dd-moment, as
\begin{equation}
(\mat{A}-\mat{B})(\mat{A}+\mat{B}) \mat{\eta}^{(0)} = \mat{\eta}^{(2)}.
\end{equation}
The above equations can be further generalized as a recursive relation to generate all higher-order moments from lower-order ones, as
\begin{align}
\mat{\eta}^{(m)} &= (\mat{A} - \mat{B})(\mat{A} + \mat{B}) \mat{\eta}^{(m-2)} \label{eq:mom_recursion_main} \\
&= [\mat{\eta}^{(0)} (\mat{A} + \mat{B})]^m \mat{\eta}^{(0)} . \label{eq:mom_recursion}
\end{align}
While these are important relations in themselves, they also illustrate that all the RPA excitations and weights in the dd-response of Eq.~\ref{eq:dd-res} are implicitly accessible without requiring the explicit solution to the Casida equation ($\mat{X}$, $\mat{Y}$ and $\mat{\Omega}$ matrices). This reformulation just requires knowledge of $\mat{\eta}^{(0)}$ as the central variable (which can be defined independently of the original equations via Eq.~\ref{eq:zero_mom}), the $\mat{A}$ and $\mat{B}$ matrices defining the system and their interactions, and the recursive relations of Eq.~\ref{eq:mom_recursion}. 
As an aside, the Tamm--Dancoff approximation (TDA) sets $\mat{B}=0$, which dramatically simplifies the resulting expressions due to the lack of correlation in the ground state, with $\eta^{(0)}=\mat{I}$, and $\eta^{(n)}=\mat{A}^n$. This reflects that the 2-RDM in the TDA is equivalent to that of mean-field theory. Finally, we note in passing that the ground state correlation energy from the RPA can be similarly formulated in terms of the zeroth-order dd-moment, as
\begin{equation}
E_{\mathrm{corr}}^{\mathrm{RPA}} = \frac{1}{2} \mathrm{Tr}[\mat{\eta}^{(0)} (\mat{A} + \mat{B}) - \mat{A}] .
\end{equation}
Related expressions for the RPA correlation energy can be found found in Eqs.~39-40 of Ref.~\onlinecite{angyan2011correlation} in terms of the quantities $\mat{\eta}^{(0)}$, $\mat{A+B}$ and $\mat{A - B}$ (where there, these quantities are referred to as $\mat{Q}^{\text{dRPA}}_{\text{I}}$, $\mat{\varepsilon}$ and $\mat{\varepsilon} + 2\mat{K}$ ).
Equivalence between these expressions (as well as the more commonly used expression found in Ref.~\onlinecite{Furche2008}) can be seen by noting (using Eqs.~\ref{eq:inverse}, \ref{eq:ApB_def} and \ref{eq:AmB_def}) that
\begin{align}
    &\text{Tr}\left[ \mat{\eta}^{(0)}(\mat{A+B})\right] \\
    = &\text{Tr}\left[ (\mat{\eta}^{(0)})^{-1} (\mat{A-B})\right] \\
    = &\text{Tr}\left[\left((\mat{A-B})^{\frac{1}{2}} (\mat{A+B}) (\mat{A-B})^{\frac{1}{2}}\right)^{\frac{1}{2}}\right] = \text{Tr}\left[ \mat{\Omega} \right].
\end{align}
Overall, this perspective on the RPA in terms of dd-moments is key to open new avenues such as the ones explored in this work.

\section{Efficient evaluation of self-energy and density response moments} \label{sec:EfficientEval}

Given the recasting of the RPA dd-response in Sec.~\ref{sec:DD_RPA} in terms of its lowest order moment (Eq.~\ref{eq:zero_mom}) and recursion to access the higher moments via Eq.~\ref{eq:mom_recursion}, we now consider the efficient $\mathcal{O}[N^4]$ evaluation of these quantities which are central to the approach in this work, thus avoiding their formal $\mathcal{O}[N^6]$ construction via Eq.~\ref{eq:formal_moms}.
The derivation here is heavily inspired by the seminal RI approach of Furche to compute RPA correlation energies \cite{Furche2010}, with key adaptations to target these dd-moments to arbitrary order, rather than the correlation energy. 

We first employ a standard low-rank decomposition of the two-electron repulsion integrals (e.g. via density fitting or Cholesky decomposition) as
\begin{equation}
(ia|jb) \simeq \sum_P V_{ia,P} V_{jb,P} = \mat{V}\mat{V}^T , \label{eq:RI_eri}
\end{equation}
where we use $P,Q,\dots$ to index elements of this auxiliary (RI) basis, whose dimension $N_{\mathrm{aux}}$ scales $\mathcal{O}[N]$ with system size. We define an intermediate quantity
\begin{equation}
{\tilde \eta}^{(n)}_{ia,P}=\sum_{jb} \eta^{(n)}_{ia,jb} V_{jb,P} \quad . \label{eq:orig}
\end{equation}
If this intermediate can be efficiently found, then the greater self-energy moment of Eq.~\ref{eq:SEMoms2_great} can be rewritten as
\begin{equation}
\Sigma_{pq}^{(n,>)}=\sum_{t=0}^n \binom{n}{t} \left( \epsilon_{c}^{n-t} V_{pc,Q} \left( V_{qc,P} \left( {\tilde \eta}_{ia,P}^{(t)} V_{ia,Q} \right) \right) \right) , \label{eq:SEMoms3_great}
\end{equation}
where the brackets indicate the order of contractions in order to preserve $\mathcal{O}[N^4]$ scaling, and einstein summation is implied. The lesser self-energy moment of Eq.~\ref{eq:SEMoms2_less} can be recast in an analogous fashion.
\footnote{We note that the derivation in this section does not rely on the $V_{jb,P}$ tensor in Eq.~\ref{eq:orig} arising from this Coulomb form specifically, but rather that it involves a more general linear transformation from a space in the particle-hole product basis to a space which scales no more than $\mathcal{O}[N]$ (in this case the auxiliary basis).}

Obtaining all dd-moments of the form of Eq.~\ref{eq:orig} up to order $n$ can be simply reduced to knowledge of the first two moments ${\tilde{\eta}}_{ia,P}^{(0)}$ and ${\tilde{\eta}}_{ia,P}^{(1)}$, via use of the recursive relationship between the moments as given by Eq.~\ref{eq:mom_recursion_main}, as for even moment orders,
\begin{equation}
\tilde{\mat{\eta}}^{(n)} = [(\mat{A}-\mat{B})(\mat{A}+\mat{B})]^{n/2} \tilde{\mat{\eta}}^{(0)} \label{eq:recursion_even}
\end{equation}
 and for odd moment orders,
\begin{equation}
\tilde{\mat{\eta}}^{(n)} = [(\mat{A}-\mat{B})(\mat{A}+\mat{B})]^{(n-1)/2} \tilde{\mat{\eta}}^{(1)} \label{eq:recursion_odd}
\end{equation}
where we have omitted explicit indices for brevity.
Ensuring that an $\mathcal{O}[N^4]$ scaling is retained in this recursion relies on $(\mat{A}-\mat{B})(\mat{A}+\mat{B})$ admitting a form where it can be written as a diagonal plus low-rank correction. 
For the RPA, this is true since (from Eqs.~\ref{eq:Casida_def_A}-\ref{eq:Casida_def_B}),
\begin{equation}
(\mat{A}-\mat{B})_{ia,jb} = (\epsilon_a - \epsilon_i) \delta_{ij} \delta_{ab} = \mat{D} \label{eq:A_min_B}
\end{equation}
is a purely diagonal matrix, while using Eq.~\ref{eq:RI_eri} we can cast $(\mat{A}+\mat{B})$ into an appropriate form as
\begin{equation}
(\mat{A} + \mat{B}) = \mat{D}+2\mathcal{K} = \mat{D} + 2 \sum_P V_{ia,P} V_{jb,P} . \label{eq:A_plus_B}
\end{equation}
We therefore express the low-rank asymmetrically decomposed form of $(\mat{A}-\mat{B})(\mat{A}+\mat{B})$ in a general fashion as a diagonal plus asymmetric low-rank part, as
\begin{equation}
(\mat{A}-\mat{B})(\mat{A}+\mat{B}) = \mat{D}^2 + \mat{S}_{L} \mat{S}_R^T , \label{eq:A_min_BA_plus_B}
\end{equation}
where for the RPA, $\mat{D}$ is defined by Eq.~\ref{eq:A_min_B}, $\mat{S}_L=\mat{D}\mat{V}$ and $\mat{S}_R=2\mat{V}$.
Future work will explore other analogous approaches where $(\mat{A}-\mat{B})(\mat{A}+\mat{B})$ can be decomposed in this way, for applicability to e.g. the Bethe-Salpeter equation or other RPA variants with (screened) exchange contributions \cite{Bintrim2022,doi:10.1021/acs.jctc.8b01247,doi:10.1021/acs.jctc.2c00366}. From this low-rank decomposition and the recursive definition of Eqs.~\ref{eq:recursion_even}-\ref{eq:recursion_odd}, a fixed number of dd-moments of the form of Eq.~\ref{eq:orig} can be found in $\mathcal{O}[N^4]$ time, provided the original $\tilde{\eta}^{(0)}$ and $\tilde{\eta}^{(1)}$ values are known.

We now consider how to obtain these initial low-order dd-moments efficiently. From the definitions of Eqs.~\ref{eq:one_mom} and \ref{eq:orig}, and specifying the standard RPA definition of Eq.~\ref{eq:A_min_B}, we find that it is straightforward to efficiently construct the first moment, as
\begin{equation}
\tilde{\eta}^{(1)} = \mat{D}\mat{V} = \mat{S}_L . \label{eq:firstmom}
\end{equation}
The zeroth-order dd-moment can be constructed via a rapidly-convergent numerical integration, as we will show. 
From Eq.~\ref{eq:zero_mom}, we can write
\begin{equation}
(\mat{A}-\mat{B})(\mat{A}+\mat{B}) = \left( \eta^{(0)} (\mat{A}+\mat{B}) \right)^2 ,
\end{equation}
from which we can find an expression for $\eta^{(0)}$ as
\begin{equation}
\eta^{(0)} = [(\mat{A}-\mat{B})(\mat{A}+\mat{B})]^{\frac{1}{2}} (\mat{A}+\mat{B})^{-1} .
\end{equation}
We note that for RPA to be well-defined with positive excitation energies, $(\mat{A}-\mat{B})$ and $(\mat{A}+\mat{B})$ must both be positive-definite matrices \cite{Furche2001}.
Using Eqs.~\ref{eq:A_plus_B} and \ref{eq:A_min_BA_plus_B}, we can write the low-rank RPA form of this as
\begin{equation}
\tilde{\eta}^{(0)} = (\mat{D}^2 + \mat{S}_L \mat{S}_R^T)^{\frac{1}{2}} (\mat{D} + 2 \mat{V} \mat{V}^T)^{-1} \mat{V} . \label{eq:tilde_mom_zero}
\end{equation}
We first consider the evaluation of the second half of this expression, which we denote as $\mat{T}$. We can use the Woodbury matrix identity to rewrite it as
\begin{align}
\mat{T} &= (\mat{D}+2\mat{V}\mat{V}^T)^{-1}\mat{V}  \\
&= \mat{D}^{-1} \mat{V} - 2\mat{D}^{-1} \mat{V}(\mat{I}+2\mat{V}^T \mat{D}^{-1} \mat{V})^{-1} \mat{V}^T \mat{D}^{-1} \mat{V} . \label{eq:Woodbury}
\end{align}
This now only requires the inversion of the diagonal matrix, $\mat{D}$, and a matrix of dimension $N_{\mathrm{aux}}$, with the overall $ov \times N_{\mathrm{aux}}$ matrix able to be constructed in $\mathcal{O}[N_{\mathrm{aux}}^3 + N_{\mathrm{aux}}^2ov]$ time. 

Having constructed $\mat{T}$, we can complete the evaluation of $\tilde{\eta}^{(0)}$ using the definition of the matrix square-root as an integration in the complex plane \cite{matrixsqrt},
\begin{equation}
\mat{M}^{\frac{1}{2}}=\frac{1}{\pi} \int_{-\infty}^{\infty} \left( \mat{I}-z^2(\mat{M}+z^2 \mat{I})^{-1} \right) dz \label{eq:matsqrt} .
\end{equation}
From Eq.~\ref{eq:tilde_mom_zero}, this results in
\begin{align}
\tilde{\mat{\eta}}^{(0)} &= (\mat{D}^2 + \mat{S}_L \mat{S}_R^T)^{\frac{1}{2}} \mat{T} \\
&= \frac{1}{\pi} \int_{-\infty}^{\infty} \left( \mat{I} - z^2 (\mat{D}^2 + \mat{S}_L \mat{S}_R^T + z^2 \mat{I})^{-1} \right) \mat{T} dz . \label{eq:NI}
\end{align}
We can modify this integrand into one more efficient for numerical integration, via another application of the Woodbury matrix identity to reduce the scaling of the matrix inverse. We also simplify the notation by introducing the intermediates,
\begin{align}
\mat{F}(z) &= (\mat{D}^2 + z^2 \mat{I})^{-1} \label{eq:F} \\
\mat{Q}(z) &= \mat{S}_R^T \mat{F}(z) \mat{S}_L , \label{eq:Q}
\end{align}
where $\mat{F}(z)$ is a diagonal matrix in the $ov$ space, and $\mat{Q}(z)$ is a $N_{\mathrm{aux}}\times N_{\mathrm{aux}}$ matrix which can be constructed in $\mathcal{O}[N_{\mathrm{aux}}^2 o v]$ time. This casts Eq.~\ref{eq:NI} into the form
\begin{equation}
\tilde{\eta}^{(0)} = \frac{1}{\pi} \int_{-\infty}^{\infty} \left[ \mat{I} - z^2 \mat{F}(z) \left(\mat{I}-\mat{S}_L(\mat{I}+\mat{Q}(z))^{-1}\mat{S}_R^T \mat{F}(z) \right) \right] \mat{T} dz . \label{eq:NI_efficient}
\end{equation}
For each value of the integration variable $z$, the integrand is a matrix of size $ov \times N_{\mathrm{aux}}$, which can be constructed in $\mathcal{O}[N^4]$ scaling, rendering it efficient for numerical quadrature. Along with the results of Eqs.~\ref{eq:firstmom}, \ref{eq:A_min_BA_plus_B}, \ref{eq:recursion_even} and \ref{eq:recursion_odd}, this therefore completes the ambition of constructing a fixed number of dd-response moments needed for the moment-truncated $GW$ method as defined in Eq.~\ref{eq:orig}, in no more than $\mathcal{O}[N^4]$ scaling (and $\mathcal{O}[N^3]$ memory). 

\begin{figure}[t]
    \centering
    \includegraphics[width=0.99\columnwidth]{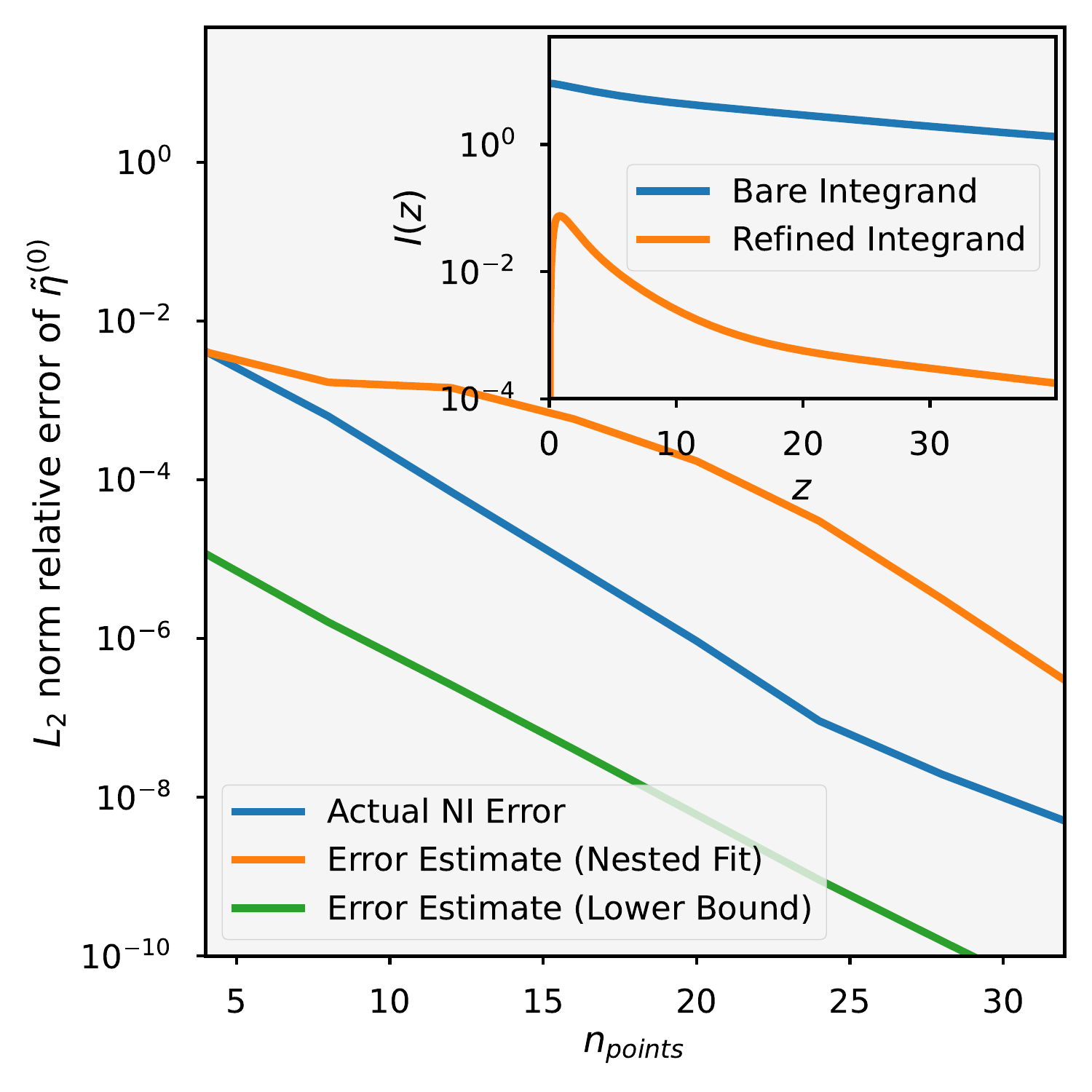}
    \caption{Exponential convergence of the numerical integration (NI) error for $\tilde{\mat{\eta}^{(0)}}$ in Eq.~\ref{eq:final_NI_zero_mom_main} with respect to integration points, for the singlet oxygen dimer in a \mbox{cc-pVTZ} basis at equilibrium (1.207 \AA)~ bond length. Also included are the two error estimates of the true NI error, used to check convergence and estimate the required number of points (see App. B, Eq.~\ref{eq:cubic_fit_err_estimate} for the `Nested Fit' and Eq.~\ref{eq:lowerbound_err} for the `Lower Bound' definitions). Inset: The originally derived integrand (Eq.~\ref{eq:NI_efficient}), and the form optimized for efficient NI given in Eq.~\ref{eq:final_NI_zero_mom_main} and derived in App. A, showing the faster decay.}
    \label{fig:NumQuadConv}
\end{figure}

However, manipulations of the resulting integrand and choice of quadrature points can further improve the efficiency of their construction by ensuring a faster decay of the integrand and separating components which can be analytically integrated. This derivation is given in Appendix~A, and results in a final $\mathcal{O}[N_{\mathrm{aux}}^2 ov + N_{\mathrm{aux}}^3]$ expression to evaluate for the zeroth-order dd-moment as
\begin{align}
\tilde{\eta}^{(0)} &= \mat{D}\mat{T} \nonumber \\
&+ \int_0^{\infty} e^{-t \mat{D}} \mat{S}_L \mat{S}_R^T e^{-t \mat{D}} \mat{T} dt \nonumber \\
&+ \frac{1}{\pi}\int_{-\infty}^{\infty} z^2 \mat{F}(z) \mat{S}_L\left((\mat{I}+\mat{Q}(z))^{-1}-\mat{I} \right) \mat{S}_R^T \mat{F}(z) \mat{T} dz . \label{eq:final_NI_zero_mom_main}
\end{align}
The first numerical integral in Eq.~\ref{eq:final_NI_zero_mom_main} (where the integrand decays exponentially) is computed via Gauss-Laguerre quadrature, while the second (where the integrand decays as $\mathcal{O}[z^{-4}]$) is evaluated via Clenshaw--Curtis quadrature. A comparison of the decay of the original and refined integrands is shown in the inset to Fig.~\ref{fig:NumQuadConv}.

The scaling of the grid spacing of both numerical quadratures is optimised to ensure exact integration of the trace of a diagonal approximation for the integrand, analogous to the grid optimization discussed in Ref.~\onlinecite{Furche2010}.
For numerical robustness, we optimize the quadrature for evaluating $\textrm{Tr}[(\mat{A}-\mat{B})(\mat{A}+\mat{B})]^{\frac{1}{2}}$, rather than the full integrand. 
We write a diagonal approximation to $(\mat{A}-\mat{B})(\mat{A}+\mat{B})$ as
\begin{equation}
    \mat{M}^D = \mat{D}^2 + \mat{S}_L^D (\mat{S}_R^D)^T , 
\end{equation}
with
\begin{align}
    (S_L^D)_{ia,R} &= (S_L)_{ia,Q}(S_R)_{ia,Q} \delta_{R,ia} \\
    (S_R^D)_{jb,R} &= \delta_{R,jb}.
\end{align}
This is the same form as Eq.~\eqref{eq:A_min_BA_plus_B}, but contains only diagonal matrices (denoted by subscript `$D$' labels) and has an auxiliary space of size $ov$.
As such, the exact square root and all quantities within the numerical integrations can be obtained in $\mathcal{O}[ov]$ computational time. We then seek to ensure the trace of the difference between the exact and numerically integrated estimate of ${\mat{M}^D}^{\frac{1}{2}}$ vanishes. This is achieved for both integrals in Eq.~\eqref{eq:final_NI_zero_mom_main}, with the analytic form for the first integral given by $\frac{1}{2}\text{diag}(\mat{D}^{-1}\mat{S}_L \mat{S}_R^T) = I^\text{offset}$ and the second numerical integral as ${\mat{M}^D}^{\frac{1}{2}}-\mat{D}-I^\text{offset}=I^\text{int}$.

Writing this explicitly, given quadrature points and weights $\{z_i, w_i\}$ for a $n_p$-point infinite or semi-infinite quadrature, we seek to scale our points by a factor $a$, which is a root the objective functions
\begin{align}
    O^\text{offset}(a) &= I^\text{offset} - \frac{1}{\pi}\sum_i^{n_p} a w_i \text{Tr} ( e^{-2\mat{D} a z_i} \mat{S}_L \mat{S}_R^T ) \label{eq:offset_obj} \\
    O^\text{int}(a) &= I^\text{int} - \frac{1}{\pi}\sum_i^{n_p} a w_i \text{Tr}(I^\text{D}(a z_i)) \label{eq:int_obj} \\
    I^\text{D}(z) &= z^2 \mat{F}^D(z) \mat{S}^D_L\left((\mat{I}+\mat{Q}^D(z))^{-1}-\mat{I} \right) (\mat{S}^D_R)^T \mat{F}^D(z),
\end{align}
where Eq.~\ref{eq:offset_obj} is minimized to optimize the grid for the first integral of Eq.~\ref{eq:final_NI_zero_mom_main}, and Eq.~\ref{eq:int_obj} minimized for the second integral.
This can be done via either simple root-finding or minimization, and gives a robust optimization of the integration grids in $\mathcal{O}[ov]$ computational cost.
The resulting exponential convergence of the zeroth dd-moment estimate with number of quadrature points, along with the error estimates derived in App.~B, are shown in Fig.~\ref{fig:NumQuadConv} for both numerical integrands. We find that as few as $12$ quadrature points are sufficient for high accuracy in the results of this work, while the number of points is expected to increase for systems with a small or vanishing spectral gap.

\section{Reduction to cubic-scaling GW} \label{sec:CubicScaling}

With this reformulation of $GW$ in terms of the moments of the self-energy, it is possible to further reduce the scaling to cubic in time and quadratic in memory with respect to system size, in common with the lowest-scaling $GW$ approaches \cite{Hutter16,PhysRevB.94.165109,Visscher20,Duchemin2021}. We stress that this is not an asymptotic scaling after exploiting screening and locality arguments, but rather a formal scaling exploiting further rank reduction of quantities. 
To do this, we employ a {\em double} factorization of the Coulomb integrals, allowing them to be written as as a product of five rank-2 tensors, as
\begin{equation}
(ia|jb) \simeq \sum_{P,Q} X_{iP} X_{aP} Z_{PQ} X_{jQ} X_{bQ} , \label{eq:THC_eri}
\end{equation}
This factorizes the orbital product into separate terms, and is also known as tensor hypercontraction or CP decomposition, used for various recent low-scaling formulations of quantum chemical methods, where the dimension of the $P$ and $Q$ space rigorously grow linearly with system size\cite{THC_V}. Use of this doubly-factorized form has also been previously suggested in the use of a reduced-scaling RPA and particle-particle RPA schemes \cite{LU2017187, Yang_THCppRPA}. This form for the integrals can be directly constructed with controllable errors in $\mathcal{O}[N^3]$ time\cite{LU2015329, doi:10.1021/acs.jctc.2c00861}.

Once found, the $Z_{PQ}$ can be symmetrically decomposed as $Z_{PQ}=Y_{PR} Y_{QR}$. By replacing the density-fitted integral tensor $V_{iaR}$ in the above expressions with the fully factorized form $X_{iP} X_{aP} Y_{RP}$, the contractions to form the moments of the $GW$ self-energy, and hence the Green's function and quasi-particle spectrum naturally follow as $\mathcal{O}[N^3]$ with formation of appropriate intermediates. We also require the numerical computation of the partially transformed dd-moment, $Z_{QS} X_{aS} X_{iS} \eta^{(0)}_{ia,jb} X_{jR} X_{bR} Z_{PR}$. Inspired by the low-scaling approach taken to RPA correlation energies in the work of Refs.~\onlinecite{Schurkus2016}, this can be achieved with the use of a contracted double-Laplace transform in the place of the original numerical integration procedure. This factorizes the squared energy denominator $\mat{F}(z)$, allowing the occupied and unoccupied indices to be contracted independently, similar to the space-time approaches to $GW$\cite{Visscher20}.
While this becomes a two-dimensional numerical integral, optimal quadrature grids can be calculated in a minimax sense \cite{HELMICHPARIS2016927,Graf2018}.

Applied to Eq.~\ref{eq:F}, this contracted double-Laplace transform takes the form
\begin{align}
    {F}_{ia,ia}(z) &= D_{ia,ia} \int_{p=0}^\infty \frac{\sin{zp}}{z} e^{-pD_{ia,ia}}dp,
\end{align}
which allows the key matrix $\mat{Q}(z)$ of Eq.~\ref{eq:Q} to be obtained as
\begin{align}
    Q_{PS}(z) &= 2\int_{p=0}^\infty \frac{\sin{zp}}{z} Y_{PQ} (A_{QR}(p) - B_{QR}(p) ) Y_{SR} dp
\end{align}
where both intermediates
\begin{align}
    A_{QR}(p) &= X_{iQ} X_{aQ} \epsilon_a e^{-\epsilon_a p} e^{\epsilon_i p}  X_{iR} X_{aR} \\
    B_{QR}(p) &= X_{iQ} X_{aQ} \epsilon_i e^{-\epsilon_a p} e^{\epsilon_i p}  X_{iR} X_{aR}
\end{align}
can be evaluated in $\mathcal{O}[N^3]$ cost. Further contractions in the evaluation of Eq.~\ref{eq:final_NI_zero_mom_main} also follow naturally with cubic scaling. 

An alternative approach to reduce the scaling (to \emph{asymptotically} linear) without requiring the doubly-factorized integrals, is to screen the atomic orbital density-fitted integral contributions (constructed with the overlap metric), along with the double-Laplace transform, exploiting locality as has been recently performed for the RPA correlation energy \cite{Schurkus2016,Graf2018}.
An explicit numerical demonstration of this reduction to cubic cost via the double factorization of the Coulomb tensor of Eq.~\ref{eq:THC_eri} will follow in forthcoming work, with numerical results in the rest of this work employing the quartic scaling algorithm described in Sec.~\ref{sec:EfficientEval}.


\section{Results}\label{sec:Results}

\subsection{Comparison to quasiparticle GW approaches}

\begin{figure}[t]
    \centering
    \includegraphics[width=0.99\columnwidth]{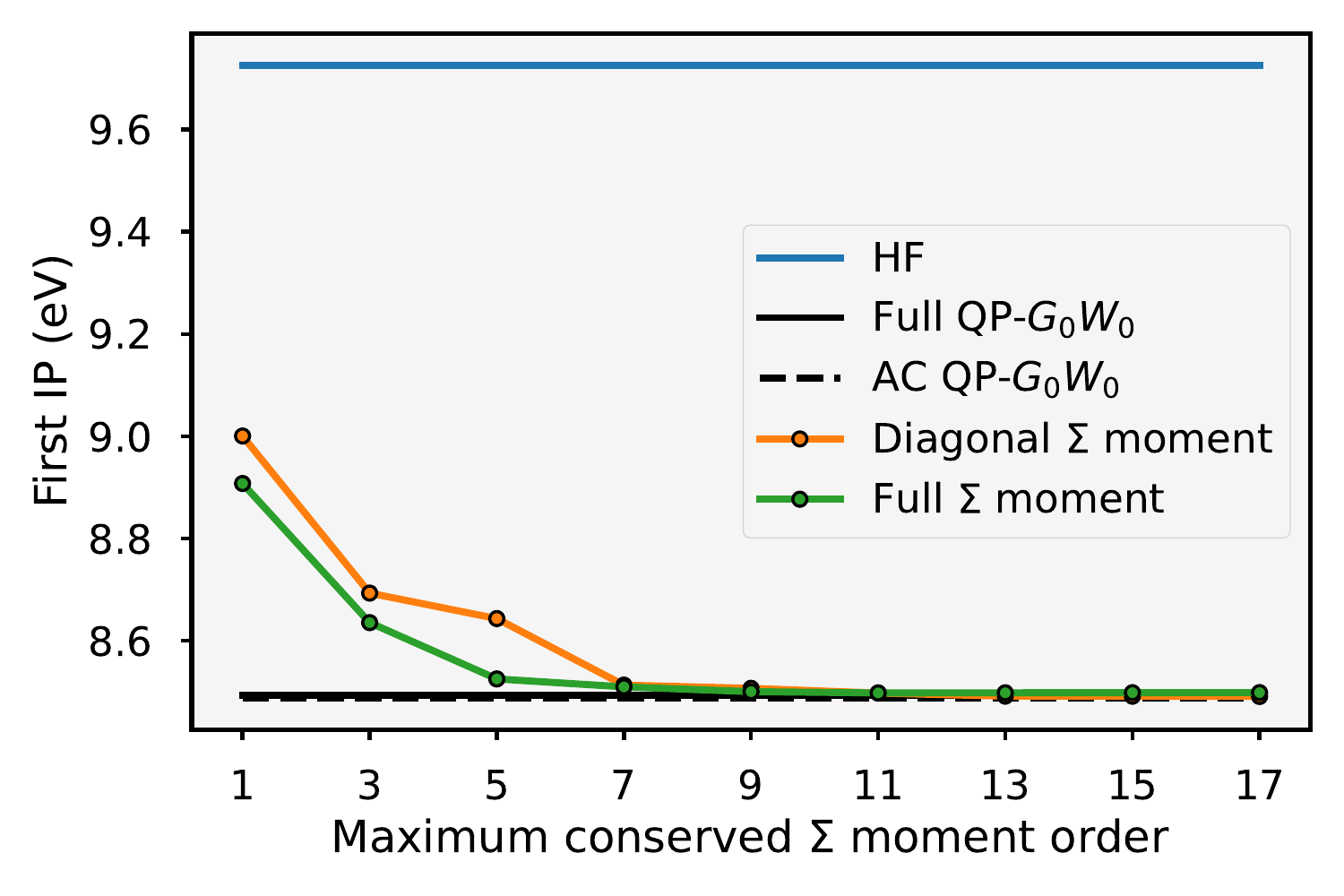}
    \caption{Convergence of the first $G_0W_0$ IP of singlet O$_2$ in a cc-pVDZ basis with respect to the number of conserved self-energy moments. Also shown is the same quasiparticle state computed from traditional $G_0W_0$ via an exact frequency integration (`Full') and analytic continuation approach to $G_0W_0$ (`AC'), albeit both imposing a diagonal approximation in their solution to the QP equation. In the moment expansion, we also consider a diagonal approximation, by explicitly removing the non-diagonal parts of our computed self-energy moments (`Diagonal $\Sigma$ moment'). All approaches find an IP of $8.49$~eV to within $10$~meV, with the difference between full-frequency and AC approaches $5$~meV, and the relaxation of diagonal approximation also accounting for small $\sim5$~meV differences, 
    with the reference Hartree--Fock IP for comparison being 9.73 eV.}
    \label{fig:O2Conv}
\end{figure}

We first consider the convergence of the moment-truncated $G_0W_0$ algorithm compared to more traditional implementations, as found in the {\tt PySCF} software package \cite{PySCF2017, PySCF2020, PhysRevX.11.021006, doi:10.1021/acs.jctc.0c00704}. As found to be more effective for molecular systems due to the importance of exact exchange, we perform all calculations on a restricted Hartree--Fock reference \cite{VanSetten2013}. In Fig.~\ref{fig:O2Conv} we consider the convergence of the first ionization potential (IP) of singlet O$_2$ with conserved self-energy moment order. We compare this to two $G_0W_0$ implementations, one of which performs an exact frequency integration (denoted `full QP-$G_0W_0$', which scales as $\mathcal{O}[N^6]$), and one which performs an analytic continuation (AC) of the imaginary frequency self-energy to real-frequencies via a fit to Pad{\'e} approximants in order to perform the convolution (denoted `AC QP-$G_0W_0$', which scales as $\mathcal{O}[N^4]$) \cite{Ren_2012,doi:10.1021/acs.jctc.6b00380,doi:10.1021/acs.jctc.0c00704}. However, both of these two implementations also solve the diagonal approximation to the quasiparticle equation in solving for each state, effectively imposing a diagonal approximation to the self-energy in the MO basis. This is avoided in our work, however we can constrain a similar diagonal approximation by simply removing the off-diagonal components of our computed self-energy moments. This does not result in a significant computational saving in our approach, and therefore is only relevant for comparison purposes when considering the effect of this neglected off-diagonal part of the self-energy.

As can be seen in Fig.~\ref{fig:O2Conv}, the IP converges rapidly with moment order, with the full self-energy moments converging slightly faster and more systematically without the diagonal approximation (something also observed in other applications). The diagonal approximation converges to the `exact' frequency integration as expected, with our more complete (non-diagonal) self-energy moment approach only very slightly different, indicating the relative unimportance of the non-diagonal self-energy components in this system and lack of significant correlation-induced coupling between the mean-field quasiparticle states. Furthermore, the analytic continuation approach is also highly accurate for this system, introducing an error of only $5$~meV compared to the full frequency integration. We have furthermore numerically verified that our approach scales as $\mathcal{O}[N^4]$, with computational cost comparable to the analytic continuation approach.

Having demonstrated correctness compared to a high-scaling exact frequency implementation, we can compare our results to AC-$G_0W_0$ across a larger test set to consider the moment truncation convergence. We use the established `$GW100$' benchmark test set, where many $GW$ (and other excited state method) implementations have been rigorously compared \cite{VanSetten2015, Caruso2016, Lange2018, Backhouse2021}.
This benchmark set contains 102 diverse systems, with the IP of the molecules in the set ranging from $\sim4 - 25$ eV, and featuring molecules with bound metal atoms (including metal diatomics and clusters), strongly ionic bonding, and molecules with a strongly delocalised electronic structure. The molecules range in size from simple atomic systems to the five canonical nucleobases and large aliphatic and aromatic hydrocarbons, providing a suitable range in system size. 

\begin{figure}[t]
    \centering
    \includegraphics[width=0.99\columnwidth]{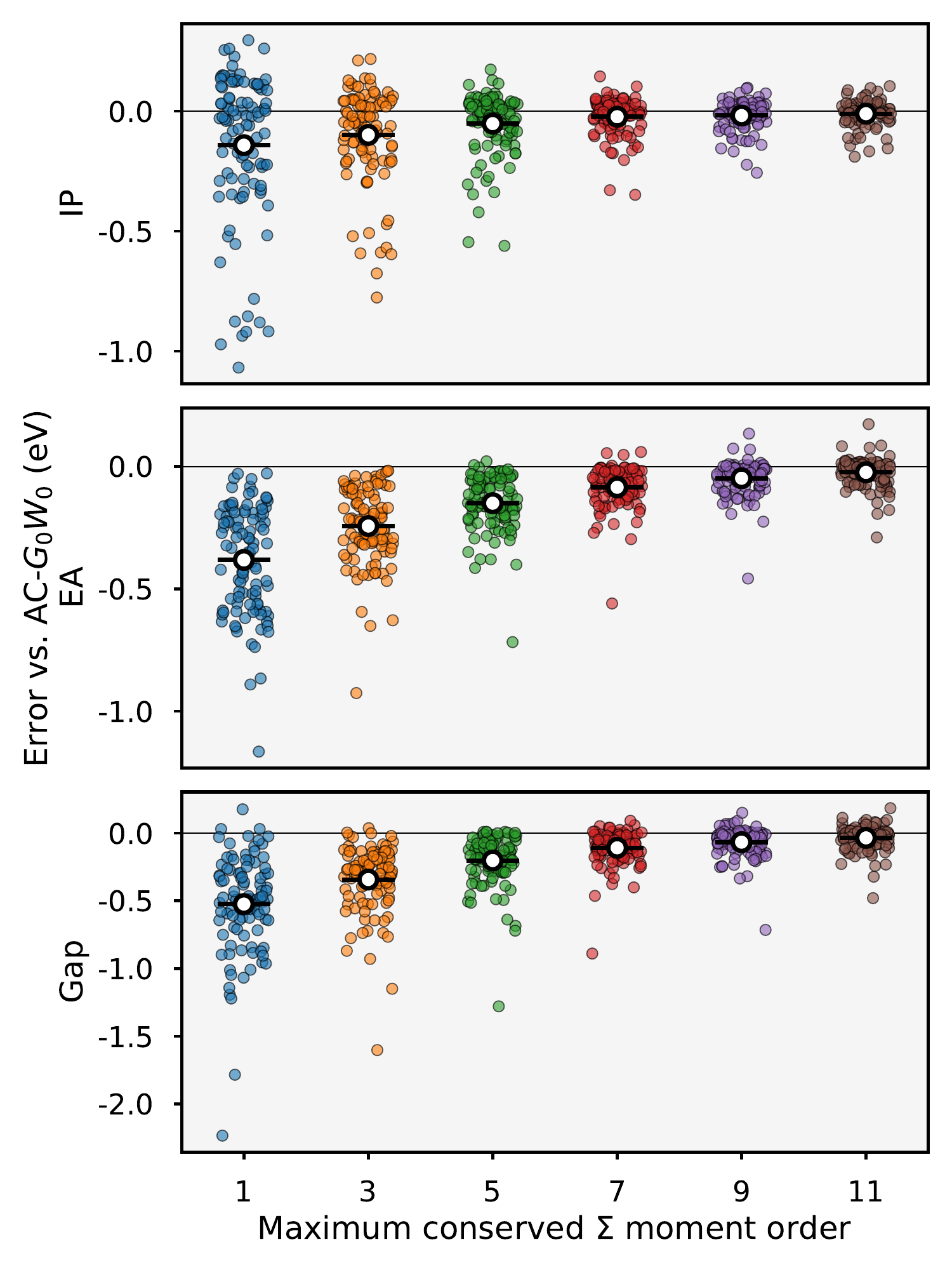}
    \caption{Errors in the IP, EA and gap for each system in the $GW100$ benchmark set, compared to AC-$G_0W_0$ in a def2-TZVPP basis, for each order of self-energy moment conservation. White circles show the mean signed error (MSE) aggregated over the test set for the given moment truncation in each quantity.}
    \label{fig:swarm_GW}
\end{figure}

In Fig.~\ref{fig:swarm_GW} we consider the discrepancy in the first IP, electron affinity (EA), and quasiparticle gap across all systems as the order of conserved self-energy moments increases in a realistic def2-TZVPP basis, again compared to AC-$G_0W_0$. Errors in individual systems, along with the mean signed error (MSE) across the set (white circle) is shown for each conserved moment order. This MSE for the first IP decreases from -0.142~eV for the lowest order moment conservation, to -11~meV when up to the $11^\textrm{th}$-order moment is conserved, with the EA errors generally a little larger. Similarly, the gap calculations converge to a MSE of -34.8~meV, with standard deviation about the AC-$G_0W_0$ result of only 91~meV across all systems. 
We note that there may also be small differences arising due to the comparison with AC-$G_0W_0$ due to the approximations of the frequency integration via analytically continued quantities, as well as the differences in whether off-diagonal parts of the self-energy are included. These will contribute to the discrepancy between the methods at each order, however while the comparison is not strictly equivalent and therefore these errors will be overestimated compared to an exact frequency and non-diagonal $G_0W_0$ limit, the general trend, convergence and level of accuracy which can be reached with moment order is likely to be similar.

\begin{figure}[t]
    \centering
    \includegraphics[width=0.99\columnwidth]{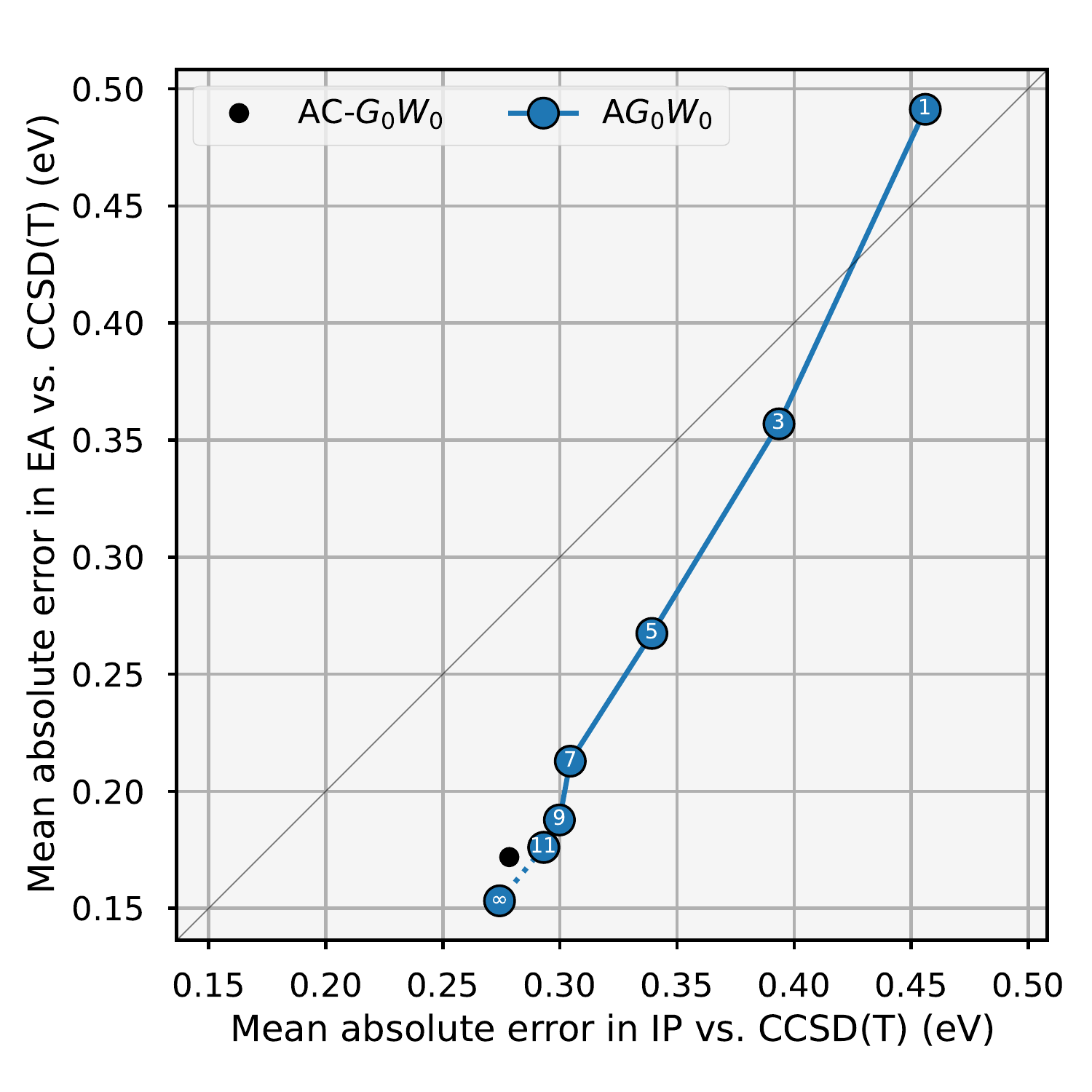}
    \caption{Mean absolute errors (eV) for the IP ($x$-axis) and EA ($y$-axis) across the $GW100$ benchmark set in a def2-TZVPP basis compared to accurate CCSD(T) values. AC-$G_0W_0$ values, as well as moment-truncated results are shown, with the number in each data point marker giving the number of conserved moments. Extrapolation of individual data points from the $9^{\mathrm{th}}$- and $11^{\mathrm{th}}$-order conserved moment values are also provided, denoted by the truncation label `$\infty$'.}
    \label{fig:GW_vs_CCSDT}
\end{figure}

It is important to put the scale of the moment truncation convergence in the context of the overall accuracy of the $G_0W_0$ method for these systems. In Fig.~\ref{fig:GW_vs_CCSDT} we therefore compare the aggregated mean absolute error (MAE) in the moment-truncated $G_0W_0$ IP and EA values over this $GW100$ test set to highly accurate CCSD(T) calculations on the separate charged and neutral systems, which is often used as a more faithful benchmark to compare against than experiment. We can therefore see the convergence of the moment truncation to the AC-$G_0W_0$ values compared to the intrinsic error in the method. This intrinsic error is found to dominate over the error due to the self-energy moment truncation for higher numbers of conserved moments.
\toadd{We note however that the mean error compared to CCSD decreases systematically with increasing numbers of moments for these systems, which contrasts with observed behaviour for moment-truncated GF2 theory (where lower order self-energy truncations were found to give rise to a more accurate overall excitations) \cite{Backhouse2020b}.}
It is natural to therefore also consider whether a simple extrapolation can improve the moment-truncated results to an infinite moment limit. We therefore apply a linear extrapolation of the excitation energies to the infinite moment limit from the two most complete moment calculations of each system. We can see from Fig.~\ref{fig:GW_vs_CCSDT} that these results continue the trend of the MAE across the test set, slightly overshooting the AC-$G_0W_0$ comparison, albeit noting the other potential sources of discrepancy between these values discussed earlier.

\subsection{Full frequency spectra}

\begin{figure}[t]
    \centering
    \includegraphics[width=0.99\columnwidth]{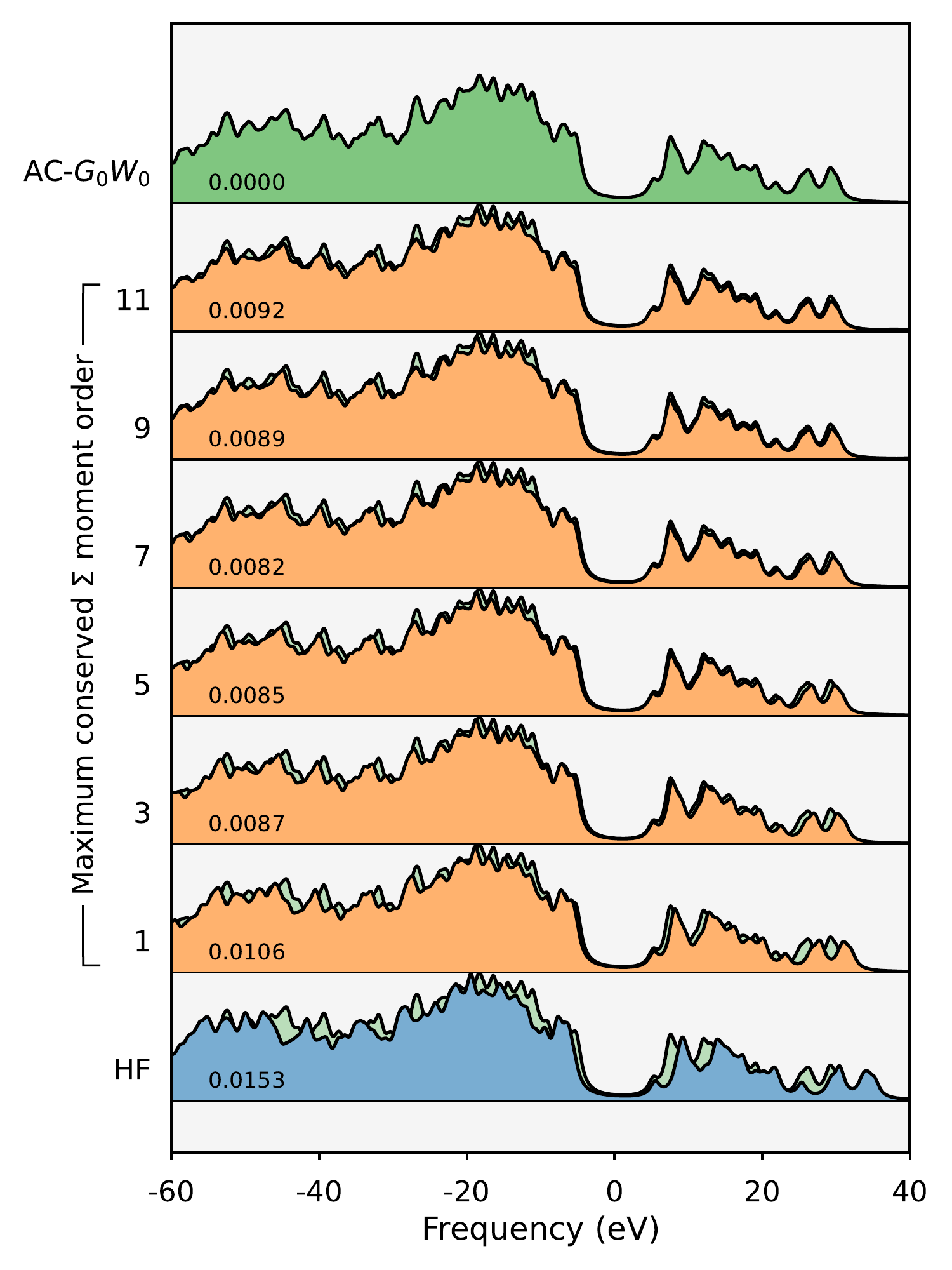}
    \caption{
        Spectral functions for guanine in a def2-TZVPP basis set, for HF, AC-$G_{0}W_{0}$,
        and the moment-conserving $G_{0}W_{0}$ approach with zero to five block Lanczos iterations, thereby conserving up to the $11^{\mathrm{th}}$-order self-energy moment.
        The values indicated in the spectra indicate the Wasserstein metric taken with respect to the AC-$G_0W_0$ spectrum,
        quantifying the difference between the spectral distributions.
        The AC-$G_0W_0$ spectrum is indicated transparently behind the other spectra to ease visualisation of the convergence,
        and was calculated using an iterative diagonal approximation to the quasiparticle equation.
    }
    \label{fig:guanine_spectra}
\end{figure}

One of the strengths of the moment-conserving approach to $GW$ of this work, is the ability to obtain all excitations from a given order of truncation in a single complete diagonalization of the effective Hamiltonian of Eq.~\ref{eq:concatenated_h}.  This allows full frequency spectra to be obtained, with the approximation not expected to bias significantly towards accuracy in any particular energy range, making it suitable for $G_0W_0$ excitations beyond frontier excitations. 
Description of low-lying states is a particular challenge for many other $GW$ approaches, with analytic continuation becoming less reliable, and alternatives like contour-deformation scaling as $\mathcal{O}[N^5]$ in general to obtain the full spectrum \cite{Duchemin2020}.
In Fig.~\ref{fig:guanine_spectra} we therefore show a series of spectra plotting on the real frequency axis for the guanine nucleobase
in a def2-TZVPP basis set over a $100$~eV energy window about the quasiparticle gap,
taken from the $GW$100 benchmark set, and compared to the AC-$G_0W_0$ full-frequency spectrum.

The convergence of the spectrum is shown for a series of conserved $G_0W_0$ moment orders, from the HF level up to the AC-$G_0W_0$ spectrum. The AC-$G_0W_0$ spectrum is also shown `behind' the other spectra, to allow the deviations to be observed for each moment. It can be seen that the full-frequency spectrum rapidly converges with conserved self-energy moment order, even for high-energy states where the HF approximation is poor. The similarity of each spectrum to the AC-$G_0W_0$ result accross the full frequency range can be rigorously quantified via the Wasserstein or `earth-mover' metric, which describes the similarity between probability distributions. This metric is shown as the value inside of each plot, indicating a rapid and robust convergence of the spectral features from the mean-field to the full $G_0W_0$ spectrum with increasing numbers of included moments.

This Wasserstein metric convergence plateaus at the $\sim7^{\mathrm{th}}$-order conserved moments, with further orders not improving this metric further. This could be due to numerical precision of the algorithm, or fundamental approximations in the AC-$G_0W_0$ such as the precision of the analytic continuation, or the diagonal approximation to the self-energy. Furthermore, it should be noted that the moment-conserving $GW$ approximation will rigorously have a larger number of poles included in its spectrum compared to those $GW$ approaches which rely on an iterative solution to the QP equation which considers the change to a single MO at a time. These additional peaks are likely very low weighted for this weakly-correlated system, yet could be contributing to this discrepancy with AC-$G_0W_0$ in the Wasserstein metric. We consider this point in more detail in the next section.

\subsection{Multiple solutions and additional spectral features}

\begin{figure*}[t]
    \centering
    \includegraphics[width=0.99\textwidth]{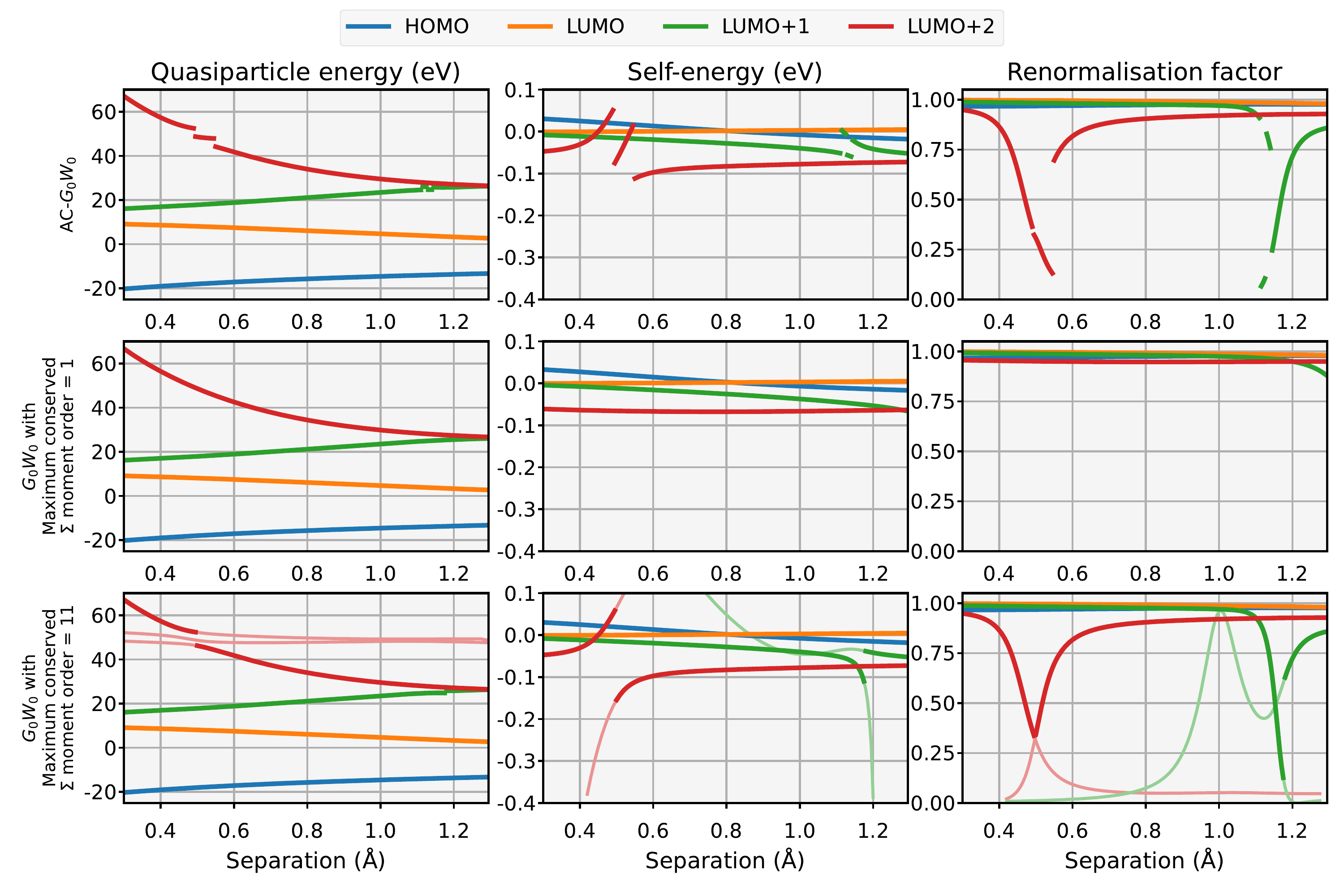}
    \caption{
        Quasiparticle energies, self-energies, and renormalisation factors for the H$_2$ dimer in a 6-31G basis set
        with varying bond lengths.
        Shown are results for AC-$G_{0}W_{0}$ and moment-conserved $G_{0}W_{0}$ with zero and five iterations,
        thereby conserving up to the $1^{\textrm{st}}$- and $11^{\mathrm{th}}$-order moments, respectively.
        The self-energy corresponds to the diagonal element corresponding to the particular orbital,
        evaluated at the quasiparticle energy.
        The renormalisation factor corresponds to $Z_{p} = (1 - \frac{\partial \Sigma_{pp}(\epsilon_p)}{\partial \omega})^{-1}$,
        with larger values indicating a more quasiparticle-like excitation.
        The transparent lines in the lower panel show the existence of multiple solutions, broadening spectral features near the self-energy poles, and where a single dominant solution can be chosen (defined here by maximum overlap with the corresponding MO) denoted by the thicker line.
    }
    \label{fig:h2_discontinuities}
\end{figure*}

This fact that many low-scaling $GW$ implementations rely on an iterative solution to solve the quasiparticle equation can be a source of error and loss of robustness. This is because when self-energy poles are found near quasiparticle energies, the $GW$ poles can split into multiple peaks, where the final excitation energy converged to can depend sensitively on the specifics of the root-finding algorithm used to solve the QP equation. This was highlighted in the Ref.~\onlinecite{VanSetten2015} as a significant source of error, where a number of simple systems were found to exhibit a number of poles close to the HOMO energy level (with these solutions spanning a range of up to 1~eV). The specifics of which pole is converged to (with undesired solutions called `spurious') then depended on initial conditions, choices of optimization method, and specifics of the self-energy construction or linearization of the QP equation. The requirement to select one of these multiple solutions to the QP equation can also manifest in undesirable discontinuities in the excitation energies as e.g. the molecular geometry or correlation strength changes, as described in Refs.~\onlinecite{doi:10.1021/acs.jctc.8b00260,10.3389/fchem.2021.751054,Monino2022}. An indicator for the presence of these `spurious' poles and multiple solutions is the magnitude of the quasi-particle weight or renomalization factor evaluated at the quasi-particle energy, defined as $Z_p = (1 - \frac{\partial \Sigma_{pp}(\epsilon_p)}{\partial \omega})^{-1}$, which indicates the approximate weight in the quasi-particle solutions.

Since the moment-conserving $GW$ approach obtains all poles in one step (including satellites and low-weighted features), all the excitations can be characterized by their quasiparticle weight, and either selected as specific excitation energies, or all excitations included in the spectrum to ensure smooth changes with molecular geometry. Points of discontinuity will therefore manifest as the presence of multiple lower-weighted solutions at a given energy, giving a smooth change to a broadened feature in the spectrum near self-energy poles. To demonstrate this, we consider the same simple system as Ref.~\onlinecite{Monino2022}, observing the $GW$ quasiparticle energies as a function of the inter-nuclear distance in the H$_2$ dimer in a 6-31G basis set. Figure~\ref{fig:h2_discontinuities} shows quasiparticle energies, self-energies and quasiparticle renormalisation factors
for AC-$G_0W_0$, and the moment conserving $G_0W_0$ approach with both up to the $1^{\mathrm{st}}$- and $11^{\mathrm{th}}$-order conserved self-energy moments in each sector.
The self-energies plotted are the diagonal elements corresponding to the particular MO, evaluated at the respective quasiparticle energy.

The figure shows the HOMO and first three unoccupied states found in this system, with the AC-$G_0W_0$ (first row) exhibiting discontinuities in the LUMO+2 state at slightly compressed geometries, and discontinuities in the LUMO+1 state at slightly stretched geometries. As discussed, these changing solutions arise from the specifics of the root-finding in the solution to the QP equation, which will generally (but not always) converge to the solution with largest quasiparticle weight between the multiple options, indicated by the renormalization factor. These discontinuous changes between states are also shown to coincide with poles in the self-energy in the second column at the MO energies for a given separation. These AC-$G_0W_0$ results are essentially the same as those found in Ref.~\onlinecite{Monino2022}.

When only the first two self-energy moments are conserved in each sector (second row), the approximation to the self-energy renders its pole structure sufficiently sparse such that their poles are pushed far from the MO energies at all geometries for these states. While this regularization removes the discontinuities, this significant approximation renders the renormalization factors close to one at all points, indicating only small changes from the original MOs. The final row represents a $G_0W_0$ calculation with up to the $11^{\mathrm{th}}$-order conserved moments. With this finer resolution of the self-energy dynamics, the structure of the self-energy closely matches the one from AC-$G_0W_0$, however, the multiple solutions are all found simultaneously, with their changing quasiparticle weights shown. The points of discontinuity are replaced by the presence of multiple solutions at similar energies and with competing quasiparticle weights, providing broad spectral features at those points. 

If a single solution is required, the specific excitation can be selected from the manifold based rigorously on e.g. the maximum overlap with the MO of interest (shown as thicker lines in the plot) or largest quasiparticle weight, all of which can easily be selected. This removes the uncertainty in the energies of the states based on the unphysical specifics of the QP solution algorithm, without incurring additional complexity or cost in the moment-conserving algorithm. Furthermore, the relaxation of the diagonal approximation of the self-energy in this approach is expected to be more significant at these points of multiple solution, where mixing between the different MOs is expected to be more pronounced.


\section{Conclusions and outlook}

In this work, we present a reformulation of the $GW$ theory of quasiparticle excitations, based around a systematic expansion and conservation of the spectral moments of the self-energy. This contrasts with other approaches designed to approximate the central frequency integration of $GW$ theory, which use e.g. grid expansions, analytic continuation or contour deformation in order to affect a scaling reduction from the exact theory. The moment expansion presented in this work has appealing features arising from the avoidance of the an iterative solution to the quasiparticle equation for each state (avoiding `spurious' solutions), diagonal self-energy approximations, or requirements for analytic continuation of dynamical quantities. It allows for all excitation energies and weights to be obtained directly in a non-iterative single diagonalization of a small effective Hamiltonian, controlled by a single parameter governing the number of conserved self-energy moments. 

Full RPA screening and particle-hole coupling in the self-energy is included, which is captured with $\mathcal{O}[N^4]$ computational scaling via a numerical one-dimensional integration, with a reduction to cubic scaling also proposed. This approach is enabled by a recasting of the RPA in terms of the moments of the density-density response function. Applied across we $GW100$ test set, we find rapid convergence to established $GW$ methodology results for both state-specific and full spectral properties, with errors due to the incompleteness of the moment expansion many times smaller than the inherent accuracy of the method. The formulation follows relatively closely from previous low-scaling approaches to RPA correlation energies, enabling these codes to be adapted to low-scaling $GW$ methods with relatively little effort.

Going forwards, we will aim to test the limits of the moment-truncated $GW$ formulation, pushing it to larger systems including the solid-state and different reference states, lower-scaling variants, and the inclusion of various self-consistent flavors of the theory \toadd{(beyond the $G_0W_0$ implementation here)}. The reformulation of RPA in terms of a recursive moment expansion also lends itself to a low-scaling implementation of the Bethe-Salpeter equation for neutral excitations, which we will explore in the future, as well as other beyond-RPA approaches. Finally, we will also explore the connections of this moment expansion to kernel polynomial approaches which expand spectral quantities in terms of Chebyshev and other orthogonal polynomial expansions \cite{RevModPhys.78.275}.

\section*{Code and data availability}

Open-source code for reproduction of all results in this paper, along with examples, can be found at \href{https://github.com/BoothGroup/momentGW}{\tt https://github.com/BoothGroup/momentGW}.
The repository also includes the data used in this paper relating to the $GW$100 benchmark.


\ifjcp\else
    \newpage
\fi

\section*{Acknowledgements}
The authors thank Filipp Furche and Johannes T{\"o}lle for useful feedback on the manuscript.
G.H.B. gratefully acknowledges support from the Royal Society via a University Research Fellowship, as well as funding from the European Union's Horizon 2020 research and innovation programme under grant agreement No. 759063. We are grateful to the UK Materials and Molecular Modelling Hub for computational resources, which is partially funded by EPSRC (EP/P020194/1 and EP/T022213/1).
Further computational resources were awarded under the embedded CSE programme of the ARCHER2 UK National Supercomputing Service (http://www.archer2.ac.uk).

\ifjcp
%
\else
    \bibliographystyle{achemso}
    \bibliography{refs.bib}

\begin{thebibliography}{114}%
\makeatletter
\providecommand \@ifxundefined [1]{%
 \@ifx{#1\undefined}
}%
\providecommand \@ifnum [1]{%
 \ifnum #1\expandafter \@firstoftwo
 \else \expandafter \@secondoftwo
 \fi
}%
\providecommand \@ifx [1]{%
 \ifx #1\expandafter \@firstoftwo
 \else \expandafter \@secondoftwo
 \fi
}%
\providecommand \natexlab [1]{#1}%
\providecommand \enquote  [1]{``#1''}%
\providecommand \bibnamefont  [1]{#1}%
\providecommand \bibfnamefont [1]{#1}%
\providecommand \citenamefont [1]{#1}%
\providecommand \href@noop [0]{\@secondoftwo}%
\providecommand \href [0]{\begingroup \@sanitize@url \@href}%
\providecommand \@href[1]{\@@startlink{#1}\@@href}%
\providecommand \@@href[1]{\endgroup#1\@@endlink}%
\providecommand \@sanitize@url [0]{\catcode `\\12\catcode `\$12\catcode
  `\&12\catcode `\#12\catcode `\^12\catcode `\_12\catcode `\%12\relax}%
\providecommand \@@startlink[1]{}%
\providecommand \@@endlink[0]{}%
\providecommand \url  [0]{\begingroup\@sanitize@url \@url }%
\providecommand \@url [1]{\endgroup\@href {#1}{\urlprefix }}%
\providecommand \urlprefix  [0]{URL }%
\providecommand \Eprint [0]{\href }%
\providecommand \doibase [0]{https://doi.org/}%
\providecommand \selectlanguage [0]{\@gobble}%
\providecommand \bibinfo  [0]{\@secondoftwo}%
\providecommand \bibfield  [0]{\@secondoftwo}%
\providecommand \translation [1]{[#1]}%
\providecommand \BibitemOpen [0]{}%
\providecommand \bibitemStop [0]{}%
\providecommand \bibitemNoStop [0]{.\EOS\space}%
\providecommand \EOS [0]{\spacefactor3000\relax}%
\providecommand \BibitemShut  [1]{\csname bibitem#1\endcsname}%
\let\auto@bib@innerbib\@empty
\bibitem [{\citenamefont {Cohen}, \citenamefont {Mori-S\'{a}nchez},\ and\
  \citenamefont {Yang}(2012)}]{doi:10.1021/cr200107z}%
  \BibitemOpen
  \bibfield  {author} {\bibinfo {author} {\bibfnamefont {A.~J.}\ \bibnamefont
  {Cohen}}, \bibinfo {author} {\bibfnamefont {P.}~\bibnamefont
  {Mori-S\'{a}nchez}},\ and\ \bibinfo {author} {\bibfnamefont {W.}~\bibnamefont
  {Yang}},\ }\bibfield  {title} {\enquote {\bibinfo {title} {{Challenges for
  Density Functional Theory}},}\ }\href {https://doi.org/10.1021/cr200107z}
  {\bibfield  {journal} {\bibinfo  {journal} {Chem. Rev.}\ }\textbf {\bibinfo
  {volume} {112}},\ \bibinfo {pages} {289--320} (\bibinfo {year}
  {2012})}\BibitemShut {NoStop}%
\bibitem [{\citenamefont {Onida}, \citenamefont {Reining},\ and\ \citenamefont
  {Rubio}(2002)}]{Onida2002}%
  \BibitemOpen
  \bibfield  {author} {\bibinfo {author} {\bibfnamefont {G.}~\bibnamefont
  {Onida}}, \bibinfo {author} {\bibfnamefont {L.}~\bibnamefont {Reining}},\
  and\ \bibinfo {author} {\bibfnamefont {A.}~\bibnamefont {Rubio}},\ }\bibfield
   {title} {\enquote {\bibinfo {title} {{Electronic excitations:
  density-functional versus many-body Green's-function approaches}},}\
  }\href@noop {} {\bibfield  {journal} {\bibinfo  {journal} {Rev. Mod. Phys.}\
  }\textbf {\bibinfo {volume} {74}},\ \bibinfo {pages} {601--659} (\bibinfo
  {year} {2002})}\BibitemShut {NoStop}%
\bibitem [{\citenamefont {Hybertsen}\ and\ \citenamefont
  {Louie}(1985)}]{Hybertsen1985}%
  \BibitemOpen
  \bibfield  {author} {\bibinfo {author} {\bibfnamefont {M.~S.}\ \bibnamefont
  {Hybertsen}}\ and\ \bibinfo {author} {\bibfnamefont {S.~G.}\ \bibnamefont
  {Louie}},\ }\bibfield  {title} {\enquote {\bibinfo {title} {{First-Principles
  Theory of Quasiparticles: Calculation of Band Gaps in Semiconductors and
  Insulators}},}\ }\href {https://link.aps.org/doi/10.1103/PhysRevLett.55.1418}
  {\bibfield  {journal} {\bibinfo  {journal} {Phys. Rev. Lett.}\ }\textbf
  {\bibinfo {volume} {55}},\ \bibinfo {pages} {1418--1421} (\bibinfo {year}
  {1985})}\BibitemShut {NoStop}%
\bibitem [{\citenamefont {Aryasetiawan}\ and\ \citenamefont
  {Gunnarsson}(1998)}]{Aryasetiawan1998}%
  \BibitemOpen
  \bibfield  {author} {\bibinfo {author} {\bibfnamefont {F.}~\bibnamefont
  {Aryasetiawan}}\ and\ \bibinfo {author} {\bibfnamefont {O.}~\bibnamefont
  {Gunnarsson}},\ }\bibfield  {title} {\enquote {\bibinfo {title} {{The GW
  method}},}\ }\href {https://doi.org/10.1088\%2F0034-4885\%2F61\%2F3\%2F002}
  {\bibfield  {journal} {\bibinfo  {journal} {Rep. Prog. Phys.}\ }\textbf
  {\bibinfo {volume} {61}},\ \bibinfo {pages} {237--312} (\bibinfo {year}
  {1998})}\BibitemShut {NoStop}%
\bibitem [{\citenamefont {Hedin}(1999)}]{Hedin1999}%
  \BibitemOpen
  \bibfield  {author} {\bibinfo {author} {\bibfnamefont {L.}~\bibnamefont
  {Hedin}},\ }\bibfield  {title} {\enquote {\bibinfo {title} {{On correlation
  effects in electron spectroscopies and the $GW$ approximation}},}\ }\href
  {https://doi.org/10.1088/0953-8984/11/42/201} {\bibfield  {journal} {\bibinfo
   {journal} {J. Phys. Condens. Matter}\ }\textbf {\bibinfo {volume} {11}},\
  \bibinfo {pages} {R489--R528} (\bibinfo {year} {1999})}\BibitemShut {NoStop}%
\bibitem [{\citenamefont {Aulbur}, \citenamefont {J\"{o}nsson},\ and\
  \citenamefont {Wilkins}(2000)}]{Aulbur2000}%
  \BibitemOpen
  \bibfield  {author} {\bibinfo {author} {\bibfnamefont {W.~G.}\ \bibnamefont
  {Aulbur}}, \bibinfo {author} {\bibfnamefont {L.}~\bibnamefont
  {J\"{o}nsson}},\ and\ \bibinfo {author} {\bibfnamefont {J.~W.}\ \bibnamefont
  {Wilkins}},\ }\bibfield  {title} {\enquote {\bibinfo {title} {{Quasiparticle
  Calculations in Solids}},}\ }in\ \href
  {https://doi.org/10.1016/s0081-1947(08)60248-9} {\emph {\bibinfo {booktitle}
  {Solid State Physics}}}\ (\bibinfo  {publisher} {Elsevier},\ \bibinfo {year}
  {2000})\ pp.\ \bibinfo {pages} {1--218}\BibitemShut {NoStop}%
\bibitem [{\citenamefont {Friedrich}\ and\ \citenamefont
  {Schindlmayr}(2006)}]{Friedrich2006}%
  \BibitemOpen
  \bibfield  {author} {\bibinfo {author} {\bibfnamefont {C.}~\bibnamefont
  {Friedrich}}\ and\ \bibinfo {author} {\bibfnamefont {A.}~\bibnamefont
  {Schindlmayr}},\ }\bibfield  {title} {\enquote {\bibinfo {title} {{Many-Body
  Perturbation Theory : The GW Approximation}},}\ }\href@noop {} {\bibfield
  {journal} {\bibinfo  {journal} {Computing}\ }\textbf {\bibinfo {volume}
  {31}},\ \bibinfo {pages} {335--355} (\bibinfo {year} {2006})}\BibitemShut
  {NoStop}%
\bibitem [{\citenamefont {Kutepov}, \citenamefont {Savrasov},\ and\
  \citenamefont {Kotliar}(2009)}]{Kutepov2009}%
  \BibitemOpen
  \bibfield  {author} {\bibinfo {author} {\bibfnamefont {A.}~\bibnamefont
  {Kutepov}}, \bibinfo {author} {\bibfnamefont {S.~Y.}\ \bibnamefont
  {Savrasov}},\ and\ \bibinfo {author} {\bibfnamefont {G.}~\bibnamefont
  {Kotliar}},\ }\bibfield  {title} {\enquote {\bibinfo {title} {{Ground-state
  properties of simple elements from GW calculations}},}\ }\href@noop {}
  {\bibfield  {journal} {\bibinfo  {journal} {Phys. Rev. B}\ }\textbf {\bibinfo
  {volume} {80}},\ \bibinfo {pages} {1--4} (\bibinfo {year}
  {2009})}\BibitemShut {NoStop}%
\bibitem [{\citenamefont {Ke}(2011)}]{Ke2011}%
  \BibitemOpen
  \bibfield  {author} {\bibinfo {author} {\bibfnamefont {S.~H.}\ \bibnamefont
  {Ke}},\ }\bibfield  {title} {\enquote {\bibinfo {title} {{All-electron GW
  methods implemented in molecular orbital space: Ionization energy and
  electron affinity of conjugated molecules}},}\ }\href@noop {} {\bibfield
  {journal} {\bibinfo  {journal} {Phys. Rev. B}\ }\textbf {\bibinfo {volume}
  {84}},\ \bibinfo {pages} {1--6} (\bibinfo {year} {2011})}\BibitemShut
  {NoStop}%
\bibitem [{\citenamefont {Bruneval}(2012)}]{Bruneval2012}%
  \BibitemOpen
  \bibfield  {author} {\bibinfo {author} {\bibfnamefont {F.}~\bibnamefont
  {Bruneval}},\ }\bibfield  {title} {\enquote {\bibinfo {title} {{Ionization
  energy of atoms obtained from GW self-energy or from random phase
  approximation total energies}},}\ }\href@noop {} {\bibfield  {journal}
  {\bibinfo  {journal} {J. Chem. Phys.}\ }\textbf {\bibinfo {volume} {136}}
  (\bibinfo {year} {2012})}\BibitemShut {NoStop}%
\bibitem [{\citenamefont {{van Setten}}, \citenamefont {Weigend},\ and\
  \citenamefont {Evers}(2013)}]{VanSetten2013}%
  \BibitemOpen
  \bibfield  {author} {\bibinfo {author} {\bibfnamefont {M.~J.}\ \bibnamefont
  {{van Setten}}}, \bibinfo {author} {\bibfnamefont {F.}~\bibnamefont
  {Weigend}},\ and\ \bibinfo {author} {\bibfnamefont {F.}~\bibnamefont
  {Evers}},\ }\bibfield  {title} {\enquote {\bibinfo {title} {{The GW-method
  for quantum chemistry applications: Theory and implementation}},}\
  }\href@noop {} {\bibfield  {journal} {\bibinfo  {journal} {J. Chem. Theory
  Comput.}\ }\textbf {\bibinfo {volume} {9}},\ \bibinfo {pages} {232--246}
  (\bibinfo {year} {2013})}\BibitemShut {NoStop}%
\bibitem [{\citenamefont {van Setten}\ \emph {et~al.}(2015)\citenamefont {van
  Setten}, \citenamefont {Caruso}, \citenamefont {Sharifzadeh}, \citenamefont
  {Ren}, \citenamefont {Scheffler}, \citenamefont {Liu}, \citenamefont
  {Lischner}, \citenamefont {Lin}, \citenamefont {Deslippe}, \citenamefont
  {Louie}, \citenamefont {Yang}, \citenamefont {Weigend}, \citenamefont
  {Neaton}, \citenamefont {Evers},\ and\ \citenamefont
  {Rinke}}]{VanSetten2015}%
  \BibitemOpen
  \bibfield  {author} {\bibinfo {author} {\bibfnamefont {M.~J.}\ \bibnamefont
  {van Setten}}, \bibinfo {author} {\bibfnamefont {F.}~\bibnamefont {Caruso}},
  \bibinfo {author} {\bibfnamefont {S.}~\bibnamefont {Sharifzadeh}}, \bibinfo
  {author} {\bibfnamefont {X.}~\bibnamefont {Ren}}, \bibinfo {author}
  {\bibfnamefont {M.}~\bibnamefont {Scheffler}}, \bibinfo {author}
  {\bibfnamefont {F.}~\bibnamefont {Liu}}, \bibinfo {author} {\bibfnamefont
  {J.}~\bibnamefont {Lischner}}, \bibinfo {author} {\bibfnamefont
  {L.}~\bibnamefont {Lin}}, \bibinfo {author} {\bibfnamefont {J.~R.}\
  \bibnamefont {Deslippe}}, \bibinfo {author} {\bibfnamefont {S.~G.}\
  \bibnamefont {Louie}}, \bibinfo {author} {\bibfnamefont {C.}~\bibnamefont
  {Yang}}, \bibinfo {author} {\bibfnamefont {F.}~\bibnamefont {Weigend}},
  \bibinfo {author} {\bibfnamefont {J.~B.}\ \bibnamefont {Neaton}}, \bibinfo
  {author} {\bibfnamefont {F.}~\bibnamefont {Evers}},\ and\ \bibinfo {author}
  {\bibfnamefont {P.}~\bibnamefont {Rinke}},\ }\bibfield  {title} {\enquote
  {\bibinfo {title} {{GW100: Benchmarking G0W0 for Molecular Systems}},}\
  }\href@noop {} {\bibfield  {journal} {\bibinfo  {journal} {J. Chem. Theory
  Comput.}\ }\textbf {\bibinfo {volume} {11}},\ \bibinfo {pages} {5665--5687}
  (\bibinfo {year} {2015})}\BibitemShut {NoStop}%
\bibitem [{\citenamefont {Reining}(2017)}]{Reining2017}%
  \BibitemOpen
  \bibfield  {author} {\bibinfo {author} {\bibfnamefont {L.}~\bibnamefont
  {Reining}},\ }\bibfield  {title} {\enquote {\bibinfo {title} {{The GW
  approximation: content, successes and limitations}},}\ }\href
  {https://doi.org/10.1002/wcms.1344} {\bibfield  {journal} {\bibinfo
  {journal} {Wiley Interdiscip. Rev. Comput. Mol. Sci.}\ }\textbf {\bibinfo
  {volume} {8}} (\bibinfo {year} {2017})}\BibitemShut {NoStop}%
\bibitem [{\citenamefont {Golze}, \citenamefont {Dvorak},\ and\ \citenamefont
  {Rinke}(2019)}]{Golze2019}%
  \BibitemOpen
  \bibfield  {author} {\bibinfo {author} {\bibfnamefont {D.}~\bibnamefont
  {Golze}}, \bibinfo {author} {\bibfnamefont {M.}~\bibnamefont {Dvorak}},\ and\
  \bibinfo {author} {\bibfnamefont {P.}~\bibnamefont {Rinke}},\ }\bibfield
  {title} {\enquote {\bibinfo {title} {{The GW Compendium: A Practical Guide to
  Theoretical Photoemission Spectroscopy}},}\ }\href@noop {} {\bibfield
  {journal} {\bibinfo  {journal} {Front. Chem.}\ }\textbf {\bibinfo {volume}
  {7}} (\bibinfo {year} {2019})}\BibitemShut {NoStop}%
\bibitem [{\citenamefont {Hedin}(1965)}]{Hedin1965}%
  \BibitemOpen
  \bibfield  {author} {\bibinfo {author} {\bibfnamefont {L.}~\bibnamefont
  {Hedin}},\ }\bibfield  {title} {\enquote {\bibinfo {title} {{New Method for
  Calculating the One-Particle Green's Function with Application to the
  Electron-Gas Problem}},}\ }\href
  {https://link.aps.org/doi/10.1103/PhysRev.139.A796} {\bibfield  {journal}
  {\bibinfo  {journal} {Phys. Rev.}\ }\textbf {\bibinfo {volume} {139}},\
  \bibinfo {pages} {A796--A823} (\bibinfo {year} {1965})}\BibitemShut {NoStop}%
\bibitem [{\citenamefont {Hedin}\ and\ \citenamefont
  {Lundqvist}(1970)}]{Hedin1970}%
  \BibitemOpen
  \bibfield  {author} {\bibinfo {author} {\bibfnamefont {L.}~\bibnamefont
  {Hedin}}\ and\ \bibinfo {author} {\bibfnamefont {S.}~\bibnamefont
  {Lundqvist}},\ }\bibfield  {title} {\enquote {\bibinfo {title} {{Effects of
  Electron-Electron and Electron-Phonon Interactions on the One-Electron States
  of Solids}},}\ }in\ \href {https://doi.org/10.1016/s0081-1947(08)60615-3}
  {\emph {\bibinfo {booktitle} {Solid State Physics}}}\ (\bibinfo  {publisher}
  {Elsevier},\ \bibinfo {year} {1970})\ pp.\ \bibinfo {pages}
  {1--181}\BibitemShut {NoStop}%
\bibitem [{\citenamefont {Langreth}\ and\ \citenamefont
  {Perdew}(1977)}]{Langreth1977}%
  \BibitemOpen
  \bibfield  {author} {\bibinfo {author} {\bibfnamefont {D.~C.}\ \bibnamefont
  {Langreth}}\ and\ \bibinfo {author} {\bibfnamefont {J.~P.}\ \bibnamefont
  {Perdew}},\ }\bibfield  {title} {\enquote {\bibinfo {title}
  {{Exchange-correlation energy of a metallic surface: Wave-vector
  analysis}},}\ }\href {https://doi.org/10.1103/physrevb.15.2884} {\bibfield
  {journal} {\bibinfo  {journal} {Phys. Rev. B}\ }\textbf {\bibinfo {volume}
  {15}},\ \bibinfo {pages} {2884--2901} (\bibinfo {year} {1977})}\BibitemShut
  {NoStop}%
\bibitem [{\citenamefont {Chong}(1995)}]{Chong1995}%
  \BibitemOpen
  \bibfield  {author} {\bibinfo {author} {\bibfnamefont {D.~P.}\ \bibnamefont
  {Chong}},\ }\href {https://doi.org/10.1142/2914} {\emph {\bibinfo {title}
  {{Recent Advances in Density Functional Methods}}}}\ (\bibinfo  {publisher}
  {{World} {Scientific}},\ \bibinfo {year} {1995})\BibitemShut {NoStop}%
\bibitem [{\citenamefont {He{\ss}elmann}\ and\ \citenamefont
  {G\"{o}rling}(2010)}]{Heselmann2010}%
  \BibitemOpen
  \bibfield  {author} {\bibinfo {author} {\bibfnamefont {A.}~\bibnamefont
  {He{\ss}elmann}}\ and\ \bibinfo {author} {\bibfnamefont {A.}~\bibnamefont
  {G\"{o}rling}},\ }\bibfield  {title} {\enquote {\bibinfo {title} {{Random
  phase approximation correlation energies with exact Kohn-Sham exchange}},}\
  }\href {https://doi.org/10.1080/00268970903476662} {\bibfield  {journal}
  {\bibinfo  {journal} {Mol. Phys.}\ }\textbf {\bibinfo {volume} {108}},\
  \bibinfo {pages} {359--372} (\bibinfo {year} {2010})}\BibitemShut {NoStop}%
\bibitem [{\citenamefont {Ren}\ \emph {et~al.}(2012{\natexlab{a}})\citenamefont
  {Ren}, \citenamefont {Rinke}, \citenamefont {Joas},\ and\ \citenamefont
  {Scheffler}}]{Ren2012}%
  \BibitemOpen
  \bibfield  {author} {\bibinfo {author} {\bibfnamefont {X.}~\bibnamefont
  {Ren}}, \bibinfo {author} {\bibfnamefont {P.}~\bibnamefont {Rinke}}, \bibinfo
  {author} {\bibfnamefont {C.}~\bibnamefont {Joas}},\ and\ \bibinfo {author}
  {\bibfnamefont {M.}~\bibnamefont {Scheffler}},\ }\bibfield  {title} {\enquote
  {\bibinfo {title} {{Random-phase approximation and its applications in
  computational chemistry and materials science}},}\ }\href
  {https://doi.org/10.1007/s10853-012-6570-4} {\bibfield  {journal} {\bibinfo
  {journal} {J. Mater. Sci.}\ }\textbf {\bibinfo {volume} {47}},\ \bibinfo
  {pages} {7447--7471} (\bibinfo {year} {2012}{\natexlab{a}})}\BibitemShut
  {NoStop}%
\bibitem [{\citenamefont {Stan}, \citenamefont {Dahlen},\ and\ \citenamefont
  {van Leeuwen}(2009)}]{Leeuwen09}%
  \BibitemOpen
  \bibfield  {author} {\bibinfo {author} {\bibfnamefont {A.}~\bibnamefont
  {Stan}}, \bibinfo {author} {\bibfnamefont {N.~E.}\ \bibnamefont {Dahlen}},\
  and\ \bibinfo {author} {\bibfnamefont {R.}~\bibnamefont {van Leeuwen}},\
  }\bibfield  {title} {\enquote {\bibinfo {title} {{Levels of self-consistency
  in the GW approximation}},}\ }\href {https://doi.org/10.1063/1.3089567}
  {\bibfield  {journal} {\bibinfo  {journal} {J. Chem. Phys.}\ }\textbf
  {\bibinfo {volume} {130}},\ \bibinfo {pages} {114105} (\bibinfo {year}
  {2009})}\BibitemShut {NoStop}%
\bibitem [{\citenamefont {Foerster}, \citenamefont {Koval},\ and\ \citenamefont
  {Snchez-Portal}(2011)}]{Foerster2011}%
  \BibitemOpen
  \bibfield  {author} {\bibinfo {author} {\bibfnamefont {D.}~\bibnamefont
  {Foerster}}, \bibinfo {author} {\bibfnamefont {P.}~\bibnamefont {Koval}},\
  and\ \bibinfo {author} {\bibfnamefont {D.}~\bibnamefont {Snchez-Portal}},\
  }\bibfield  {title} {\enquote {\bibinfo {title} {{An O(N3) implementation of
  Hedins GW approximation for molecules}},}\ }\href@noop {} {\bibfield
  {journal} {\bibinfo  {journal} {J. Chem. Phys.}\ }\textbf {\bibinfo {volume}
  {135}},\ \bibinfo {pages} {1--19} (\bibinfo {year} {2011})}\BibitemShut
  {NoStop}%
\bibitem [{\citenamefont {Bruneval}\ and\ \citenamefont
  {Marques}(2013)}]{Bruneval2013}%
  \BibitemOpen
  \bibfield  {author} {\bibinfo {author} {\bibfnamefont {F.}~\bibnamefont
  {Bruneval}}\ and\ \bibinfo {author} {\bibfnamefont {M.~A.~L.}\ \bibnamefont
  {Marques}},\ }\bibfield  {title} {\enquote {\bibinfo {title} {{Benchmarking
  the Starting Points of the GW Approximation for Molecules}},}\ }\href@noop {}
  {\bibfield  {journal} {\bibinfo  {journal} {J. Chem. Theory Comput.}\
  }\textbf {\bibinfo {volume} {9}},\ \bibinfo {pages} {324--329} (\bibinfo
  {year} {2013})}\BibitemShut {NoStop}%
\bibitem [{\citenamefont {Liu}\ \emph {et~al.}(2016)\citenamefont {Liu},
  \citenamefont {Kaltak}, \citenamefont {Klime\ifmmode~\check{s}\else
  \v{s}\fi{}},\ and\ \citenamefont {Kresse}}]{PhysRevB.94.165109}%
  \BibitemOpen
  \bibfield  {author} {\bibinfo {author} {\bibfnamefont {P.}~\bibnamefont
  {Liu}}, \bibinfo {author} {\bibfnamefont {M.}~\bibnamefont {Kaltak}},
  \bibinfo {author} {\bibfnamefont {J.}~\bibnamefont
  {Klime\ifmmode~\check{s}\else \v{s}\fi{}}},\ and\ \bibinfo {author}
  {\bibfnamefont {G.}~\bibnamefont {Kresse}},\ }\bibfield  {title} {\enquote
  {\bibinfo {title} {{Cubic scaling $GW$: Towards fast quasiparticle
  calculations}},}\ }\href
  {https://link.aps.org/doi/10.1103/PhysRevB.94.165109} {\bibfield  {journal}
  {\bibinfo  {journal} {Phys. Rev. B}\ }\textbf {\bibinfo {volume} {94}},\
  \bibinfo {pages} {165109} (\bibinfo {year} {2016})}\BibitemShut {NoStop}%
\bibitem [{\citenamefont {Knight}\ \emph {et~al.}(2016)\citenamefont {Knight},
  \citenamefont {Wang}, \citenamefont {Gallandi}, \citenamefont
  {Dolgounitcheva}, \citenamefont {Ren}, \citenamefont {Ortiz}, \citenamefont
  {Rinke}, \citenamefont {K{\"{o}}rzd{\"{o}}rfer},\ and\ \citenamefont
  {Marom}}]{Knight2016}%
  \BibitemOpen
  \bibfield  {author} {\bibinfo {author} {\bibfnamefont {J.~W.}\ \bibnamefont
  {Knight}}, \bibinfo {author} {\bibfnamefont {X.}~\bibnamefont {Wang}},
  \bibinfo {author} {\bibfnamefont {L.}~\bibnamefont {Gallandi}}, \bibinfo
  {author} {\bibfnamefont {O.}~\bibnamefont {Dolgounitcheva}}, \bibinfo
  {author} {\bibfnamefont {X.}~\bibnamefont {Ren}}, \bibinfo {author}
  {\bibfnamefont {J.~V.}\ \bibnamefont {Ortiz}}, \bibinfo {author}
  {\bibfnamefont {P.}~\bibnamefont {Rinke}}, \bibinfo {author} {\bibfnamefont
  {T.}~\bibnamefont {K{\"{o}}rzd{\"{o}}rfer}},\ and\ \bibinfo {author}
  {\bibfnamefont {N.}~\bibnamefont {Marom}},\ }\bibfield  {title} {\enquote
  {\bibinfo {title} {{Accurate Ionization Potentials and Electron Affinities of
  Acceptor Molecules III: A Benchmark of GW Methods}},}\ }\href@noop {}
  {\bibfield  {journal} {\bibinfo  {journal} {J. Chem. Theory Comput.}\
  }\textbf {\bibinfo {volume} {12}},\ \bibinfo {pages} {615--626} (\bibinfo
  {year} {2016})}\BibitemShut {NoStop}%
\bibitem [{\citenamefont {Maggio}\ \emph {et~al.}(2017)\citenamefont {Maggio},
  \citenamefont {Liu}, \citenamefont {van Setten},\ and\ \citenamefont
  {Kresse}}]{Maggio2017}%
  \BibitemOpen
  \bibfield  {author} {\bibinfo {author} {\bibfnamefont {E.}~\bibnamefont
  {Maggio}}, \bibinfo {author} {\bibfnamefont {P.}~\bibnamefont {Liu}},
  \bibinfo {author} {\bibfnamefont {M.~J.}\ \bibnamefont {van Setten}},\ and\
  \bibinfo {author} {\bibfnamefont {G.}~\bibnamefont {Kresse}},\ }\bibfield
  {title} {\enquote {\bibinfo {title} {{$GW_100$: A Plane Wave Perspective for
  Small Molecules}},}\ }\href {https://doi.org/10.1021/acs.jctc.6b01150}
  {\bibfield  {journal} {\bibinfo  {journal} {J. Chem. Theory Comput.}\
  }\textbf {\bibinfo {volume} {13}},\ \bibinfo {pages} {635--648} (\bibinfo
  {year} {2017})}\BibitemShut {NoStop}%
\bibitem [{\citenamefont {Bruneval}, \citenamefont {Dattani},\ and\
  \citenamefont {van Setten}(2021)}]{Bruneval2021}%
  \BibitemOpen
  \bibfield  {author} {\bibinfo {author} {\bibfnamefont {F.}~\bibnamefont
  {Bruneval}}, \bibinfo {author} {\bibfnamefont {N.}~\bibnamefont {Dattani}},\
  and\ \bibinfo {author} {\bibfnamefont {M.~J.}\ \bibnamefont {van Setten}},\
  }\bibfield  {title} {\enquote {\bibinfo {title} {{The GW Miracle in Many-Body
  Perturbation Theory for the Ionization Potential of Molecules}},}\ }\href
  {https://doi.org/10.3389/fchem.2021.749779} {\bibfield  {journal} {\bibinfo
  {journal} {Front. Chem.}\ }\textbf {\bibinfo {volume} {9}} (\bibinfo {year}
  {2021})}\BibitemShut {NoStop}%
\bibitem [{\citenamefont {von Barth}\ and\ \citenamefont
  {Holm}(1996)}]{vonBarth1996}%
  \BibitemOpen
  \bibfield  {author} {\bibinfo {author} {\bibfnamefont {U.}~\bibnamefont {von
  Barth}}\ and\ \bibinfo {author} {\bibfnamefont {B.}~\bibnamefont {Holm}},\
  }\bibfield  {title} {\enquote {\bibinfo {title} {{Self-consistent
  $\mathit{GW}_0$ results for the electron gas: Fixed screened potential
  $\mathit{W}_0$ within the random-phase approximation}},}\ }\href
  {https://link.aps.org/doi/10.1103/PhysRevB.54.8411} {\bibfield  {journal}
  {\bibinfo  {journal} {Phys. Rev. B}\ }\textbf {\bibinfo {volume} {54}},\
  \bibinfo {pages} {8411--8419} (\bibinfo {year} {1996})}\BibitemShut {NoStop}%
\bibitem [{\citenamefont {Holm}\ and\ \citenamefont {von
  Barth}(1998)}]{Holm1998}%
  \BibitemOpen
  \bibfield  {author} {\bibinfo {author} {\bibfnamefont {B.}~\bibnamefont
  {Holm}}\ and\ \bibinfo {author} {\bibfnamefont {U.}~\bibnamefont {von
  Barth}},\ }\bibfield  {title} {\enquote {\bibinfo {title} {{Fully
  self-consistent $\mathrm{GW}$ self-energy of the electron gas}},}\
  }\href@noop {} {\bibfield  {journal} {\bibinfo  {journal} {Phys. Rev. B}\
  }\textbf {\bibinfo {volume} {57}},\ \bibinfo {pages} {2108--2117} (\bibinfo
  {year} {1998})}\BibitemShut {NoStop}%
\bibitem [{\citenamefont {Sch\"one}\ and\ \citenamefont
  {Eguiluz}(1998)}]{Schone1998}%
  \BibitemOpen
  \bibfield  {author} {\bibinfo {author} {\bibfnamefont {W.-D.}\ \bibnamefont
  {Sch\"one}}\ and\ \bibinfo {author} {\bibfnamefont {A.~G.}\ \bibnamefont
  {Eguiluz}},\ }\bibfield  {title} {\enquote {\bibinfo {title}
  {{Self-Consistent Calculations of Quasiparticle States in Metals and
  Semiconductors}},}\ }\href
  {https://link.aps.org/doi/10.1103/PhysRevLett.81.1662} {\bibfield  {journal}
  {\bibinfo  {journal} {Phys. Rev. Lett.}\ }\textbf {\bibinfo {volume} {81}},\
  \bibinfo {pages} {1662--1665} (\bibinfo {year} {1998})}\BibitemShut {NoStop}%
\bibitem [{\citenamefont {Garc{\'{\i}}a-Gonz{\'{a}}lez}\ and\ \citenamefont
  {Godby}(2001)}]{GarciaGonzalez2001}%
  \BibitemOpen
  \bibfield  {author} {\bibinfo {author} {\bibfnamefont {P.}~\bibnamefont
  {Garc{\'{\i}}a-Gonz{\'{a}}lez}}\ and\ \bibinfo {author} {\bibfnamefont
  {R.~W.}\ \bibnamefont {Godby}},\ }\bibfield  {title} {\enquote {\bibinfo
  {title} {{Self-consistent calculation of total energies of the electron gas
  using many-body perturbation theory}},}\ }\href
  {https://doi.org/10.1103/physrevb.63.075112} {\bibfield  {journal} {\bibinfo
  {journal} {Phys. Rev. B}\ }\textbf {\bibinfo {volume} {63}} (\bibinfo {year}
  {2001})}\BibitemShut {NoStop}%
\bibitem [{\citenamefont {Faleev}, \citenamefont {{van Schilfgaarde}},\ and\
  \citenamefont {Kotani}(2004)}]{Faleev2004}%
  \BibitemOpen
  \bibfield  {author} {\bibinfo {author} {\bibfnamefont {S.~V.}\ \bibnamefont
  {Faleev}}, \bibinfo {author} {\bibfnamefont {M.}~\bibnamefont {{van
  Schilfgaarde}}},\ and\ \bibinfo {author} {\bibfnamefont {T.}~\bibnamefont
  {Kotani}},\ }\bibfield  {title} {\enquote {\bibinfo {title} {{All-electron
  self-consistent GW approximation: Application to Si, MnO, and NiO}},}\
  }\href@noop {} {\bibfield  {journal} {\bibinfo  {journal} {Phys. Rev. Lett.}\
  }\textbf {\bibinfo {volume} {93}},\ \bibinfo {pages} {12--15} (\bibinfo
  {year} {2004})}\BibitemShut {NoStop}%
\bibitem [{\citenamefont {van Schilfgaarde}, \citenamefont {Kotani},\ and\
  \citenamefont {Faleev}(2006)}]{VanSchilfgaarde2005}%
  \BibitemOpen
  \bibfield  {author} {\bibinfo {author} {\bibfnamefont {M.}~\bibnamefont {van
  Schilfgaarde}}, \bibinfo {author} {\bibfnamefont {T.}~\bibnamefont
  {Kotani}},\ and\ \bibinfo {author} {\bibfnamefont {S.}~\bibnamefont
  {Faleev}},\ }\bibfield  {title} {\enquote {\bibinfo {title} {{Quasiparticle
  Self-Consistent $GW$ Theory}},}\ }\href@noop {} {\bibfield  {journal}
  {\bibinfo  {journal} {Phys. Rev. Lett.}\ }\textbf {\bibinfo {volume} {96}},\
  \bibinfo {pages} {226402} (\bibinfo {year} {2006})}\BibitemShut {NoStop}%
\bibitem [{\citenamefont {Stan}, \citenamefont {Dahlen},\ and\ \citenamefont
  {{van Leeuwen}}(2006)}]{Stan2006}%
  \BibitemOpen
  \bibfield  {author} {\bibinfo {author} {\bibfnamefont {A.}~\bibnamefont
  {Stan}}, \bibinfo {author} {\bibfnamefont {N.~E.}\ \bibnamefont {Dahlen}},\
  and\ \bibinfo {author} {\bibfnamefont {R.}~\bibnamefont {{van Leeuwen}}},\
  }\bibfield  {title} {\enquote {\bibinfo {title} {{Fully self-consistent GW
  calculations for atoms and molecules}},}\ }\href@noop {} {\bibfield
  {journal} {\bibinfo  {journal} {Europhys. Lett.}\ }\textbf {\bibinfo {volume}
  {76}},\ \bibinfo {pages} {298--304} (\bibinfo {year} {2006})}\BibitemShut
  {NoStop}%
\bibitem [{\citenamefont {Kotani}, \citenamefont {{van Schilfgaarde}},\ and\
  \citenamefont {Faleev}(2007)}]{Kotani2007}%
  \BibitemOpen
  \bibfield  {author} {\bibinfo {author} {\bibfnamefont {T.}~\bibnamefont
  {Kotani}}, \bibinfo {author} {\bibfnamefont {M.}~\bibnamefont {{van
  Schilfgaarde}}},\ and\ \bibinfo {author} {\bibfnamefont {S.~V.}\ \bibnamefont
  {Faleev}},\ }\bibfield  {title} {\enquote {\bibinfo {title} {{Quasiparticle
  self-consistent GW method: A basis for the independent-particle
  approximation}},}\ }\href@noop {} {\bibfield  {journal} {\bibinfo  {journal}
  {Phys. Rev. B}\ }\textbf {\bibinfo {volume} {76}},\ \bibinfo {pages} {1--24}
  (\bibinfo {year} {2007})}\BibitemShut {NoStop}%
\bibitem [{\citenamefont {Shishkin}\ and\ \citenamefont
  {Kresse}(2007)}]{Shishkin2007}%
  \BibitemOpen
  \bibfield  {author} {\bibinfo {author} {\bibfnamefont {M.}~\bibnamefont
  {Shishkin}}\ and\ \bibinfo {author} {\bibfnamefont {G.}~\bibnamefont
  {Kresse}},\ }\bibfield  {title} {\enquote {\bibinfo {title} {{Self-consistent
  $\mathit{GW}$ calculations for semiconductors and insulators}},}\ }\href
  {https://doi.org/10.1103/physrevb.75.235102} {\bibfield  {journal} {\bibinfo
  {journal} {Phys. Rev. B}\ }\textbf {\bibinfo {volume} {75}} (\bibinfo {year}
  {2007})}\BibitemShut {NoStop}%
\bibitem [{\citenamefont {Caruso}\ \emph {et~al.}(2012)\citenamefont {Caruso},
  \citenamefont {Rinke}, \citenamefont {Ren}, \citenamefont {Scheffler},\ and\
  \citenamefont {Rubio}}]{Caruso2012}%
  \BibitemOpen
  \bibfield  {author} {\bibinfo {author} {\bibfnamefont {F.}~\bibnamefont
  {Caruso}}, \bibinfo {author} {\bibfnamefont {P.}~\bibnamefont {Rinke}},
  \bibinfo {author} {\bibfnamefont {X.}~\bibnamefont {Ren}}, \bibinfo {author}
  {\bibfnamefont {M.}~\bibnamefont {Scheffler}},\ and\ \bibinfo {author}
  {\bibfnamefont {A.}~\bibnamefont {Rubio}},\ }\bibfield  {title} {\enquote
  {\bibinfo {title} {{Unified description of ground and excited states of
  finite systems: The self-consistent $GW$ approach}},}\ }\href
  {https://link.aps.org/doi/10.1103/PhysRevB.86.081102} {\bibfield  {journal}
  {\bibinfo  {journal} {Phys. Rev. B}\ }\textbf {\bibinfo {volume} {86}},\
  \bibinfo {pages} {81102} (\bibinfo {year} {2012})}\BibitemShut {NoStop}%
\bibitem [{\citenamefont {Bruneval}\ \emph {et~al.}(2016)\citenamefont
  {Bruneval}, \citenamefont {Rangel}, \citenamefont {Hamed}, \citenamefont
  {Shao}, \citenamefont {Yang},\ and\ \citenamefont {Neaton}}]{Bruneval2016}%
  \BibitemOpen
  \bibfield  {author} {\bibinfo {author} {\bibfnamefont {F.}~\bibnamefont
  {Bruneval}}, \bibinfo {author} {\bibfnamefont {T.}~\bibnamefont {Rangel}},
  \bibinfo {author} {\bibfnamefont {S.~M.}\ \bibnamefont {Hamed}}, \bibinfo
  {author} {\bibfnamefont {M.}~\bibnamefont {Shao}}, \bibinfo {author}
  {\bibfnamefont {C.}~\bibnamefont {Yang}},\ and\ \bibinfo {author}
  {\bibfnamefont {J.~B.}\ \bibnamefont {Neaton}},\ }\bibfield  {title}
  {\enquote {\bibinfo {title} {{MOLGW 1: Many-body perturbation theory software
  for atoms, molecules, and clusters}},}\ }\href
  {http://dx.doi.org/10.1016/j.cpc.2016.06.019} {\bibfield  {journal} {\bibinfo
   {journal} {Comput. Phys. Commun.}\ }\textbf {\bibinfo {volume} {208}},\
  \bibinfo {pages} {149--161} (\bibinfo {year} {2016})}\BibitemShut {NoStop}%
\bibitem [{\citenamefont {Kaplan}\ \emph {et~al.}(2016)\citenamefont {Kaplan},
  \citenamefont {Harding}, \citenamefont {Seiler}, \citenamefont {Weigend},
  \citenamefont {Evers},\ and\ \citenamefont {van Setten}}]{Kaplan2016}%
  \BibitemOpen
  \bibfield  {author} {\bibinfo {author} {\bibfnamefont {F.}~\bibnamefont
  {Kaplan}}, \bibinfo {author} {\bibfnamefont {M.~E.}\ \bibnamefont {Harding}},
  \bibinfo {author} {\bibfnamefont {C.}~\bibnamefont {Seiler}}, \bibinfo
  {author} {\bibfnamefont {F.}~\bibnamefont {Weigend}}, \bibinfo {author}
  {\bibfnamefont {F.}~\bibnamefont {Evers}},\ and\ \bibinfo {author}
  {\bibfnamefont {M.~J.}\ \bibnamefont {van Setten}},\ }\bibfield  {title}
  {\enquote {\bibinfo {title} {{Quasi-Particle Self-Consistent GW for
  Molecules}},}\ }\href@noop {} {\bibfield  {journal} {\bibinfo  {journal} {J.
  Chem. Theory Comput.}\ }\textbf {\bibinfo {volume} {12}},\ \bibinfo {pages}
  {2528--2541} (\bibinfo {year} {2016})}\BibitemShut {NoStop}%
\bibitem [{\citenamefont {Jin}, \citenamefont {Su},\ and\ \citenamefont
  {Yang}(2019)}]{Jin2019}%
  \BibitemOpen
  \bibfield  {author} {\bibinfo {author} {\bibfnamefont {Y.}~\bibnamefont
  {Jin}}, \bibinfo {author} {\bibfnamefont {N.~Q.}\ \bibnamefont {Su}},\ and\
  \bibinfo {author} {\bibfnamefont {W.}~\bibnamefont {Yang}},\ }\bibfield
  {title} {\enquote {\bibinfo {title} {{Renormalized Singles Green's Function
  for Quasi-Particle Calculations beyond the G0W0 Approximation}},}\
  }\href@noop {} {\bibfield  {journal} {\bibinfo  {journal} {J. Phys. Chem.
  Lett.}\ }\textbf {\bibinfo {volume} {10}},\ \bibinfo {pages} {447--452}
  (\bibinfo {year} {2019})}\BibitemShut {NoStop}%
\bibitem [{\citenamefont {Duchemin}\ and\ \citenamefont
  {Blase}(2020)}]{Duchemin2020}%
  \BibitemOpen
  \bibfield  {author} {\bibinfo {author} {\bibfnamefont {I.}~\bibnamefont
  {Duchemin}}\ and\ \bibinfo {author} {\bibfnamefont {X.}~\bibnamefont
  {Blase}},\ }\bibfield  {title} {\enquote {\bibinfo {title} {{Robust
  Analytic-Continuation Approach to Many-Body GW Calculations}},}\ }\href@noop
  {} {\bibfield  {journal} {\bibinfo  {journal} {J. Chem. Theory Comput.}\
  }\textbf {\bibinfo {volume} {16}},\ \bibinfo {pages} {1742--1756} (\bibinfo
  {year} {2020})}\BibitemShut {NoStop}%
\bibitem [{\citenamefont {Duchemin}\ and\ \citenamefont
  {Blase}(2021)}]{Duchemin2021}%
  \BibitemOpen
  \bibfield  {author} {\bibinfo {author} {\bibfnamefont {I.}~\bibnamefont
  {Duchemin}}\ and\ \bibinfo {author} {\bibfnamefont {X.}~\bibnamefont
  {Blase}},\ }\bibfield  {title} {\enquote {\bibinfo {title} {{Cubic-Scaling
  All-Electron GW Calculations with a Separable Density-Fitting Space-Time
  Approach}},}\ }\href@noop {} {\bibfield  {journal} {\bibinfo  {journal} {J.
  Chem. Theory Comput.}\ }\textbf {\bibinfo {volume} {17}},\ \bibinfo {pages}
  {2383--2393} (\bibinfo {year} {2021})}\BibitemShut {NoStop}%
\bibitem [{\citenamefont {Yeh}\ \emph {et~al.}(2022)\citenamefont {Yeh},
  \citenamefont {Iskakov}, \citenamefont {Zgid},\ and\ \citenamefont
  {Gull}}]{PhysRevB.106.235104}%
  \BibitemOpen
  \bibfield  {author} {\bibinfo {author} {\bibfnamefont {C.-N.}\ \bibnamefont
  {Yeh}}, \bibinfo {author} {\bibfnamefont {S.}~\bibnamefont {Iskakov}},
  \bibinfo {author} {\bibfnamefont {D.}~\bibnamefont {Zgid}},\ and\ \bibinfo
  {author} {\bibfnamefont {E.}~\bibnamefont {Gull}},\ }\bibfield  {title}
  {\enquote {\bibinfo {title} {{Fully self-consistent finite-temperature $GW$
  in Gaussian Bloch orbitals for solids}},}\ }\href
  {https://link.aps.org/doi/10.1103/PhysRevB.106.235104} {\bibfield  {journal}
  {\bibinfo  {journal} {Phys. Rev. B}\ }\textbf {\bibinfo {volume} {106}},\
  \bibinfo {pages} {235104} (\bibinfo {year} {2022})}\BibitemShut {NoStop}%
\bibitem [{\citenamefont {Ren}\ \emph {et~al.}(2012{\natexlab{b}})\citenamefont
  {Ren}, \citenamefont {Rinke}, \citenamefont {Blum}, \citenamefont
  {Wieferink}, \citenamefont {Tkatchenko}, \citenamefont {Sanfilippo},
  \citenamefont {Reuter},\ and\ \citenamefont {Scheffler}}]{Ren_2012}%
  \BibitemOpen
  \bibfield  {author} {\bibinfo {author} {\bibfnamefont {X.}~\bibnamefont
  {Ren}}, \bibinfo {author} {\bibfnamefont {P.}~\bibnamefont {Rinke}}, \bibinfo
  {author} {\bibfnamefont {V.}~\bibnamefont {Blum}}, \bibinfo {author}
  {\bibfnamefont {J.}~\bibnamefont {Wieferink}}, \bibinfo {author}
  {\bibfnamefont {A.}~\bibnamefont {Tkatchenko}}, \bibinfo {author}
  {\bibfnamefont {A.}~\bibnamefont {Sanfilippo}}, \bibinfo {author}
  {\bibfnamefont {K.}~\bibnamefont {Reuter}},\ and\ \bibinfo {author}
  {\bibfnamefont {M.}~\bibnamefont {Scheffler}},\ }\bibfield  {title} {\enquote
  {\bibinfo {title} {{Resolution-of-identity approach to Hartree–Fock, hybrid
  density functionals, RPA, MP2 and GW with numeric atom-centered orbital basis
  functions}},}\ }\href {https://dx.doi.org/10.1088/1367-2630/14/5/053020}
  {\bibfield  {journal} {\bibinfo  {journal} {New J. Phys.}\ }\textbf {\bibinfo
  {volume} {14}},\ \bibinfo {pages} {053020} (\bibinfo {year}
  {2012}{\natexlab{b}})}\BibitemShut {NoStop}%
\bibitem [{\citenamefont {Engel}\ \emph {et~al.}(1991)\citenamefont {Engel},
  \citenamefont {Farid}, \citenamefont {Nex},\ and\ \citenamefont
  {March}}]{PhysRevB.44.13356}%
  \BibitemOpen
  \bibfield  {author} {\bibinfo {author} {\bibfnamefont {G.~E.}\ \bibnamefont
  {Engel}}, \bibinfo {author} {\bibfnamefont {B.}~\bibnamefont {Farid}},
  \bibinfo {author} {\bibfnamefont {C.~M.~M.}\ \bibnamefont {Nex}},\ and\
  \bibinfo {author} {\bibfnamefont {N.~H.}\ \bibnamefont {March}},\ }\bibfield
  {title} {\enquote {\bibinfo {title} {{Calculation of the GW self-energy in
  semiconducting crystals}},}\ }\href
  {https://doi.org/10.1103/PhysRevB.44.13356} {\bibfield  {journal} {\bibinfo
  {journal} {Phys. Rev. B}\ }\textbf {\bibinfo {volume} {44}},\ \bibinfo
  {pages} {13356--13373} (\bibinfo {year} {1991})}\BibitemShut {NoStop}%
\bibitem [{\citenamefont {Backhouse}\ and\ \citenamefont
  {Booth}(2020)}]{Backhouse2020b}%
  \BibitemOpen
  \bibfield  {author} {\bibinfo {author} {\bibfnamefont {O.~J.}\ \bibnamefont
  {Backhouse}}\ and\ \bibinfo {author} {\bibfnamefont {G.~H.}\ \bibnamefont
  {Booth}},\ }\bibfield  {title} {\enquote {\bibinfo {title} {{Efficient
  Excitations and Spectra within a Perturbative Renormalization Approach}},}\
  }\href {https://doi.org/10.1021/acs.jctc.0c00701} {\bibfield  {journal}
  {\bibinfo  {journal} {J. Chem. Theory Comput.}\ }\textbf {\bibinfo {volume}
  {16}},\ \bibinfo {pages} {6294--6304} (\bibinfo {year} {2020})}\BibitemShut
  {NoStop}%
\bibitem [{\citenamefont {Backhouse}, \citenamefont {Santana-Bonilla},\ and\
  \citenamefont {Booth}(2021)}]{Backhouse2021}%
  \BibitemOpen
  \bibfield  {author} {\bibinfo {author} {\bibfnamefont {O.~J.}\ \bibnamefont
  {Backhouse}}, \bibinfo {author} {\bibfnamefont {A.}~\bibnamefont
  {Santana-Bonilla}},\ and\ \bibinfo {author} {\bibfnamefont {G.~H.}\
  \bibnamefont {Booth}},\ }\bibfield  {title} {\enquote {\bibinfo {title}
  {{Scalable and Predictive Spectra of Correlated Molecules with Moment
  Truncated Iterated Perturbation Theory}},}\ }\href@noop {} {\bibfield
  {journal} {\bibinfo  {journal} {J. Phys. Chem. Lett.}\ }\textbf {\bibinfo
  {volume} {12}},\ \bibinfo {pages} {7650--7658} (\bibinfo {year}
  {2021})}\BibitemShut {NoStop}%
\bibitem [{\citenamefont {Backhouse}\ and\ \citenamefont
  {Booth}(2022)}]{Backhouse2022}%
  \BibitemOpen
  \bibfield  {author} {\bibinfo {author} {\bibfnamefont {O.~J.}\ \bibnamefont
  {Backhouse}}\ and\ \bibinfo {author} {\bibfnamefont {G.~H.}\ \bibnamefont
  {Booth}},\ }\bibfield  {title} {\enquote {\bibinfo {title} {{Constructing
  \textquotedblleft Full-Frequency\textquotedblright Spectra via Moment
  Constraints for Coupled Cluster Green's Functions}},}\ }\href
  {https://doi.org/10.1021/acs.jctc.2c00670} {\bibfield  {journal} {\bibinfo
  {journal} {J. Chem. Theory Comput.}\ }\textbf {\bibinfo {volume} {18}},\
  \bibinfo {pages} {6622--6636} (\bibinfo {year} {2022})}\BibitemShut {NoStop}%
\bibitem [{\citenamefont {Sriluckshmy}\ \emph {et~al.}(2021)\citenamefont
  {Sriluckshmy}, \citenamefont {Nusspickel}, \citenamefont {Fertitta},\ and\
  \citenamefont {Booth}}]{Sriluckshmy2021}%
  \BibitemOpen
  \bibfield  {author} {\bibinfo {author} {\bibfnamefont {P.~V.}\ \bibnamefont
  {Sriluckshmy}}, \bibinfo {author} {\bibfnamefont {M.}~\bibnamefont
  {Nusspickel}}, \bibinfo {author} {\bibfnamefont {E.}~\bibnamefont
  {Fertitta}},\ and\ \bibinfo {author} {\bibfnamefont {G.~H.}\ \bibnamefont
  {Booth}},\ }\bibfield  {title} {\enquote {\bibinfo {title} {{Fully algebraic
  and self-consistent effective dynamics in a static quantum embedding}},}\
  }\href@noop {} {\bibfield  {journal} {\bibinfo  {journal} {Phys. Rev. B}\
  }\textbf {\bibinfo {volume} {103}},\ \bibinfo {pages} {085131} (\bibinfo
  {year} {2021})}\BibitemShut {NoStop}%
\bibitem [{\citenamefont {Haydock}(1980)}]{HAYDOCK198011}%
  \BibitemOpen
  \bibfield  {author} {\bibinfo {author} {\bibfnamefont {R.}~\bibnamefont
  {Haydock}},\ }\bibfield  {title} {\enquote {\bibinfo {title} {{The recursive
  solution of the Schrödinger equation}},}\ }\href
  {https://doi.org/https://doi.org/10.1016/0010-4655(80)90101-0} {\bibfield
  {journal} {\bibinfo  {journal} {Comp. Phys. Comms.}\ }\textbf {\bibinfo
  {volume} {20}},\ \bibinfo {pages} {11--16} (\bibinfo {year}
  {1980})}\BibitemShut {NoStop}%
\bibitem [{\citenamefont {Adachi}\ and\ \citenamefont
  {Lipparini}(1988)}]{ADACHI1988445}%
  \BibitemOpen
  \bibfield  {author} {\bibinfo {author} {\bibfnamefont {S.}~\bibnamefont
  {Adachi}}\ and\ \bibinfo {author} {\bibfnamefont {E.}~\bibnamefont
  {Lipparini}},\ }\bibfield  {title} {\enquote {\bibinfo {title} {{Sum rules in
  extended RPA theories}},}\ }\href
  {https://doi.org/https://doi.org/10.1016/0375-9474(88)90006-1} {\bibfield
  {journal} {\bibinfo  {journal} {Nuc. Phys. A}\ }\textbf {\bibinfo {volume}
  {489}},\ \bibinfo {pages} {445--460} (\bibinfo {year} {1988})}\BibitemShut
  {NoStop}%
\bibitem [{\citenamefont {Karlsson}\ and\ \citenamefont {van
  Leeuwen}(2016)}]{Karlsson2016PartialSA}%
  \BibitemOpen
  \bibfield  {author} {\bibinfo {author} {\bibfnamefont {D.}~\bibnamefont
  {Karlsson}}\ and\ \bibinfo {author} {\bibfnamefont {R.~A.}\ \bibnamefont {van
  Leeuwen}},\ }\bibfield  {title} {\enquote {\bibinfo {title} {{Partial
  self-consistency and analyticity in many-body perturbation theory: Particle
  number conservation and a generalized sum rule}},}\ }\href@noop {} {\bibfield
   {journal} {\bibinfo  {journal} {Physical Review B}\ }\textbf {\bibinfo
  {volume} {94}},\ \bibinfo {pages} {125124} (\bibinfo {year}
  {2016})}\BibitemShut {NoStop}%
\bibitem [{\citenamefont {Sabin}\ \emph {et~al.}(2010)\citenamefont {Sabin},
  \citenamefont {Oddershede}, \citenamefont {Cabrera-Trujillo}, \citenamefont
  {Sauer}, \citenamefont {Deumens},\ and\ \citenamefont
  {Öhrn}}]{doi:10.1080/00268976.2010.508753}%
  \BibitemOpen
  \bibfield  {author} {\bibinfo {author} {\bibfnamefont {J.~R.}\ \bibnamefont
  {Sabin}}, \bibinfo {author} {\bibfnamefont {J.}~\bibnamefont {Oddershede}},
  \bibinfo {author} {\bibfnamefont {R.}~\bibnamefont {Cabrera-Trujillo}},
  \bibinfo {author} {\bibfnamefont {S.~P.}\ \bibnamefont {Sauer}}, \bibinfo
  {author} {\bibfnamefont {E.}~\bibnamefont {Deumens}},\ and\ \bibinfo {author}
  {\bibfnamefont {Y.}~\bibnamefont {Öhrn}},\ }\bibfield  {title} {\enquote
  {\bibinfo {title} {{Stopping power of molecules for fast ions}},}\ }\href
  {https://doi.org/10.1080/00268976.2010.508753} {\bibfield  {journal}
  {\bibinfo  {journal} {Mol. Phys.}\ }\textbf {\bibinfo {volume} {108}},\
  \bibinfo {pages} {2891--2897} (\bibinfo {year} {2010})},\ \Eprint
  {https://arxiv.org/abs/https://doi.org/10.1080/00268976.2010.508753}
  {https://doi.org/10.1080/00268976.2010.508753} \BibitemShut {NoStop}%
\bibitem [{\citenamefont {{Van Caillie}}\ and\ \citenamefont
  {Amos}(2000)}]{VANCAILLIE2000446}%
  \BibitemOpen
  \bibfield  {author} {\bibinfo {author} {\bibfnamefont {C.}~\bibnamefont {{Van
  Caillie}}}\ and\ \bibinfo {author} {\bibfnamefont {R.~D.}\ \bibnamefont
  {Amos}},\ }\bibfield  {title} {\enquote {\bibinfo {title} {{Static and
  dynamic polarisabilities, Cauchy coefficients and their anisotropies: an
  evaluation of DFT functionals}},}\ }\href
  {https://doi.org/https://doi.org/10.1016/S0009-2614(00)00942-8} {\bibfield
  {journal} {\bibinfo  {journal} {Chem. Phys. Lett.}\ }\textbf {\bibinfo
  {volume} {328}},\ \bibinfo {pages} {446--452} (\bibinfo {year}
  {2000})}\BibitemShut {NoStop}%
\bibitem [{\citenamefont {Kalugina}\ and\ \citenamefont
  {Thakkar}(2016)}]{KALUGINA201620}%
  \BibitemOpen
  \bibfield  {author} {\bibinfo {author} {\bibfnamefont {Y.~N.}\ \bibnamefont
  {Kalugina}}\ and\ \bibinfo {author} {\bibfnamefont {A.~J.}\ \bibnamefont
  {Thakkar}},\ }\bibfield  {title} {\enquote {\bibinfo {title} {{Ab initio
  calculations of static dipole polarizabilities and Cauchy moments for the
  halomethanes}},}\ }\href
  {https://doi.org/https://doi.org/10.1016/j.cplett.2015.11.044} {\bibfield
  {journal} {\bibinfo  {journal} {Chem. Phys. Lett.}\ }\textbf {\bibinfo
  {volume} {644}},\ \bibinfo {pages} {20--24} (\bibinfo {year}
  {2016})}\BibitemShut {NoStop}%
\bibitem [{\citenamefont {Eshuis}, \citenamefont {Yarkony},\ and\ \citenamefont
  {Furche}(2010)}]{Furche2010}%
  \BibitemOpen
  \bibfield  {author} {\bibinfo {author} {\bibfnamefont {H.}~\bibnamefont
  {Eshuis}}, \bibinfo {author} {\bibfnamefont {J.}~\bibnamefont {Yarkony}},\
  and\ \bibinfo {author} {\bibfnamefont {F.}~\bibnamefont {Furche}},\
  }\bibfield  {title} {\enquote {\bibinfo {title} {{Fast computation of
  molecular random phase approximation correlation energies using resolution of
  the identity and imaginary frequency integration}},}\ }\href
  {https://doi.org/10.1063/1.3442749} {\bibfield  {journal} {\bibinfo
  {journal} {J. Chem. Phys.}\ }\textbf {\bibinfo {volume} {132}},\ \bibinfo
  {pages} {234114} (\bibinfo {year} {2010})}\BibitemShut {NoStop}%
\bibitem [{\citenamefont {Backhouse}, \citenamefont {Nusspickel},\ and\
  \citenamefont {Booth}(2020)}]{Backhouse2020a}%
  \BibitemOpen
  \bibfield  {author} {\bibinfo {author} {\bibfnamefont {O.~J.}\ \bibnamefont
  {Backhouse}}, \bibinfo {author} {\bibfnamefont {M.}~\bibnamefont
  {Nusspickel}},\ and\ \bibinfo {author} {\bibfnamefont {G.~H.}\ \bibnamefont
  {Booth}},\ }\bibfield  {title} {\enquote {\bibinfo {title} {{Wave Function
  Perspective and Efficient Truncation of Renormalized Second-Order
  Perturbation Theory}},}\ }\href {https://doi.org/10.1021/acs.jctc.9b01182}
  {\bibfield  {journal} {\bibinfo  {journal} {J. Chem. Theory Comput.}\
  }\textbf {\bibinfo {volume} {16}},\ \bibinfo {pages} {1090--1104} (\bibinfo
  {year} {2020})}\BibitemShut {NoStop}%
\bibitem [{\citenamefont {Rebolini}\ and\ \citenamefont
  {Toulouse}(2016)}]{Rebolini2016}%
  \BibitemOpen
  \bibfield  {author} {\bibinfo {author} {\bibfnamefont {E.}~\bibnamefont
  {Rebolini}}\ and\ \bibinfo {author} {\bibfnamefont {J.}~\bibnamefont
  {Toulouse}},\ }\bibfield  {title} {\enquote {\bibinfo {title}
  {{Range-separated time-dependent density-functional theory with a
  frequency-dependent second-order Bethe-Salpeter correlation kernel}},}\
  }\href {http://dx.doi.org/10.1063/1.4943003} {\bibfield  {journal} {\bibinfo
  {journal} {J. Chem. Phys.}\ }\textbf {\bibinfo {volume} {144}} (\bibinfo
  {year} {2016})}\BibitemShut {NoStop}%
\bibitem [{\citenamefont {V{\'e}ril}\ \emph {et~al.}(2018)\citenamefont
  {V{\'e}ril}, \citenamefont {Romaniello}, \citenamefont {Berger},\ and\
  \citenamefont {Loos}}]{Loos2018}%
  \BibitemOpen
  \bibfield  {author} {\bibinfo {author} {\bibfnamefont {M.}~\bibnamefont
  {V{\'e}ril}}, \bibinfo {author} {\bibfnamefont {P.}~\bibnamefont
  {Romaniello}}, \bibinfo {author} {\bibfnamefont {J.~A.}\ \bibnamefont
  {Berger}},\ and\ \bibinfo {author} {\bibfnamefont {P.-F.}\ \bibnamefont
  {Loos}},\ }\bibfield  {title} {\enquote {\bibinfo {title} {{Unphysical
  Discontinuities in GW Methods}},}\ }\href@noop {} {\bibfield  {journal}
  {\bibinfo  {journal} {J. Chem. Theory Comput.}\ }\textbf {\bibinfo {volume}
  {14}},\ \bibinfo {pages} {5220--5228} (\bibinfo {year} {2018})}\BibitemShut
  {NoStop}%
\bibitem [{\citenamefont {Quintero-Monsebaiz}\ \emph
  {et~al.}(2022)\citenamefont {Quintero-Monsebaiz}, \citenamefont {Monino},
  \citenamefont {Marie},\ and\ \citenamefont {Loos}}]{doi:10.1063/5.0130837}%
  \BibitemOpen
  \bibfield  {author} {\bibinfo {author} {\bibfnamefont {R.}~\bibnamefont
  {Quintero-Monsebaiz}}, \bibinfo {author} {\bibfnamefont {E.}~\bibnamefont
  {Monino}}, \bibinfo {author} {\bibfnamefont {A.}~\bibnamefont {Marie}},\ and\
  \bibinfo {author} {\bibfnamefont {P.-F.}\ \bibnamefont {Loos}},\ }\bibfield
  {title} {\enquote {\bibinfo {title} {{Connections between many-body
  perturbation and coupled-cluster theories}},}\ }\href
  {https://doi.org/10.1063/5.0130837} {\bibfield  {journal} {\bibinfo
  {journal} {J. Chem. Phys.}\ }\textbf {\bibinfo {volume} {157}},\ \bibinfo
  {pages} {231102} (\bibinfo {year} {2022})},\ \Eprint
  {https://arxiv.org/abs/https://doi.org/10.1063/5.0130837}
  {https://doi.org/10.1063/5.0130837} \BibitemShut {NoStop}%
\bibitem [{\citenamefont {Bintrim}\ and\ \citenamefont
  {Berkelbach}(2021)}]{Bintrim2021}%
  \BibitemOpen
  \bibfield  {author} {\bibinfo {author} {\bibfnamefont {S.~J.}\ \bibnamefont
  {Bintrim}}\ and\ \bibinfo {author} {\bibfnamefont {T.~C.}\ \bibnamefont
  {Berkelbach}},\ }\bibfield  {title} {\enquote {\bibinfo {title}
  {{Full-frequency GW without frequency}},}\ }\href
  {http://arxiv.org/abs/2009.14315} {\bibfield  {journal} {\bibinfo  {journal}
  {J. Chem. Phys.}\ }\textbf {\bibinfo {volume} {154}},\ \bibinfo {pages}
  {041101} (\bibinfo {year} {2021})}\BibitemShut {NoStop}%
\bibitem [{\citenamefont {Bintrim}\ and\ \citenamefont
  {Berkelbach}(2022)}]{Bintrim2022}%
  \BibitemOpen
  \bibfield  {author} {\bibinfo {author} {\bibfnamefont {S.~J.}\ \bibnamefont
  {Bintrim}}\ and\ \bibinfo {author} {\bibfnamefont {T.~C.}\ \bibnamefont
  {Berkelbach}},\ }\bibfield  {title} {\enquote {\bibinfo {title}
  {{Full-frequency dynamical Bethe--Salpeter equation without frequency and a
  study of double excitations}},}\ }\href@noop {} {\bibfield  {journal}
  {\bibinfo  {journal} {J. Chem. Phys.}\ }\textbf {\bibinfo {volume} {156}},\
  \bibinfo {pages} {044114} (\bibinfo {year} {2022})}\BibitemShut {NoStop}%
\bibitem [{\citenamefont {T\"{o}lle}\ and\ \citenamefont
  {Chan}(2022)}]{Tolle2022}%
  \BibitemOpen
  \bibfield  {author} {\bibinfo {author} {\bibfnamefont {J.}~\bibnamefont
  {T\"{o}lle}}\ and\ \bibinfo {author} {\bibfnamefont {G.~K.-L.}\ \bibnamefont
  {Chan}},\ }\href {https://arxiv.org/abs/2212.08982} {\enquote {\bibinfo
  {title} {{Exact relationships between the GW approximation and
  equation-of-motion coupled-cluster theories through the quasi-boson
  formalism}},}\ } (\bibinfo {year} {2022})\BibitemShut {NoStop}%
\bibitem [{\citenamefont {Meyer}\ and\ \citenamefont {Pal}(1989)}]{Meyer1989}%
  \BibitemOpen
  \bibfield  {author} {\bibinfo {author} {\bibfnamefont {H.-D.}\ \bibnamefont
  {Meyer}}\ and\ \bibinfo {author} {\bibfnamefont {S.}~\bibnamefont {Pal}},\
  }\bibfield  {title} {\enquote {\bibinfo {title} {{A band-Lanczos method for
  computing matrix elements of a resolvent}},}\ }\href
  {https://doi.org/10.1063/1.457438} {\bibfield  {journal} {\bibinfo  {journal}
  {J. Chem. Phys.}\ }\textbf {\bibinfo {volume} {91}},\ \bibinfo {pages}
  {6195--6204} (\bibinfo {year} {1989})}\BibitemShut {NoStop}%
\bibitem [{\citenamefont {Weikert}\ \emph {et~al.}(1996)\citenamefont
  {Weikert}, \citenamefont {Meyer}, \citenamefont {Cederbaum},\ and\
  \citenamefont {Tarantelli}}]{Weikert1996}%
  \BibitemOpen
  \bibfield  {author} {\bibinfo {author} {\bibfnamefont {H.-G.}\ \bibnamefont
  {Weikert}}, \bibinfo {author} {\bibfnamefont {H.-D.}\ \bibnamefont {Meyer}},
  \bibinfo {author} {\bibfnamefont {L.~S.}\ \bibnamefont {Cederbaum}},\ and\
  \bibinfo {author} {\bibfnamefont {F.}~\bibnamefont {Tarantelli}},\ }\bibfield
   {title} {\enquote {\bibinfo {title} {{Block Lanczos and many-body theory:
  Application to the one-particle Green's function}},}\ }\href@noop {}
  {\bibfield  {journal} {\bibinfo  {journal} {J. Chem. Phys.}\ }\textbf
  {\bibinfo {volume} {104}},\ \bibinfo {pages} {7122--7138} (\bibinfo {year}
  {1996})}\BibitemShut {NoStop}%
\bibitem [{\citenamefont {Haydock}\ and\ \citenamefont
  {Nex}(1985)}]{Haydock_1985}%
  \BibitemOpen
  \bibfield  {author} {\bibinfo {author} {\bibfnamefont {R.}~\bibnamefont
  {Haydock}}\ and\ \bibinfo {author} {\bibfnamefont {C.~M.~M.}\ \bibnamefont
  {Nex}},\ }\bibfield  {title} {\enquote {\bibinfo {title} {{A general
  terminator for the recursion method}},}\ }\href
  {https://doi.org/10.1088/0022-3719/18/11/007} {\bibfield  {journal} {\bibinfo
   {journal} {J. Phys. C: Solid State Physics}\ }\textbf {\bibinfo {volume}
  {18}},\ \bibinfo {pages} {2235} (\bibinfo {year} {1985})}\BibitemShut
  {NoStop}%
\bibitem [{\citenamefont {Engel}\ and\ \citenamefont
  {Farid}(1992)}]{PhysRevB.46.15812}%
  \BibitemOpen
  \bibfield  {author} {\bibinfo {author} {\bibfnamefont {G.~E.}\ \bibnamefont
  {Engel}}\ and\ \bibinfo {author} {\bibfnamefont {B.}~\bibnamefont {Farid}},\
  }\bibfield  {title} {\enquote {\bibinfo {title} {{Calculation of the
  dielectric properties of semiconductors}},}\ }\href
  {https://doi.org/10.1103/PhysRevB.46.15812} {\bibfield  {journal} {\bibinfo
  {journal} {Phys. Rev. B}\ }\textbf {\bibinfo {volume} {46}},\ \bibinfo
  {pages} {15812--15827} (\bibinfo {year} {1992})}\BibitemShut {NoStop}%
\bibitem [{\citenamefont {Farid}(2002)}]{doi:10.1080/13642810208222682}%
  \BibitemOpen
  \bibfield  {author} {\bibinfo {author} {\bibfnamefont {B.}~\bibnamefont
  {Farid}},\ }\bibfield  {title} {\enquote {\bibinfo {title} {{Dynamical
  correlation functions expressed in terms of many-particle ground-state
  wavefunction; the dynamical self-energy operator}},}\ }\href
  {https://doi.org/10.1080/13642810208222682} {\bibfield  {journal} {\bibinfo
  {journal} {Phil. Mag. B}\ }\textbf {\bibinfo {volume} {82}},\ \bibinfo
  {pages} {1413--1610} (\bibinfo {year} {2002})},\ \Eprint
  {https://arxiv.org/abs/https://doi.org/10.1080/13642810208222682}
  {https://doi.org/10.1080/13642810208222682} \BibitemShut {NoStop}%
\bibitem [{\citenamefont {Fukaya}\ \emph {et~al.}(2014)\citenamefont {Fukaya},
  \citenamefont {Nakatsukasa}, \citenamefont {Yanagisawa},\ and\ \citenamefont
  {Yamamoto}}]{Fukaya2014}%
  \BibitemOpen
  \bibfield  {author} {\bibinfo {author} {\bibfnamefont {T.}~\bibnamefont
  {Fukaya}}, \bibinfo {author} {\bibfnamefont {Y.}~\bibnamefont {Nakatsukasa}},
  \bibinfo {author} {\bibfnamefont {Y.}~\bibnamefont {Yanagisawa}},\ and\
  \bibinfo {author} {\bibfnamefont {Y.}~\bibnamefont {Yamamoto}},\ }\bibfield
  {title} {\enquote {\bibinfo {title} {{CholeskyQR2: A Simple and
  Communication-Avoiding Algorithm for Computing a Tall-Skinny QR Factorization
  on a Large-Scale Parallel System}},}\ }in\ \href@noop {} {\emph {\bibinfo
  {booktitle} {2014 5th Workshop on Latest Advances in Scalable Algorithms for
  Large-Scale Systems}}}\ (\bibinfo {year} {2014})\ pp.\ \bibinfo {pages}
  {31--38}\BibitemShut {NoStop}%
\bibitem [{\citenamefont {Fukaya}\ \emph {et~al.}(2020)\citenamefont {Fukaya},
  \citenamefont {Kannan}, \citenamefont {Nakatsukasa}, \citenamefont
  {Yamamoto},\ and\ \citenamefont {Yanagisawa}}]{Fukaya2020}%
  \BibitemOpen
  \bibfield  {author} {\bibinfo {author} {\bibfnamefont {T.}~\bibnamefont
  {Fukaya}}, \bibinfo {author} {\bibfnamefont {R.}~\bibnamefont {Kannan}},
  \bibinfo {author} {\bibfnamefont {Y.}~\bibnamefont {Nakatsukasa}}, \bibinfo
  {author} {\bibfnamefont {Y.}~\bibnamefont {Yamamoto}},\ and\ \bibinfo
  {author} {\bibfnamefont {Y.}~\bibnamefont {Yanagisawa}},\ }\bibfield  {title}
  {\enquote {\bibinfo {title} {{Shifted Cholesky QR for Computing the QR
  Factorization of Ill-Conditioned Matrices}},}\ }\href@noop {} {\bibfield
  {journal} {\bibinfo  {journal} {SIAM J. Sci. Comput.}\ }\textbf {\bibinfo
  {volume} {42}},\ \bibinfo {pages} {A477--a503} (\bibinfo {year}
  {2020})}\BibitemShut {NoStop}%
\bibitem [{\citenamefont {Schirmer}, \citenamefont {Trofimov},\ and\
  \citenamefont {Stelter}(1998)}]{Schirmer1998}%
  \BibitemOpen
  \bibfield  {author} {\bibinfo {author} {\bibfnamefont {J.}~\bibnamefont
  {Schirmer}}, \bibinfo {author} {\bibfnamefont {A.~B.}\ \bibnamefont
  {Trofimov}},\ and\ \bibinfo {author} {\bibfnamefont {G.}~\bibnamefont
  {Stelter}},\ }\bibfield  {title} {\enquote {\bibinfo {title} {{A non-Dyson
  third-order approximation scheme for the electron propagator}},}\ }\href@noop
  {} {\bibfield  {journal} {\bibinfo  {journal} {J. Chem. Phys.}\ }\textbf
  {\bibinfo {volume} {109}},\ \bibinfo {pages} {4734--4744} (\bibinfo {year}
  {1998})}\BibitemShut {NoStop}%
\bibitem [{\citenamefont {Schirmer}(2018)}]{Schirmer18}%
  \BibitemOpen
  \bibfield  {author} {\bibinfo {author} {\bibfnamefont {J.}~\bibnamefont
  {Schirmer}},\ }\href
  {https://doi.org/https://doi.org/10.1007/978-3-319-93602-4} {\emph {\bibinfo
  {title} {{Many-Body Methods for Atoms, Molecules and Clusters}}}}\ (\bibinfo
  {publisher} {Springer Chem},\ \bibinfo {address} {Switzerland},\ \bibinfo
  {year} {2018})\BibitemShut {NoStop}%
\bibitem [{\citenamefont {Berkelbach}(2018)}]{doi:10.1063/1.5032314}%
  \BibitemOpen
  \bibfield  {author} {\bibinfo {author} {\bibfnamefont {T.~C.}\ \bibnamefont
  {Berkelbach}},\ }\bibfield  {title} {\enquote {\bibinfo {title}
  {Communication: Random-phase approximation excitation energies from
  approximate equation-of-motion coupled-cluster doubles},}\ }\href
  {https://doi.org/10.1063/1.5032314} {\bibfield  {journal} {\bibinfo
  {journal} {J. Chem. Phys.}\ }\textbf {\bibinfo {volume} {149}},\ \bibinfo
  {pages} {041103} (\bibinfo {year} {2018})},\ \Eprint
  {https://arxiv.org/abs/https://doi.org/10.1063/1.5032314}
  {https://doi.org/10.1063/1.5032314} \BibitemShut {NoStop}%
\bibitem [{\citenamefont {Rishi}, \citenamefont {Perera},\ and\ \citenamefont
  {Bartlett}(2020)}]{doi:10.1063/5.0023862}%
  \BibitemOpen
  \bibfield  {author} {\bibinfo {author} {\bibfnamefont {V.}~\bibnamefont
  {Rishi}}, \bibinfo {author} {\bibfnamefont {A.}~\bibnamefont {Perera}},\ and\
  \bibinfo {author} {\bibfnamefont {R.~J.}\ \bibnamefont {Bartlett}},\
  }\bibfield  {title} {\enquote {\bibinfo {title} {A route to improving rpa
  excitation energies through its connection to equation-of-motion coupled
  cluster theory},}\ }\href {https://doi.org/10.1063/5.0023862} {\bibfield
  {journal} {\bibinfo  {journal} {The Journal of Chemical Physics}\ }\textbf
  {\bibinfo {volume} {153}},\ \bibinfo {pages} {234101} (\bibinfo {year}
  {2020})},\ \Eprint {https://arxiv.org/abs/https://doi.org/10.1063/5.0023862}
  {https://doi.org/10.1063/5.0023862} \BibitemShut {NoStop}%
\bibitem [{\citenamefont {Dempwolff}\ \emph {et~al.}(2019)\citenamefont
  {Dempwolff}, \citenamefont {Schneider}, \citenamefont {Hodecker},\ and\
  \citenamefont {Dreuw}}]{Dempwolff2019}%
  \BibitemOpen
  \bibfield  {author} {\bibinfo {author} {\bibfnamefont {A.~L.}\ \bibnamefont
  {Dempwolff}}, \bibinfo {author} {\bibfnamefont {M.}~\bibnamefont
  {Schneider}}, \bibinfo {author} {\bibfnamefont {M.}~\bibnamefont
  {Hodecker}},\ and\ \bibinfo {author} {\bibfnamefont {A.}~\bibnamefont
  {Dreuw}},\ }\bibfield  {title} {\enquote {\bibinfo {title} {{Efficient
  implementation of the non-Dyson third-order algebraic diagrammatic
  construction approximation for the electron propagator for closed- and
  open-shell molecules}},}\ }\href {http://dx.doi.org/10.1063/1.5081674}
  {\bibfield  {journal} {\bibinfo  {journal} {J. Chem. Phys.}\ }\textbf
  {\bibinfo {volume} {150}} (\bibinfo {year} {2019})}\BibitemShut {NoStop}%
\bibitem [{\citenamefont {Trofimov}\ and\ \citenamefont
  {Schirmer}(2005)}]{Trofimov2005}%
  \BibitemOpen
  \bibfield  {author} {\bibinfo {author} {\bibfnamefont {A.~B.}\ \bibnamefont
  {Trofimov}}\ and\ \bibinfo {author} {\bibfnamefont {J.}~\bibnamefont
  {Schirmer}},\ }\bibfield  {title} {\enquote {\bibinfo {title} {{Molecular
  ionization energies and ground- and ionic-state properties using a non-Dyson
  electron propagator approach}},}\ }\href {https://doi.org/10.1063/1.2047550}
  {\bibfield  {journal} {\bibinfo  {journal} {J. Chem. Phys.}\ }\textbf
  {\bibinfo {volume} {123}},\ \bibinfo {pages} {144115} (\bibinfo {year}
  {2005})}\BibitemShut {NoStop}%
\bibitem [{\citenamefont {Scott}\ and\ \citenamefont
  {Booth}(2021)}]{PhysRevB.104.245114}%
  \BibitemOpen
  \bibfield  {author} {\bibinfo {author} {\bibfnamefont {C.~J.~C.}\
  \bibnamefont {Scott}}\ and\ \bibinfo {author} {\bibfnamefont {G.~H.}\
  \bibnamefont {Booth}},\ }\bibfield  {title} {\enquote {\bibinfo {title}
  {{Extending density matrix embedding: A static two-particle theory}},}\
  }\href {https://link.aps.org/doi/10.1103/PhysRevB.104.245114} {\bibfield
  {journal} {\bibinfo  {journal} {Phys. Rev. B}\ }\textbf {\bibinfo {volume}
  {104}},\ \bibinfo {pages} {245114} (\bibinfo {year} {2021})}\BibitemShut
  {NoStop}%
\bibitem [{\citenamefont {Langreth}(1970)}]{PhysRevB.1.471}%
  \BibitemOpen
  \bibfield  {author} {\bibinfo {author} {\bibfnamefont {D.~C.}\ \bibnamefont
  {Langreth}},\ }\bibfield  {title} {\enquote {\bibinfo {title} {Singularities
  in the x-ray spectra of metals},}\ }\href
  {https://doi.org/10.1103/PhysRevB.1.471} {\bibfield  {journal} {\bibinfo
  {journal} {Phys. Rev. B}\ }\textbf {\bibinfo {volume} {1}},\ \bibinfo {pages}
  {471--477} (\bibinfo {year} {1970})}\BibitemShut {NoStop}%
\bibitem [{\citenamefont {Nozi\`eres}\ and\ \citenamefont
  {De~Dominicis}(1969)}]{PhysRev.178.1097}%
  \BibitemOpen
  \bibfield  {author} {\bibinfo {author} {\bibfnamefont {P.}~\bibnamefont
  {Nozi\`eres}}\ and\ \bibinfo {author} {\bibfnamefont {C.~T.}\ \bibnamefont
  {De~Dominicis}},\ }\bibfield  {title} {\enquote {\bibinfo {title}
  {Singularities in the x-ray absorption and emission of metals. iii. one-body
  theory exact solution},}\ }\href {https://doi.org/10.1103/PhysRev.178.1097}
  {\bibfield  {journal} {\bibinfo  {journal} {Phys. Rev.}\ }\textbf {\bibinfo
  {volume} {178}},\ \bibinfo {pages} {1097--1107} (\bibinfo {year}
  {1969})}\BibitemShut {NoStop}%
\bibitem [{\citenamefont {Lundqvist}\ and\ \citenamefont
  {Samathiyakanit}(1969)}]{Lundqvist69}%
  \BibitemOpen
  \bibfield  {author} {\bibinfo {author} {\bibfnamefont {B.}~\bibnamefont
  {Lundqvist}}\ and\ \bibinfo {author} {\bibfnamefont {V.}~\bibnamefont
  {Samathiyakanit}},\ }\bibfield  {title} {\enquote {\bibinfo {title}
  {Single-particle spectrum of the degenerate electron gas iv. ground state
  energy.}}\ }\href {https://doi.org/https://doi.org/10.1007/BF02422566}
  {\bibfield  {journal} {\bibinfo  {journal} {Phys. kondens. Materie.}\ ,\
  \bibinfo {pages} {231–235}} (\bibinfo {year} {1969})}\BibitemShut {NoStop}%
\bibitem [{\citenamefont {He\ss{}elmann}\ and\ \citenamefont
  {G\"{o}rling}(2011)}]{Hesselmann2011}%
  \BibitemOpen
  \bibfield  {author} {\bibinfo {author} {\bibfnamefont {A.}~\bibnamefont
  {He\ss{}elmann}}\ and\ \bibinfo {author} {\bibfnamefont {A.}~\bibnamefont
  {G\"{o}rling}},\ }\bibfield  {title} {\enquote {\bibinfo {title}
  {{Random-phase approximation correlation methods for molecules and
  solids}},}\ }\href {https://doi.org/10.1080/00268976.2011.614282} {\bibfield
  {journal} {\bibinfo  {journal} {Mol. Phys.}\ }\textbf {\bibinfo {volume}
  {109}},\ \bibinfo {pages} {2473--2500} (\bibinfo {year} {2011})}\BibitemShut
  {NoStop}%
\bibitem [{\citenamefont {Chen}\ \emph {et~al.}(2017)\citenamefont {Chen},
  \citenamefont {Voora}, \citenamefont {Agee}, \citenamefont {Balasubramani},\
  and\ \citenamefont {Furche}}]{doi:10.1146/annurev-physchem-040215-112308}%
  \BibitemOpen
  \bibfield  {author} {\bibinfo {author} {\bibfnamefont {G.~P.}\ \bibnamefont
  {Chen}}, \bibinfo {author} {\bibfnamefont {V.~K.}\ \bibnamefont {Voora}},
  \bibinfo {author} {\bibfnamefont {M.~M.}\ \bibnamefont {Agee}}, \bibinfo
  {author} {\bibfnamefont {S.~G.}\ \bibnamefont {Balasubramani}},\ and\
  \bibinfo {author} {\bibfnamefont {F.}~\bibnamefont {Furche}},\ }\bibfield
  {title} {\enquote {\bibinfo {title} {{Random-Phase Approximation Methods}},}\
  }\href {https://doi.org/10.1146/annurev-physchem-040215-112308} {\bibfield
  {journal} {\bibinfo  {journal} {Annu. Rev. Phys. Chem.}\ }\textbf {\bibinfo
  {volume} {68}},\ \bibinfo {pages} {421--445} (\bibinfo {year}
  {2017})}\BibitemShut {NoStop}%
\bibitem [{\citenamefont {Parkinson}\ and\ \citenamefont
  {Zerner}(1989)}]{doi:10.1063/1.456413}%
  \BibitemOpen
  \bibfield  {author} {\bibinfo {author} {\bibfnamefont {W.~A.}\ \bibnamefont
  {Parkinson}}\ and\ \bibinfo {author} {\bibfnamefont {M.~C.}\ \bibnamefont
  {Zerner}},\ }\bibfield  {title} {\enquote {\bibinfo {title} {{The calculation
  of dynamic molecular polarizability}},}\ }\href
  {https://doi.org/10.1063/1.456413} {\bibfield  {journal} {\bibinfo  {journal}
  {J. Chem. Phys.}\ }\textbf {\bibinfo {volume} {90}},\ \bibinfo {pages}
  {5606--5611} (\bibinfo {year} {1989})}\BibitemShut {NoStop}%
\bibitem [{Note1()}]{Note1}%
  \BibitemOpen
  \bibinfo {note} {We also note that a related expansion and recursive relation
  can also be constructed in terms of the inverse of these moments courtesy of
  the equivalence $(\protect \mathbf {X+Y})^{-1}=(\protect \mathbf {X - Y})^T$,
  that is $(\protect \mathbf {\eta }^{(n)})^{-1} = (\protect \mathbf
  {X}-\protect \mathbf {Y}) \protect \mathbf {\Omega }^{-n} (\protect \mathbf
  {X}-\protect \mathbf {Y})^T$.}\BibitemShut {Stop}%
\bibitem [{\citenamefont {Furche}(2001)}]{Furche2001}%
  \BibitemOpen
  \bibfield  {author} {\bibinfo {author} {\bibfnamefont {F.}~\bibnamefont
  {Furche}},\ }\bibfield  {title} {\enquote {\bibinfo {title} {{On the density
  matrix based approach to time-dependent density functional response
  theory}},}\ }\href {https://doi.org/10.1063/1.1353585} {\bibfield  {journal}
  {\bibinfo  {journal} {J. Chem. Phys.}\ }\textbf {\bibinfo {volume} {114}},\
  \bibinfo {pages} {5982--5992} (\bibinfo {year} {2001})},\ \Eprint
  {https://arxiv.org/abs/https://doi.org/10.1063/1.1353585}
  {https://doi.org/10.1063/1.1353585} \BibitemShut {NoStop}%
\bibitem [{\citenamefont {{\'A}ngy{\'a}n}\ \emph {et~al.}(2011)\citenamefont
  {{\'A}ngy{\'a}n}, \citenamefont {Liu}, \citenamefont {Toulouse},\ and\
  \citenamefont {Jansen}}]{angyan2011correlation}%
  \BibitemOpen
  \bibfield  {author} {\bibinfo {author} {\bibfnamefont {J.~G.}\ \bibnamefont
  {{\'A}ngy{\'a}n}}, \bibinfo {author} {\bibfnamefont {R.-F.}\ \bibnamefont
  {Liu}}, \bibinfo {author} {\bibfnamefont {J.}~\bibnamefont {Toulouse}},\ and\
  \bibinfo {author} {\bibfnamefont {G.}~\bibnamefont {Jansen}},\ }\bibfield
  {title} {\enquote {\bibinfo {title} {{Correlation energy expressions from the
  adiabatic-connection fluctuation--dissipation theorem approach}},}\ }\href
  {https://doi.org/10.1021/ct200501r} {\bibfield  {journal} {\bibinfo
  {journal} {J. Chem. Theory Comput.}\ }\textbf {\bibinfo {volume} {7}},\
  \bibinfo {pages} {3116--3130} (\bibinfo {year} {2011})},\ \Eprint
  {https://arxiv.org/abs/https://doi.org/10.1021/ct200501r}
  {https://doi.org/10.1021/ct200501r} \BibitemShut {NoStop}%
\bibitem [{\citenamefont {Furche}(2008)}]{Furche2008}%
  \BibitemOpen
  \bibfield  {author} {\bibinfo {author} {\bibfnamefont {F.}~\bibnamefont
  {Furche}},\ }\bibfield  {title} {\enquote {\bibinfo {title} {{Developing the
  random phase approximation into a practical post-Kohn--Sham correlation
  model}},}\ }\href {https://doi.org/10.1063/1.2977789} {\bibfield  {journal}
  {\bibinfo  {journal} {J. Chem. Phys.}\ }\textbf {\bibinfo {volume} {129}},\
  \bibinfo {pages} {114105} (\bibinfo {year} {2008})}\BibitemShut {NoStop}%
\bibitem [{Note2()}]{Note2}%
  \BibitemOpen
  \bibinfo {note} {We note that the derivation in this section does not rely on
  the $V_{jb,P}$ tensor in Eq.~\ref {eq:orig} arising from this Coulomb form
  specifically, but rather that it involves a more general linear
  transformation from a space in the particle-hole product basis to a space
  which scales no more than $\protect \mathcal {O}[N]$ (in this case the
  auxiliary basis).}\BibitemShut {Stop}%
\bibitem [{\citenamefont {Hummel}\ \emph {et~al.}(2019)\citenamefont {Hummel},
  \citenamefont {Gr\"{u}neis}, \citenamefont {Kresse},\ and\ \citenamefont
  {Ziesche}}]{doi:10.1021/acs.jctc.8b01247}%
  \BibitemOpen
  \bibfield  {author} {\bibinfo {author} {\bibfnamefont {F.}~\bibnamefont
  {Hummel}}, \bibinfo {author} {\bibfnamefont {A.}~\bibnamefont {Gr\"{u}neis}},
  \bibinfo {author} {\bibfnamefont {G.}~\bibnamefont {Kresse}},\ and\ \bibinfo
  {author} {\bibfnamefont {P.}~\bibnamefont {Ziesche}},\ }\bibfield  {title}
  {\enquote {\bibinfo {title} {{Screened Exchange Corrections to the Random
  Phase Approximation from Many-Body Perturbation Theory}},}\ }\href
  {https://doi.org/10.1021/acs.jctc.8b01247} {\bibfield  {journal} {\bibinfo
  {journal} {J. Chem. Theory Comput.}\ }\textbf {\bibinfo {volume} {15}},\
  \bibinfo {pages} {3223--3236} (\bibinfo {year} {2019})}\BibitemShut {NoStop}%
\bibitem [{\citenamefont {F\"{o}rster}(2022)}]{doi:10.1021/acs.jctc.2c00366}%
  \BibitemOpen
  \bibfield  {author} {\bibinfo {author} {\bibfnamefont {A.}~\bibnamefont
  {F\"{o}rster}},\ }\bibfield  {title} {\enquote {\bibinfo {title} {{Assessment
  of the Second-Order Statically Screened Exchange Correction to the Random
  Phase Approximation for Correlation Energies}},}\ }\href
  {https://doi.org/10.1021/acs.jctc.2c00366} {\bibfield  {journal} {\bibinfo
  {journal} {J. Chem. Theory Comput.}\ }\textbf {\bibinfo {volume} {18}},\
  \bibinfo {pages} {5948--5965} (\bibinfo {year} {2022})}\BibitemShut {NoStop}%
\bibitem [{\citenamefont {Hale}, \citenamefont {Higham},\ and\ \citenamefont
  {Trefethen}(2008)}]{matrixsqrt}%
  \BibitemOpen
  \bibfield  {author} {\bibinfo {author} {\bibfnamefont {N.}~\bibnamefont
  {Hale}}, \bibinfo {author} {\bibfnamefont {N.~J.}\ \bibnamefont {Higham}},\
  and\ \bibinfo {author} {\bibfnamefont {L.~N.}\ \bibnamefont {Trefethen}},\
  }\bibfield  {title} {\enquote {\bibinfo {title} {{Computing $A^\alpha,
  \log(A)$, and Related Matrix Functions by Contour Integrals}},}\ }\href
  {https://doi.org/10.1137/070700607} {\bibfield  {journal} {\bibinfo
  {journal} {SIAM J. Numer. Anal.}\ }\textbf {\bibinfo {volume} {46}},\
  \bibinfo {pages} {2505--2523} (\bibinfo {year} {2008})}\BibitemShut {NoStop}%
\bibitem [{\citenamefont {Wilhelm}\ \emph {et~al.}(2016)\citenamefont
  {Wilhelm}, \citenamefont {Seewald}, \citenamefont {Del~Ben},\ and\
  \citenamefont {Hutter}}]{Hutter16}%
  \BibitemOpen
  \bibfield  {author} {\bibinfo {author} {\bibfnamefont {J.}~\bibnamefont
  {Wilhelm}}, \bibinfo {author} {\bibfnamefont {P.}~\bibnamefont {Seewald}},
  \bibinfo {author} {\bibfnamefont {M.}~\bibnamefont {Del~Ben}},\ and\ \bibinfo
  {author} {\bibfnamefont {J.}~\bibnamefont {Hutter}},\ }\bibfield  {title}
  {\enquote {\bibinfo {title} {{Large-Scale Cubic-Scaling Random Phase
  Approximation Correlation Energy Calculations Using a Gaussian Basis}},}\
  }\href {https://doi.org/10.1021/acs.jctc.6b00840} {\bibfield  {journal}
  {\bibinfo  {journal} {J. Chem. Theory Comput.}\ }\textbf {\bibinfo {volume}
  {12}},\ \bibinfo {pages} {5851--5859} (\bibinfo {year} {2016})},\ \bibinfo
  {note} {pMID: 27779863}\BibitemShut {NoStop}%
\bibitem [{\citenamefont {F\"{o}rster}\ and\ \citenamefont
  {Visscher}(2020)}]{Visscher20}%
  \BibitemOpen
  \bibfield  {author} {\bibinfo {author} {\bibfnamefont {A.}~\bibnamefont
  {F\"{o}rster}}\ and\ \bibinfo {author} {\bibfnamefont {L.}~\bibnamefont
  {Visscher}},\ }\bibfield  {title} {\enquote {\bibinfo {title} {{Low-Order
  Scaling G0W0 by Pair Atomic Density Fitting}},}\ }\href
  {https://doi.org/10.1021/acs.jctc.0c00693} {\bibfield  {journal} {\bibinfo
  {journal} {J. Chem. Theory Comput.}\ }\textbf {\bibinfo {volume} {16}},\
  \bibinfo {pages} {7381--7399} (\bibinfo {year} {2020})},\ \bibinfo {note}
  {pMID: 33174743}\BibitemShut {NoStop}%
\bibitem [{\citenamefont {Parrish}\ \emph {et~al.}(2012)\citenamefont
  {Parrish}, \citenamefont {Hohenstein}, \citenamefont {Mart\'{\i}nez},\ and\
  \citenamefont {Sherrill}}]{THC_V}%
  \BibitemOpen
  \bibfield  {author} {\bibinfo {author} {\bibfnamefont {R.~M.}\ \bibnamefont
  {Parrish}}, \bibinfo {author} {\bibfnamefont {E.~G.}\ \bibnamefont
  {Hohenstein}}, \bibinfo {author} {\bibfnamefont {T.~J.}\ \bibnamefont
  {Mart\'{\i}nez}},\ and\ \bibinfo {author} {\bibfnamefont {C.~D.}\
  \bibnamefont {Sherrill}},\ }\bibfield  {title} {\enquote {\bibinfo {title}
  {{Tensor hypercontraction. II. Least-squares renormalization}},}\ }\href
  {https://doi.org/10.1063/1.4768233} {\bibfield  {journal} {\bibinfo
  {journal} {J. Chem. Phys.}\ }\textbf {\bibinfo {volume} {137}},\ \bibinfo
  {pages} {224106} (\bibinfo {year} {2012})}\BibitemShut {NoStop}%
\bibitem [{\citenamefont {Lu}\ and\ \citenamefont {Thicke}(2017)}]{LU2017187}%
  \BibitemOpen
  \bibfield  {author} {\bibinfo {author} {\bibfnamefont {J.}~\bibnamefont
  {Lu}}\ and\ \bibinfo {author} {\bibfnamefont {K.}~\bibnamefont {Thicke}},\
  }\bibfield  {title} {\enquote {\bibinfo {title} {{Cubic scaling algorithms
  for RPA correlation using interpolative separable density fitting}},}\ }\href
  {https://doi.org/https://doi.org/10.1016/j.jcp.2017.09.012} {\bibfield
  {journal} {\bibinfo  {journal} {J. Comput. Phys.}\ }\textbf {\bibinfo
  {volume} {351}},\ \bibinfo {pages} {187--202} (\bibinfo {year}
  {2017})}\BibitemShut {NoStop}%
\bibitem [{\citenamefont {Shenvi}\ \emph {et~al.}(2014)\citenamefont {Shenvi},
  \citenamefont {van Aggelen}, \citenamefont {Yang},\ and\ \citenamefont
  {Yang}}]{Yang_THCppRPA}%
  \BibitemOpen
  \bibfield  {author} {\bibinfo {author} {\bibfnamefont {N.}~\bibnamefont
  {Shenvi}}, \bibinfo {author} {\bibfnamefont {H.}~\bibnamefont {van Aggelen}},
  \bibinfo {author} {\bibfnamefont {Y.}~\bibnamefont {Yang}},\ and\ \bibinfo
  {author} {\bibfnamefont {W.}~\bibnamefont {Yang}},\ }\bibfield  {title}
  {\enquote {\bibinfo {title} {{Tensor hypercontracted ppRPA: Reducing the cost
  of the particle-particle random phase approximation from O($r^6$) to
  O($r^4$)}},}\ }\href {https://doi.org/10.1063/1.4886584} {\bibfield
  {journal} {\bibinfo  {journal} {J. Chem. Phys.}\ }\textbf {\bibinfo {volume}
  {141}},\ \bibinfo {pages} {024119} (\bibinfo {year} {2014})}\BibitemShut
  {NoStop}%
\bibitem [{\citenamefont {Lu}\ and\ \citenamefont {Ying}(2015)}]{LU2015329}%
  \BibitemOpen
  \bibfield  {author} {\bibinfo {author} {\bibfnamefont {J.}~\bibnamefont
  {Lu}}\ and\ \bibinfo {author} {\bibfnamefont {L.}~\bibnamefont {Ying}},\
  }\bibfield  {title} {\enquote {\bibinfo {title} {{Compression of the electron
  repulsion integral tensor in tensor hypercontraction format with cubic
  scaling cost}},}\ }\href
  {https://www.sciencedirect.com/science/article/pii/S0021999115006075}
  {\bibfield  {journal} {\bibinfo  {journal} {J. Comput. Phys}\ }\textbf
  {\bibinfo {volume} {302}},\ \bibinfo {pages} {329--335} (\bibinfo {year}
  {2015})}\BibitemShut {NoStop}%
\bibitem [{\citenamefont {Pierce}\ and\ \citenamefont
  {Valeev}(2023)}]{doi:10.1021/acs.jctc.2c00861}%
  \BibitemOpen
  \bibfield  {author} {\bibinfo {author} {\bibfnamefont {K.}~\bibnamefont
  {Pierce}}\ and\ \bibinfo {author} {\bibfnamefont {E.~F.}\ \bibnamefont
  {Valeev}},\ }\bibfield  {title} {\enquote {\bibinfo {title} {{Efficient
  Construction of Canonical Polyadic Approximations of Tensor Networks}},}\
  }\href {https://doi.org/10.1021/acs.jctc.2c00861} {\bibfield  {journal}
  {\bibinfo  {journal} {J. Chem. Theory Comput.}\ }\textbf {\bibinfo {volume}
  {19}},\ \bibinfo {pages} {71--81} (\bibinfo {year} {2023})},\ \Eprint
  {https://arxiv.org/abs/https://doi.org/10.1021/acs.jctc.2c00861}
  {https://doi.org/10.1021/acs.jctc.2c00861} \BibitemShut {NoStop}%
\bibitem [{\citenamefont {Schurkus}\ and\ \citenamefont
  {Ochsenfeld}(2016)}]{Schurkus2016}%
  \BibitemOpen
  \bibfield  {author} {\bibinfo {author} {\bibfnamefont {H.~F.}\ \bibnamefont
  {Schurkus}}\ and\ \bibinfo {author} {\bibfnamefont {C.}~\bibnamefont
  {Ochsenfeld}},\ }\bibfield  {title} {\enquote {\bibinfo {title}
  {{Communication: An effective linear-scaling atomic-orbital reformulation of
  the random-phase approximation using a contracted double-Laplace
  transformation}},}\ }\href {https://doi.org/10.1063/1.4939841} {\bibfield
  {journal} {\bibinfo  {journal} {J. Chem. Phys.}\ }\textbf {\bibinfo {volume}
  {144}},\ \bibinfo {pages} {31101} (\bibinfo {year} {2016})}\BibitemShut
  {NoStop}%
\bibitem [{\citenamefont {Helmich-Paris}\ and\ \citenamefont
  {Visscher}(2016)}]{HELMICHPARIS2016927}%
  \BibitemOpen
  \bibfield  {author} {\bibinfo {author} {\bibfnamefont {B.}~\bibnamefont
  {Helmich-Paris}}\ and\ \bibinfo {author} {\bibfnamefont {L.}~\bibnamefont
  {Visscher}},\ }\bibfield  {title} {\enquote {\bibinfo {title} {{Improvements
  on the minimax algorithm for the Laplace transformation of orbital energy
  denominators}},}\ }\href
  {https://www.sciencedirect.com/science/article/pii/S0021999116302364}
  {\bibfield  {journal} {\bibinfo  {journal} {J. Comput. Phys}\ }\textbf
  {\bibinfo {volume} {321}},\ \bibinfo {pages} {927--931} (\bibinfo {year}
  {2016})}\BibitemShut {NoStop}%
\bibitem [{\citenamefont {Graf}\ \emph {et~al.}(2018)\citenamefont {Graf},
  \citenamefont {Beuerle}, \citenamefont {Schurkus}, \citenamefont {Luenser},
  \citenamefont {Savasci},\ and\ \citenamefont {Ochsenfeld}}]{Graf2018}%
  \BibitemOpen
  \bibfield  {author} {\bibinfo {author} {\bibfnamefont {D.}~\bibnamefont
  {Graf}}, \bibinfo {author} {\bibfnamefont {M.}~\bibnamefont {Beuerle}},
  \bibinfo {author} {\bibfnamefont {H.~F.}\ \bibnamefont {Schurkus}}, \bibinfo
  {author} {\bibfnamefont {A.}~\bibnamefont {Luenser}}, \bibinfo {author}
  {\bibfnamefont {G.}~\bibnamefont {Savasci}},\ and\ \bibinfo {author}
  {\bibfnamefont {C.}~\bibnamefont {Ochsenfeld}},\ }\bibfield  {title}
  {\enquote {\bibinfo {title} {{Accurate and Efficient Parallel Implementation
  of an Effective Linear-Scaling Direct Random Phase Approximation Method}},}\
  }\href {https://pubs.acs.org/sharingguidelines} {\bibfield  {journal}
  {\bibinfo  {journal} {J. Chem. Theory Comput.}\ }\textbf {\bibinfo {volume}
  {14}},\ \bibinfo {pages} {2505--2515} (\bibinfo {year} {2018})}\BibitemShut
  {NoStop}%
\bibitem [{\citenamefont {Sun}\ \emph {et~al.}(2017)\citenamefont {Sun},
  \citenamefont {Berkelbach}, \citenamefont {Blunt}, \citenamefont {Booth},
  \citenamefont {Guo}, \citenamefont {Li}, \citenamefont {Liu}, \citenamefont
  {McClain}, \citenamefont {Sayfutyarova}, \citenamefont {Sharma},
  \citenamefont {Wouters},\ and\ \citenamefont {Chan}}]{PySCF2017}%
  \BibitemOpen
  \bibfield  {author} {\bibinfo {author} {\bibfnamefont {Q.}~\bibnamefont
  {Sun}}, \bibinfo {author} {\bibfnamefont {T.~C.}\ \bibnamefont {Berkelbach}},
  \bibinfo {author} {\bibfnamefont {N.~S.}\ \bibnamefont {Blunt}}, \bibinfo
  {author} {\bibfnamefont {G.~H.}\ \bibnamefont {Booth}}, \bibinfo {author}
  {\bibfnamefont {S.}~\bibnamefont {Guo}}, \bibinfo {author} {\bibfnamefont
  {Z.}~\bibnamefont {Li}}, \bibinfo {author} {\bibfnamefont {J.}~\bibnamefont
  {Liu}}, \bibinfo {author} {\bibfnamefont {J.~D.}\ \bibnamefont {McClain}},
  \bibinfo {author} {\bibfnamefont {E.~R.}\ \bibnamefont {Sayfutyarova}},
  \bibinfo {author} {\bibfnamefont {S.}~\bibnamefont {Sharma}}, \bibinfo
  {author} {\bibfnamefont {S.}~\bibnamefont {Wouters}},\ and\ \bibinfo {author}
  {\bibfnamefont {G.~K.-L.}\ \bibnamefont {Chan}},\ }\bibfield  {title}
  {\enquote {\bibinfo {title} {{PySCF: the Python-based simulations of
  chemistry framework}},}\ }\href@noop {} {\bibfield  {journal} {\bibinfo
  {journal} {Wiley Interdiscip. Rev. Comput. Mol. Sci.}\ }\textbf {\bibinfo
  {volume} {8}},\ \bibinfo {pages} {e1340} (\bibinfo {year}
  {2017})}\BibitemShut {NoStop}%
\bibitem [{\citenamefont {Sun}\ \emph {et~al.}(2020)\citenamefont {Sun},
  \citenamefont {Zhang}, \citenamefont {Banerjee}, \citenamefont {Bao},
  \citenamefont {Barbry}, \citenamefont {Blunt}, \citenamefont {Bogdanov},
  \citenamefont {Booth}, \citenamefont {Chen}, \citenamefont {Cui},
  \citenamefont {Eriksen}, \citenamefont {Gao}, \citenamefont {Guo},
  \citenamefont {Hermann}, \citenamefont {Hermes}, \citenamefont {Koh},
  \citenamefont {Koval}, \citenamefont {Lehtola}, \citenamefont {Li},
  \citenamefont {Liu}, \citenamefont {Mardirossian}, \citenamefont {McClain},
  \citenamefont {Motta}, \citenamefont {Mussard}, \citenamefont {Pham},
  \citenamefont {Pulkin}, \citenamefont {Purwanto}, \citenamefont {Robinson},
  \citenamefont {Ronca}, \citenamefont {Sayfutyarova}, \citenamefont
  {Scheurer}, \citenamefont {Schurkus}, \citenamefont {Smith}, \citenamefont
  {Sun}, \citenamefont {Sun}, \citenamefont {Upadhyay}, \citenamefont {Wagner},
  \citenamefont {Wang}, \citenamefont {White}, \citenamefont {Whitfield},
  \citenamefont {Williamson}, \citenamefont {Wouters}, \citenamefont {Yang},
  \citenamefont {Yu}, \citenamefont {Zhu}, \citenamefont {Berkelbach},
  \citenamefont {Sharma}, \citenamefont {Sokolov},\ and\ \citenamefont
  {Chan}}]{PySCF2020}%
  \BibitemOpen
  \bibfield  {author} {\bibinfo {author} {\bibfnamefont {Q.}~\bibnamefont
  {Sun}}, \bibinfo {author} {\bibfnamefont {X.}~\bibnamefont {Zhang}}, \bibinfo
  {author} {\bibfnamefont {S.}~\bibnamefont {Banerjee}}, \bibinfo {author}
  {\bibfnamefont {P.}~\bibnamefont {Bao}}, \bibinfo {author} {\bibfnamefont
  {M.}~\bibnamefont {Barbry}}, \bibinfo {author} {\bibfnamefont {N.~S.}\
  \bibnamefont {Blunt}}, \bibinfo {author} {\bibfnamefont {N.~A.}\ \bibnamefont
  {Bogdanov}}, \bibinfo {author} {\bibfnamefont {G.~H.}\ \bibnamefont {Booth}},
  \bibinfo {author} {\bibfnamefont {J.}~\bibnamefont {Chen}}, \bibinfo {author}
  {\bibfnamefont {Z.-H.}\ \bibnamefont {Cui}}, \bibinfo {author} {\bibfnamefont
  {J.~J.}\ \bibnamefont {Eriksen}}, \bibinfo {author} {\bibfnamefont
  {Y.}~\bibnamefont {Gao}}, \bibinfo {author} {\bibfnamefont {S.}~\bibnamefont
  {Guo}}, \bibinfo {author} {\bibfnamefont {J.}~\bibnamefont {Hermann}},
  \bibinfo {author} {\bibfnamefont {M.~R.}\ \bibnamefont {Hermes}}, \bibinfo
  {author} {\bibfnamefont {K.}~\bibnamefont {Koh}}, \bibinfo {author}
  {\bibfnamefont {P.}~\bibnamefont {Koval}}, \bibinfo {author} {\bibfnamefont
  {S.}~\bibnamefont {Lehtola}}, \bibinfo {author} {\bibfnamefont
  {Z.}~\bibnamefont {Li}}, \bibinfo {author} {\bibfnamefont {J.}~\bibnamefont
  {Liu}}, \bibinfo {author} {\bibfnamefont {N.}~\bibnamefont {Mardirossian}},
  \bibinfo {author} {\bibfnamefont {J.~D.}\ \bibnamefont {McClain}}, \bibinfo
  {author} {\bibfnamefont {M.}~\bibnamefont {Motta}}, \bibinfo {author}
  {\bibfnamefont {B.}~\bibnamefont {Mussard}}, \bibinfo {author} {\bibfnamefont
  {H.~Q.}\ \bibnamefont {Pham}}, \bibinfo {author} {\bibfnamefont
  {A.}~\bibnamefont {Pulkin}}, \bibinfo {author} {\bibfnamefont
  {W.}~\bibnamefont {Purwanto}}, \bibinfo {author} {\bibfnamefont {P.~J.}\
  \bibnamefont {Robinson}}, \bibinfo {author} {\bibfnamefont {E.}~\bibnamefont
  {Ronca}}, \bibinfo {author} {\bibfnamefont {E.~R.}\ \bibnamefont
  {Sayfutyarova}}, \bibinfo {author} {\bibfnamefont {M.}~\bibnamefont
  {Scheurer}}, \bibinfo {author} {\bibfnamefont {H.~F.}\ \bibnamefont
  {Schurkus}}, \bibinfo {author} {\bibfnamefont {J.~E.~T.}\ \bibnamefont
  {Smith}}, \bibinfo {author} {\bibfnamefont {C.}~\bibnamefont {Sun}}, \bibinfo
  {author} {\bibfnamefont {S.-N.}\ \bibnamefont {Sun}}, \bibinfo {author}
  {\bibfnamefont {S.}~\bibnamefont {Upadhyay}}, \bibinfo {author}
  {\bibfnamefont {L.~K.}\ \bibnamefont {Wagner}}, \bibinfo {author}
  {\bibfnamefont {X.}~\bibnamefont {Wang}}, \bibinfo {author} {\bibfnamefont
  {A.}~\bibnamefont {White}}, \bibinfo {author} {\bibfnamefont {J.~D.}\
  \bibnamefont {Whitfield}}, \bibinfo {author} {\bibfnamefont {M.~J.}\
  \bibnamefont {Williamson}}, \bibinfo {author} {\bibfnamefont
  {S.}~\bibnamefont {Wouters}}, \bibinfo {author} {\bibfnamefont
  {J.}~\bibnamefont {Yang}}, \bibinfo {author} {\bibfnamefont {J.~M.}\
  \bibnamefont {Yu}}, \bibinfo {author} {\bibfnamefont {T.}~\bibnamefont
  {Zhu}}, \bibinfo {author} {\bibfnamefont {T.~C.}\ \bibnamefont {Berkelbach}},
  \bibinfo {author} {\bibfnamefont {S.}~\bibnamefont {Sharma}}, \bibinfo
  {author} {\bibfnamefont {A.~Y.}\ \bibnamefont {Sokolov}},\ and\ \bibinfo
  {author} {\bibfnamefont {G.~K.-L.}\ \bibnamefont {Chan}},\ }\bibfield
  {title} {\enquote {\bibinfo {title} {{Recent developments in the PySCF
  program package}},}\ }\href@noop {} {\bibfield  {journal} {\bibinfo
  {journal} {J. Chem. Phys.}\ }\textbf {\bibinfo {volume} {153}},\ \bibinfo
  {pages} {024109} (\bibinfo {year} {2020})}\BibitemShut {NoStop}%
\bibitem [{\citenamefont {Zhu}\ and\ \citenamefont
  {Chan}(2021{\natexlab{a}})}]{PhysRevX.11.021006}%
  \BibitemOpen
  \bibfield  {author} {\bibinfo {author} {\bibfnamefont {T.}~\bibnamefont
  {Zhu}}\ and\ \bibinfo {author} {\bibfnamefont {G.~K.-L.}\ \bibnamefont
  {Chan}},\ }\bibfield  {title} {\enquote {\bibinfo {title} {{Ab Initio Full
  Cell $GW+\mathrm{DMFT}$ for Correlated Materials}},}\ }\href
  {https://link.aps.org/doi/10.1103/PhysRevX.11.021006} {\bibfield  {journal}
  {\bibinfo  {journal} {Phys. Rev. X}\ }\textbf {\bibinfo {volume} {11}},\
  \bibinfo {pages} {021006} (\bibinfo {year} {2021}{\natexlab{a}})}\BibitemShut
  {NoStop}%
\bibitem [{\citenamefont {Zhu}\ and\ \citenamefont
  {Chan}(2021{\natexlab{b}})}]{doi:10.1021/acs.jctc.0c00704}%
  \BibitemOpen
  \bibfield  {author} {\bibinfo {author} {\bibfnamefont {T.}~\bibnamefont
  {Zhu}}\ and\ \bibinfo {author} {\bibfnamefont {G.~K.-L.}\ \bibnamefont
  {Chan}},\ }\bibfield  {title} {\enquote {\bibinfo {title} {{All-Electron
  Gaussian-Based G0W0 for Valence and Core Excitation Energies of Periodic
  Systems}},}\ }\href {https://doi.org/10.1021/acs.jctc.0c00704} {\bibfield
  {journal} {\bibinfo  {journal} {J. Chem. Theory Comput.}\ }\textbf {\bibinfo
  {volume} {17}},\ \bibinfo {pages} {727--741} (\bibinfo {year}
  {2021}{\natexlab{b}})}\BibitemShut {NoStop}%
\bibitem [{\citenamefont {Wilhelm}, \citenamefont {Del~Ben},\ and\
  \citenamefont {Hutter}(2016)}]{doi:10.1021/acs.jctc.6b00380}%
  \BibitemOpen
  \bibfield  {author} {\bibinfo {author} {\bibfnamefont {J.}~\bibnamefont
  {Wilhelm}}, \bibinfo {author} {\bibfnamefont {M.}~\bibnamefont {Del~Ben}},\
  and\ \bibinfo {author} {\bibfnamefont {J.}~\bibnamefont {Hutter}},\
  }\bibfield  {title} {\enquote {\bibinfo {title} {{GW in the Gaussian and
  Plane Waves Scheme with Application to Linear Acenes}},}\ }\href
  {https://doi.org/10.1021/acs.jctc.6b00380} {\bibfield  {journal} {\bibinfo
  {journal} {J. Chem. Theory Comput.}\ }\textbf {\bibinfo {volume} {12}},\
  \bibinfo {pages} {3623--3635} (\bibinfo {year} {2016})}\BibitemShut {NoStop}%
\bibitem [{\citenamefont {Caruso}\ \emph {et~al.}(2016)\citenamefont {Caruso},
  \citenamefont {Dauth}, \citenamefont {van Setten},\ and\ \citenamefont
  {Rinke}}]{Caruso2016}%
  \BibitemOpen
  \bibfield  {author} {\bibinfo {author} {\bibfnamefont {F.}~\bibnamefont
  {Caruso}}, \bibinfo {author} {\bibfnamefont {M.}~\bibnamefont {Dauth}},
  \bibinfo {author} {\bibfnamefont {M.~J.}\ \bibnamefont {van Setten}},\ and\
  \bibinfo {author} {\bibfnamefont {P.}~\bibnamefont {Rinke}},\ }\bibfield
  {title} {\enquote {\bibinfo {title} {{Benchmark of GW Approaches for the
  GW100 Test Set}},}\ }\href@noop {} {\bibfield  {journal} {\bibinfo  {journal}
  {J. Chem. Theory Comput.}\ }\textbf {\bibinfo {volume} {12}},\ \bibinfo
  {pages} {5076--5087} (\bibinfo {year} {2016})}\BibitemShut {NoStop}%
\bibitem [{\citenamefont {Lange}\ and\ \citenamefont
  {Berkelbach}(2018)}]{Lange2018}%
  \BibitemOpen
  \bibfield  {author} {\bibinfo {author} {\bibfnamefont {M.~F.}\ \bibnamefont
  {Lange}}\ and\ \bibinfo {author} {\bibfnamefont {T.~C.}\ \bibnamefont
  {Berkelbach}},\ }\bibfield  {title} {\enquote {\bibinfo {title} {{On the
  Relation between Equation-of-Motion Coupled-Cluster Theory and the GW
  Approximation}},}\ }\href@noop {} {\bibfield  {journal} {\bibinfo  {journal}
  {J. Chem. Theory Comput.}\ }\textbf {\bibinfo {volume} {14}},\ \bibinfo
  {pages} {4224--4236} (\bibinfo {year} {2018})}\BibitemShut {NoStop}%
\bibitem [{\citenamefont {Loos}, \citenamefont {Romaniello},\ and\
  \citenamefont {Berger}(2018)}]{doi:10.1021/acs.jctc.8b00260}%
  \BibitemOpen
  \bibfield  {author} {\bibinfo {author} {\bibfnamefont {P.-F.}\ \bibnamefont
  {Loos}}, \bibinfo {author} {\bibfnamefont {P.}~\bibnamefont {Romaniello}},\
  and\ \bibinfo {author} {\bibfnamefont {J.~A.}\ \bibnamefont {Berger}},\
  }\bibfield  {title} {\enquote {\bibinfo {title} {Green functions and
  self-consistency: Insights from the spherium model},}\ }\href
  {https://doi.org/10.1021/acs.jctc.8b00260} {\bibfield  {journal} {\bibinfo
  {journal} {J. Chem. Theory Comput.}\ }\textbf {\bibinfo {volume} {14}},\
  \bibinfo {pages} {3071--3082} (\bibinfo {year} {2018})},\ \bibinfo {note}
  {pMID: 29746773},\ \Eprint
  {https://arxiv.org/abs/https://doi.org/10.1021/acs.jctc.8b00260}
  {https://doi.org/10.1021/acs.jctc.8b00260} \BibitemShut {NoStop}%
\bibitem [{\citenamefont {Di~Sabatino}, \citenamefont {Loos},\ and\
  \citenamefont {Romaniello}(2021)}]{10.3389/fchem.2021.751054}%
  \BibitemOpen
  \bibfield  {author} {\bibinfo {author} {\bibfnamefont {S.}~\bibnamefont
  {Di~Sabatino}}, \bibinfo {author} {\bibfnamefont {P.-F.}\ \bibnamefont
  {Loos}},\ and\ \bibinfo {author} {\bibfnamefont {P.}~\bibnamefont
  {Romaniello}},\ }\bibfield  {title} {\enquote {\bibinfo {title} {Scrutinizing
  gw-based methods using the hubbard dimer},}\ }\href
  {https://doi.org/10.3389/fchem.2021.751054} {\bibfield  {journal} {\bibinfo
  {journal} {Front. Chem.}\ }\textbf {\bibinfo {volume} {9}} (\bibinfo {year}
  {2021}),\ 10.3389/fchem.2021.751054}\BibitemShut {NoStop}%
\bibitem [{\citenamefont {Monino}\ and\ \citenamefont
  {Loos}(2022)}]{Monino2022}%
  \BibitemOpen
  \bibfield  {author} {\bibinfo {author} {\bibfnamefont {E.}~\bibnamefont
  {Monino}}\ and\ \bibinfo {author} {\bibfnamefont {P.-F.}\ \bibnamefont
  {Loos}},\ }\bibfield  {title} {\enquote {\bibinfo {title} {{Unphysical
  discontinuities, intruder states and regularization in $GW$ methods}},}\
  }\href {https://doi.org/10.1063/5.0089317} {\bibfield  {journal} {\bibinfo
  {journal} {J. Chem. Phys.}\ }\textbf {\bibinfo {volume} {156}},\ \bibinfo
  {pages} {231101} (\bibinfo {year} {2022})}\BibitemShut {NoStop}%
\bibitem [{\citenamefont {Wei\ss{}e}\ \emph {et~al.}(2006)\citenamefont
  {Wei\ss{}e}, \citenamefont {Wellein}, \citenamefont {Alvermann},\ and\
  \citenamefont {Fehske}}]{RevModPhys.78.275}%
  \BibitemOpen
  \bibfield  {author} {\bibinfo {author} {\bibfnamefont {A.}~\bibnamefont
  {Wei\ss{}e}}, \bibinfo {author} {\bibfnamefont {G.}~\bibnamefont {Wellein}},
  \bibinfo {author} {\bibfnamefont {A.}~\bibnamefont {Alvermann}},\ and\
  \bibinfo {author} {\bibfnamefont {H.}~\bibnamefont {Fehske}},\ }\bibfield
  {title} {\enquote {\bibinfo {title} {{The kernel polynomial method}},}\
  }\href {https://link.aps.org/doi/10.1103/RevModPhys.78.275} {\bibfield
  {journal} {\bibinfo  {journal} {Rev. Mod. Phys.}\ }\textbf {\bibinfo {volume}
  {78}},\ \bibinfo {pages} {275--306} (\bibinfo {year} {2006})}\BibitemShut
  {NoStop}%
\bibitem [{\citenamefont {Bhatia}\ and\ \citenamefont
  {Rosenthal}(1997)}]{SylvesterSoln}%
  \BibitemOpen
  \bibfield  {author} {\bibinfo {author} {\bibfnamefont {R.}~\bibnamefont
  {Bhatia}}\ and\ \bibinfo {author} {\bibfnamefont {P.}~\bibnamefont
  {Rosenthal}},\ }\bibfield  {title} {\enquote {\bibinfo {title} {{How and Why
  to Solve the Operator Equation AX-XB = Y}},}\ }\href
  {https://londmathsoc.onlinelibrary.wiley.com/doi/abs/10.1112/S0024609396001828}
  {\bibfield  {journal} {\bibinfo  {journal} {Bulletin of the London
  Mathematical Society}\ }\textbf {\bibinfo {volume} {29}},\ \bibinfo {pages}
  {1--21} (\bibinfo {year} {1997})}\BibitemShut {NoStop}%
\bibitem [{\citenamefont {H\"{a}ser}\ and\ \citenamefont
  {Alml\"{o}f}(1992)}]{LaplaceMP2}%
  \BibitemOpen
  \bibfield  {author} {\bibinfo {author} {\bibfnamefont {M.}~\bibnamefont
  {H\"{a}ser}}\ and\ \bibinfo {author} {\bibfnamefont {J.}~\bibnamefont
  {Alml\"{o}f}},\ }\bibfield  {title} {\enquote {\bibinfo {title} {{Laplace
  transform techniques in M{\o}ller–Plesset perturbation theory}},}\ }\href
  {https://doi.org/10.1063/1.462485} {\bibfield  {journal} {\bibinfo  {journal}
  {J. Chem. Phys.}\ }\textbf {\bibinfo {volume} {96}},\ \bibinfo {pages}
  {489--494} (\bibinfo {year} {1992})}\BibitemShut {NoStop}%
\end{thebibliography}
\fi

\appendix
\setcounter{equation}{0}
\section*{Appendix A: Improved Numerical Quadrature}
\label{app:NumQuad}
\renewcommand{\theequation}{A\arabic{equation}}

While the integral expression for $\tilde{\eta}^{(0)}$ derived in Eq.~\ref{eq:NI_efficient} in the main text is sufficient for numerical quadrature in $\mathcal{O}[N^4]$ scaling, it can be refined by deducting large and/or slowly-decaying contributions which can be efficiently integrated separately, in order to minimize the number of quadrature points which are required to obtain a given overall accuracy. This again follows many of the developments in the NI-RPA approaches for the correlation energy in the literature due to the commonalities in their forms \cite{Furche2010}, however there are also important differences. 

We can first consider a mean-field contribution to the integral component of $\tilde{\eta}^{(0)}$, i.e. setting the Coulomb interaction to zero in the expression for $[(\mat{A}-\mat{B})(\mat{A}+\mat{B})]^{\frac{1}{2}}$. This just leaves a contribution from the irreducible polarizability, which can be analytically integrated, as
\begin{align}
\mat{D}\mat{T} &= (\mat{D}^2)^{\frac{1}{2}} \mat{T}
&= \frac{1}{\pi} \int_{-\infty}^\infty \left(\mat{I}-z^2(\mat{D}^2+z^2\mat{I})^{-1}\right) \mat{T} dz ,
\end{align}
where we have used the integral form of the matrix square root again (Eq.~\ref{eq:matsqrt}). We can add and subtract these different forms within Eq.~\ref{eq:NI_efficient}, to obtain
\begin{equation}
\tilde{\eta}^{(0)} = \mat{D}\mat{T} + \frac{1}{\pi}\int_{-\infty}^{\infty} z^2 \mat{F}(z) \mat{S}_L(\mat{I}+\mat{Q}(z))^{-1} \mat{S}_R^T \mat{F}(z) \mat{T} dz , \label{eq:NI_removeMF}
\end{equation}
which substantially reduces the magnitude of the numerically integrated component of $\tilde{\eta}^{(0)}$. However, we can further improve on this, by increasing the rate of decay of the remaining integrand with respect to $z$. It can be seen from Eqs~\ref{eq:F}-\ref{eq:Q} that both $\mat{F}(z)$ and $\mat{Q}(z)$ decay as $z^{-2}$ in the large-$z$ limit. Series expanding $[\mat{I}+\mat{Q}(z)]^{-1} \sim \mat{I}-\mat{Q}(z) + \mat{Q}^2(z) - \dots$, we find that the leading order decay of the integrand in Eq.~\ref{eq:NI_removeMF} is $\mathcal{O}[z^{-2}]$. We consider this contribution in isolation, noting that if it can be removed from the numerical integration, the next leading order will result in the integrand decaying at an improved $\mathcal{O}[z^{-4}]$ rate.

This leading-order contribution to the integral of Eq.~\ref{eq:NI_removeMF} can be written as
\begin{equation}
\mat{C}_2 = \frac{1}{\pi} \int_{-\infty}^{\infty} z^2 \mat{F}(z)\mat{S}_L \mat{S}_R^T \mat{F}(z) \mat{T} dz .
\end{equation}
This can be analytically integrated via the residue theorem, to give 
\begin{equation}
\mat{C}_2 = \left( (\mat{S}_L \mat{S}_R^T) \circ \mat{E} \right) \mat{T} , \label{eq:C2_analytic}
\end{equation}
where $\circ$ denotes the Hadamard or element-wise product, and we define $\mat{E}$ as
\begin{equation}
\mat{E}_{ia,jb} = (\mat{D}_{ia,ia} + \mat{D}_{jb,jb})^{-1} . \label{eq:E}
\end{equation}
Writing Eq.~\ref{eq:C2_analytic} with explicit indices for clarity it can be seen as a second-order direct-MP2-like contribution of
\begin{equation}
(\mat{C}_{2})_{ia,P} = \sum_{jb} \frac{\sum_Q (\mat{S}_L)_{ia,Q} (\mat{S}_R)_{jb,Q}}{\epsilon_a+\epsilon_b-\epsilon_i-\epsilon_j} \mat{T}_{jb,P} . \label{eq:C2_analytic_2}
\end{equation}
However, in this form it is unfortunately unable to be efficiently evaluated in $\mathcal{O}[N^4]$ time (instead scaling as $\mathcal{O}[N^5]$). We note that a similar subtraction of this leading-order term in the computation of the RPA correlation energy is in contrast efficiently computable, due to the presence of a cyclically-invariant trace operation \cite{Furche2010}.

We therefore take a different approach, by noting that we can form an algebraic Sylvester equation, as
\begin{equation}
((\mat{S}_L \mat{S}_R^T) \circ \mat{E}) \mat{D} + \mat{D} ((\mat{S}_L \mat{S}_R^T) \circ \mat{E}) = \mat{S}_L \mat{S}_R^T ,  \label{eq:Sylvester}
\end{equation}
which can be verified by substitution of Eq.~\ref{eq:E}. Writing the matrix of interest we wish to efficiently compute as $\mat{N}=(\mat{S}_L \mat{S}_R^T) \circ \mat{E}$, this Sylvester equation has an integral solution of the form \cite{SylvesterSoln}
\begin{equation}
\mat{N} = \int_0^{\infty} e^{-t \mat{D}} \mat{S}_L \mat{S}_R^T e^{-t \mat{D}} dt . \label{eq:SylvesterSoln}
\end{equation}
This can be verified via substitution into Eq.~\ref{eq:Sylvester}, as
\begin{align}
\mat{D}\mat{N}+\mat{N}\mat{D} &= \int_0^{\infty} \left( \mat{D} e^{-t \mat{D}} \mat{S}_L \mat{S}_R^T e^{-t \mat{D}} \right. \nonumber \\ 
&\qquad + \left. e^{-t \mat{D}} \mat{S}_L \mat{S}_R^T e^{-t \mat{D}} \mat{D} \right) dt \\
&= - \int_0^{\infty} \frac{d}{dt} \left( e^{-t \mat{D}} \mat{S}_L \mat{S}_R^T e^{-t \mat{D}} \right) dt \\
&= - \left[e^{-t \mat{D}} \mat{S}_L \mat{S}_R^T e^{-t \mat{D}} \right]_0^{\infty} \\
&= \mat{S}_L \mat{S}_R^T .
\end{align}
The numerical integration of Eq.~\ref{eq:SylvesterSoln} can therefore be used to compute the leading order term in the numerical integration of Eq.~\ref{eq:NI_removeMF}. We note that this is essentially analogous to a matrix-generalization of the Laplace transform to the direct second-order diagram, which in frequency space involves energy denominators of the type given in Eq.~\ref{eq:E} \cite{LaplaceMP2}. It may appear as though we have swapped one numerical integral for another, however the transformation to the form given in Eq.~\ref{eq:SylvesterSoln} gives an integrand which is exponentially rather than quadratically decaying, leading to a form which can be very effectively integrated via Gauss--Laguerre quadrature with exponential convergence and only a small number of samples in practice.

The final expression for the numerical integration of the RPA zeroth-order dd-response moment is
\begin{align}
\tilde{\eta}^{(0)} &= \mat{D}\mat{T} \nonumber \\
&+ \int_0^{\infty} e^{-t \mat{D}} \mat{S}_L \mat{S}_R^T e^{-t \mat{D}} \mat{T} dt \nonumber \\
&+ \frac{1}{\pi}\int_{-\infty}^{\infty} z^2 \mat{F}(z) \mat{S}_L\left((\mat{I}+\mat{Q}(z))^{-1}-\mat{I} \right) \mat{S}_R^T \mat{F}(z) \mat{T} dz . \label{eq:final_NI_zero_mom}
\end{align}
The integrand of the first integral decays exponentially and is efficiently computed with Gauss-Laguerre quadrature, with the second integral (decaying as $\mathcal{O}[z^{-4}]$) is evaluated with Clenshaw--Curtis quadrature.

\vspace{.5cm}
\setcounter{equation}{0}
\section*{Appendix B: Numerical Integration Error Estimates}
\label{app:NumQuadErr}
\renewcommand{\theequation}{B\arabic{equation}}

We make use of two separate approaches to estimate the error in $\tilde{\eta}^{(0)}$ evaluated through numerical integration, which we can use as a check for convergence of this key intermediate.
These arise from quite different considerations, and provide complementary estimates.
In the following we will denote the moment estimate resulting from a numerical integration with $n_p$ points as $\tilde{\eta}^{(0)}_{n_p}=\tilde{\eta}^{(0)}_{\infty} + \Delta\tilde{\eta}^{(0)}_{n_p}$.

For our first estimate, we begin by noting from Eq.~\ref{eq:zero_mom} that the exact $\tilde{\eta}^{(0)}$ will satisfy the relation
\begin{equation}
    (\tilde{\eta}^{(0)}_\infty)^T(\mat{A} + \mat{B})\tilde{\eta}^{(0)}_\infty = \mat{V}^T (\mat{A} - \mat{B}) \mat{V}.
\end{equation}
We can then take the deviation from this relation at finite integration samples as an error estimate, as $X(n_p)$. Expanding the estimated quantity in terms of the exact result and error gives
\begin{align}
    X(n_p) =& (\tilde{\eta}^{(0)}_{n_p})^T(\mat{A} + \mat{B})\tilde{\eta}^{(0)}_{n_p} - \mat{V}^T (\mat{A} - \mat{B}) \mat{V} \\
    =& (\Delta\tilde{\eta}^{(0)}_{n_p})^T(\mat{A} + \mat{B})\tilde{\eta}^{(0)}_{n_p} + (\tilde{\eta}^{(0)}_{n_p})^T(\mat{A}+ \mat{B})\Delta\tilde{\eta}^{(0)}_{n_p} \nonumber\\
    &- (\Delta\tilde{\eta}^{(0)}_{n_p})^T(\mat{A} + \mat{B})\Delta\tilde{\eta}^{(0)}_{n_p}.
\end{align}
We can then explicitly evaluate $||X(n_p)||_2$ in $\mathcal{O}[N^4]$ time under the same conditions as the rest of this manuscript, and apply the properties of the Frobenius norm to write
\begin{equation}
    \begin{aligned}
        ||X(n_p)||_2 \leq & 2 ||\Delta\tilde{\eta}^{(0)}_{n_p}||_2 ||(\mat{A} + \mat{B})\tilde{\eta}^{(0)}_{n_p}||_2  \nonumber\\
        &- ||\Delta\tilde{\eta}^{(0)}_{n_p}||_2^2 ||(\mat{A} + \mat{B})||_2.
    \end{aligned}
\end{equation}
Writing $||\Delta\tilde{\eta}^{(0)}_{n_p}||_2=x$, the error is therefore rigorously bound by the solutions satisfying the quadratic inequality,
\begin{equation}
    x^2 - 2 \frac{||(\mat{A} + \mat{B})\tilde{\eta}^{(0)}_{n_p}||_2}{||(\mat{A} + \mat{B})||_2} x + \frac{||X(n_p)||_2}{||(\mat{A} + \mat{B})||_2} \leq 0. \label{eq:lowerbound_err}
\end{equation}
In practice, while we find that the two roots to this equation rigorously bound the error in the quadrature, their values are not a tight bound on the true error (in particular, the upper bound is excessively large to be used as an estimate, though the lower bound on the error that it provides is reasonable, shown in Fig.~\ref{fig:NumQuadConv} of the main text as `Lower Bound'). Combined with the $\mathcal{O}[N^4]$ scaling of evaluating $X(n_p)$ this makes this estimate of limited practical use, and we seek an improved estimate, that provides a tighter upper bound on the true error.

To do this we take advantage of the nested nature of the Clenshaw--Curtis quadrature within the integral providing the dominant error contribution.
Using this, we can obtain estimates at both one half and one quarter of the current number of integration points with no additional cost.
We show how this can give an estimate of the error in $\tilde{\eta}^{(0)}_{n_p}\left(\mat{A} + \mat{B}\right)$, which we can relate to the error in $\tilde{\eta}^{(0)}_{n_p}$.
Parameterising the exponential convergence of the $L_2$ norm error as $||\Delta\tilde{\eta}^{(0)}_{n_p}||_2 = \alpha e^{-\beta n_p}$, we can rigorously relate the differences between estimates at different levels of sampling, as
\begin{align}
    \Delta_{n_1, n_2} &= ||\tilde{\eta}^{(0)}_{n_1} - \tilde{\eta}^{(0)}_{n_2}||_2 \\
    &\leq ||\Delta\tilde{\eta}^{(0)}_{n_1}||_2 + ||\Delta\tilde{\eta}^{(0)}_{n_2}||_2 \label{eq:l2_model_v1_bound} \\
    &\approx \alpha  \left(e^{-\beta n_1} + e^{-\beta n_2} \right).
    \label{eq:l2_err_model_quadratic}
\end{align}
We seek to estimate the value of $||\Delta\tilde{\eta}^{(0)}_{4n_p}||_2$ given the values $\Delta_{4n_p, 2n_p}$ and $\Delta_{4n_p, n_p}$ which are computed as intermediates in the nested quadrature of $4n_p$ points.

Setting $x=e^{-\beta n_p}$, and using the approximation defined in Eq.~\ref{eq:l2_err_model_quadratic} for $\Delta_{4n_p, 2n_p}=\alpha(x^4+x^2)$ and $\Delta_{4n_p, n_p}=\alpha(x^4+x)$, eliminating the factor of $\alpha$ we find
\begin{equation}
    (\Delta_{4n_p, n_p} - \Delta_{4n_p, 2n_p}) x^3 + \Delta_{4n_p, n_p} x - \Delta_{4n_p, 2n_p} = 0.
\end{equation}
The smallest positive, real solution to this is used to estimate the error (given as $\alpha x^4$) as
\begin{equation}
    ||\Delta\tilde{\eta}^{(0)}_{4n_p}||_2 \approx \Delta_{4n_p, 2n_p} (1 + x^{-2})^{-1} . \label{eq:cubic_fit_err_estimate}
\end{equation}
This estimator provides a consistent and systematic overestimate of the error, allowing us to use this as a reliable upper bound to the true error of the quadrature. This property, combined with only requiring the quantities $\Delta_{4n_p, 2n_p}$ and $\Delta_{4n_p, n_p}$, which are unavoidable intermediates of the nested integration, makes this essentially a computationally free estimate of the error for a given number of grid points, and therefore as a reliable check for convergence. If the resulting error estimate is too large, the number of grid points can be increased by a factor of two, and all points already evaluated can be reused in the next estimate. The convergence of both error estimates can be seen in Fig.~\ref{fig:NumQuadConv} of the main text, with the result of Eq.~\ref{eq:cubic_fit_err_estimate} denoted `Nested Fit', while the lower bound from the solution of Eq.~\ref{eq:lowerbound_err} denoted `Lower Bound'.

\end{document}